\documentclass[11pt]{article}

\usepackage[utf8]{inputenc}

\usepackage{amsmath}

\usepackage{amssymb}
\usepackage{color,soul}

\usepackage{amsthm}
\usepackage[authoryear]{natbib}
\usepackage{booktabs}
\usepackage{xcolor}
\usepackage{comment}
\usepackage{float}
\bibpunct{(}{)}{,}{a}{}{;}
\usepackage{fullpage}
\usepackage{graphicx}
\usepackage{subcaption}
\usepackage{tikz}
\usetikzlibrary{positioning}
\usepackage{tabularx}

\usepackage{smartdiagram}
\usepackage{multirow}
\usepackage{titling} % repeat title
\usepackage{multirow}
\usepackage[unicode=true,psdextra,hidelinks]{hyperref}
\usepackage{geometry}
\usepackage{listings}
\usepackage{color}
\usepackage{adjustbox}
\usepackage{listings}
\usepackage{color} %red, green, blue, yellow, cyan, magenta, black, white
\usepackage{multirow}
\usepackage{soul}
\usepackage{tocloft}
\usepackage{titlesec}
\newcommand{\sym}[1]{\rlap{#1}}
\definecolor{mygreen}{RGB}{28,172,0} % color values Red, Green, Blue
\definecolor{mylilas}{RGB}{170,55,241}
\usepackage{xurl}
\usepackage{booktabs}
\usepackage{rotating}

\usepackage{xcolor}

\definecolor{codegreen}{rgb}{0,0.6,0}
\definecolor{codegray}{rgb}{0.5,0.5,0.5}
\definecolor{codepurple}{rgb}{0.58,0,0.82}
\definecolor{backcolour}{rgb}{0.95,0.95,0.92}

\lstdefinestyle{mystyle}{
    backgroundcolor=\color{backcolour},   
    commentstyle=\color{codegreen},
    keywordstyle=\color{magenta},
    numberstyle=\tiny\color{codegray},
    stringstyle=\color{codepurple},
    basicstyle=\ttfamily\footnotesize,
    breakatwhitespace=false,         
    breaklines=true,      
    captionpos=b,         
    keepspaces=true,      
    numbers=left,         
    numbersep=5pt,       
    showspaces=false,     
    showstringspaces=false,
    showtabs=false,       
    tabsize=2
}

\usepackage{setspace}

\begin{document}

\title{Strategic Response of News Publishers to Generative AI}
% For anonymous submission, don't add author info.

% \title{The Impact of LLMs on Online News Consumption and Production 
% %\thanks{We thank for their helpful comments. All errors are our own. }
% }
\vspace{-40pt}
\author{Hangcheng Zhao\thanks{Rutgers Business School. \href{mailto:hangcheng.zhao@rutgers.edu}{hangcheng.zhao@rutgers.edu}} 
\and Ron Berman\thanks{The Wharton School of the University of Pennsylvania.  \href{mailto:ronber@wharton.upenn.edu}{ronber@wharton.upenn.edu}} }
%\href{https://www.dropbox.com/scl/fi/hkwgngwrefyir53q3cyg7/Impact\_of\_LLM\_on\_Publishers.pdf?rlkey=5tl815poeb8aheldmm36bf9wb\&st=x5iplxis\&dl=0}{\textcolor{blue}{Click Here for the Latest Version}}
% }
% \date{December 2025\\[2ex]  % Adds vertical space after the date}
% }
\lstset{language=Matlab,%
    %basicstyle=\color{red},
    breaklines=true,%
    morekeywords={matlab2tikz},
    keywordstyle=\color{blue},%
    morekeywords=[2]{1}, keywordstyle=[2]{\color{black}},
    identifierstyle=\color{black},%
    stringstyle=\color{mylilas},
    commentstyle=\color{mygreen},%
    showstringspaces=false,%without this there will be a symbol in the places where there is a space
    numbers=left,%
    numberstyle={\tiny \color{black}},% size of the numbers
    numbersep=9pt, % this defines how far the numbers are from the text
    emph=[1]{for,end,break},emphstyle=[1]\color{red}, %some words to emphasise
    %emph=[2]{word1,word2}, emphstyle=[2]{style},    
}

\begin{titlepage}
\maketitle
% Optionally include a table of contents
\setcounter{tocdepth}{1} % adjust to 1 if desired

\end{titlepage}

\setcounter{page}{1}
\begin{center}
\Large
\thetitle
\end{center}
\thispagestyle{empty}
\setcounter{page}{1}

% % \vspace{1cm}

% % % \tableofcontents

% % \maketitle
% % \thispagestyle{empty}

% % \newpage
% % \setcounter{page}{1}
% % \begin{center}
% % \Large
% % \thetitle
% % \end{center}

% \setcounter{page}{1}
% \vspace{-20pt}
\begin{abstract}
%Large language models (LLMs) change how consumers acquire information online; their bots also crawl news publishers' websites for training data and to answer consumer queries; and they provide tools that can lower the cost of content creation. These changes lead to predictions of adverse impact on news publishers in the form of lowered consumer demand, reduced demand for newsroom employees, and an increase in news ``slop.'' Consequently, some publishers strategically responded by blocking LLM access to their websites using the \textit{robots.txt} file standard. 
Generative AI %affects how consumers acquire information online; LLM bots also crawl news publishers' websites for training data and to answer consumer queries; and they provide tools that can lower the cost of content creation. These changes
can adversely impact news publishers by lowering consumer demand. It can also reduce demand for newsroom employees, and increase the creation of news ``slop.'' However, it can also form a source of traffic referrals and an information-discovery channel that increases demand.
%News publishers can respond strategically in several ways, including blocking LLM access, reducing headcount, or changing their content.
We use high-frequency granular data to analyze the strategic response of news publishers to the introduction of Generative AI. Many publishers strategically blocked LLM access to their websites using the \textit{robots.txt} file standard. 
Using a difference-in-differences approach, we find that large publishers who block GenAI bots experience reduced website traffic compared to not blocking. In addition, we find that large publishers shift toward richer content that is harder for LLMs to replicate, without increasing text volume. Finally, we find that the share of new editorial and content-production job postings rises over time.
Together, these findings illustrate the levers that publishers choose to use to strategically respond to competitive Generative AI threats, and their consequences.
\end{abstract}
% \clearpage

\setcounter{page}{1}
\doublespacing
\section{Introduction}
\label{sec:intro}
Generative AI (GenAI) and large language models (LLMs) are reshaping how consumers discover and consume information online. Unlike search-based discovery intermediaries (e.g., Classic Google Search) that primarily redirect users to publishers through links, LLM-mediated interfaces can directly provide answers and summaries, potentially substituting away from click-through visits.\footnote{The Economist, ``AI is killing the web. Can anything save it?'' \url{https://www.economist.com/business/2025/07/14/ai-is-killing-the-web-can-anything-save-it}, accessed Dec 18, 2025.} Recent evidence shows that LLM adoption reduces traditional search activity and downstream browsing to smaller sites, and users click external links less often when AI summaries appear in search results.\footnote{Pew Research Center 2025, ``Google users are less likely to click on links when an AI summary appears in the results,'' \url{https://www.pewresearch.org/short-reads/2025/07/22/google-users-are-less-likely-to-click-on-links-when-an-ai-summary-appears-in-the-results/}, accessed Dec 18, 2025.} These models also require vast amounts of diverse data for model training and inference purposes. Since news publishers are a primary source of training data for LLMs and their online business model depends heavily on discovery and referral from intermediaries, they are at the forefront of online content providers that experience the impact of GenAI.

In this paper, we ask how news publishers respond to Generative AI threats and what the impact of their response is. There are a variety of ways news publishers can respond. Examples include blocking LLM access using the ``robots.txt'' standard,\footnote{RFC 9309 \url{https://www.rfc-editor.org/rfc/rfc9309.html} accessed Dec 18, 2025.} with the hope that LLMs will not copy their content. Publishers could alter how much content they produce in order to attract more customers or appear more often in LLM results, they could also alter the amount of rich media in their content, making it harder to replicate,\footnote{Reuters Journalism, media, and technology trends and predictions 2026, \url{https://reutersinstitute.politics.ox.ac.uk/journalism-media-and-technology-trends-and-predictions-2026}, accessed April 5, 2026.} and they could also control their editorial staff headcount to lower costs.
We empirically analyze these strategic responses of online news publishers to the introduction of LLMs from November 2022 (the introduction of OpenAI's ChatGPT) to May 2024 (the introduction of Google's AI Overview). We focus on this period for a few reasons: (1) There was an apparent traffic decline to news publishers, making it likely that they will respond; (2) many large publishers experimented with blocking LLMs and changing content structure; and (3) the large effect from Google's introduction of AI Overview does not contaminate the data.\footnote{Google introduced AI Overview in May 2024 and expanded it substantially in August 2024. Many content providers documented a decrease in website visits, mostly due to decrease in referrals from Google after AI Overview's introduction. For example see: \url{https://growtika.com/blog/tech-media-collapse}, accessed April 5, 2026. } 

%
%most news publishers have made substantial 

%Publishers can choose to block LLM access to their content, hoping that 

%, some publishers have been quick to respond by blocking LLM access to their websites using the \textit{robots.txt} standard. The impact of this blocking can be three-fold: it can reduce bot traffic from LLMs used to scan the websites to train models; it can reduce bot traffic by LLMs used to answer user queries; and it can reduce referral traffic by LLMs sending users to visit the news publisher. Publishers can also try to respond by lowering headcount which would save costs, or by increasing content production using LLMs to attract more consumers.

We note that it is unclear a priori how to best respond to the introduction of a technology such as LLMs. Blocking LLMs, for example,  can have multiple effects. Benefits include saving on costs because of reduced bot traffic and increased visits from consumers who are looking for original and fresh content that LLMs might miss as a result of blocking. Disadvantages include not being mentioned as a source of LLM query response, which would lower the brand equity of the news source, leading to a long-term decline in traffic. Not appearing in results can also lower direct referral traffic by allowing other competitors to take these referrals.
%that scans the website, but also has the disadvantage of not being mentioned as a source of query results which will lower brand exposure to the content website and will allow competitors to be mentioned more often. 

Our paper asks the following research questions: (1) How do news publishers respond to the introduction of Generative AI?
(2) Is blocking an effective strategy and what is its impact?
(3) Do news publishers choose to compete by producing more content, or differentiate by altering how content is being delivered and displayed?

%whether blocking access is an effective strategic response; and (3) how LLM tools affect the supply of news content by publishers through hiring decisions and changes in the quantity and format of content they produce.%is an economically meaningful strategic action, and how publishers adjust along labor and content format and quantity production.}

%Newspaper publishers provide a particularly informative setting because their demand and business models depend heavily on discovery and referral. A large literature on news aggregation, intermediation, and platform competition shows that shifts in discovery intermediaries can materially affect publishers’ traffic and strategic outcomes \citep{chiou2017content,calzada2020news,athey2021impact,jeon2016news,seamans2014responses,djourelova2025impact}. The LLM shock adds a new short-run strategic response: technical access control through \textit{robots.txt} (the Robots Exclusion Protocol). This choice is at the center of current debates about content reuse, licensing, and welfare under alternative copyright regimes \citep{cage2020production,gans2024copyright,yang2024generative} despite evidence indicates that bot compliance with \textit{robots.txt} is imperfect and heterogeneous \citep{kim2025scrapers,fletcher2024many}.

We address these questions by constructing a high-frequency, publisher panel that links daily traffic to strategic blocking rules for LLM crawlers, publishers' page structure, and publishers' hiring. We combine daily domain-level visits from two website traffic tracking services, SimilarWeb and Semrush, with historical \textit{robots.txt} files and HTML snapshots from the HTTP Archive and employer-linked job postings from Revelio. We also use household-level browsing history data from the Comscore Web-Behavior Panel.\footnote{\url{https://wrds-www.wharton.upenn.edu/pages/about/data-vendors/comscore/}} 

Below, we summarize our findings:
\paragraph{Publisher traffic declines.} The introduction of ChatGPT in November 2022 elicited predictions of a substantial drop in publisher traffic. However, it is unclear whether news publishers experienced actual traffic declines and how quickly. We use the Pruned Exact Linear Time (PELT) algorithm \citep{killick2012optimal} to analyze the time series of log-daily visits tracked by SimilarWeb and referral counts from Semrush\footnote{Semrush, \url{https://www.semrush.com}}. We document a moderate decline in direct traffic before May 2024, while organic search referrals remain stable. After Google introduced AI Overviews in May 2024 and a core algorithm update in August 2024,\footnote{Google Search Central, ``August 2024 core update,'' \url{https://developers.google.com/search/blog/2024/08/august-2024-core-update}, accessed February 7, 2026.} both direct and organic referral traffic decline. Because blocking adoption was concentrated in mid-to-late 2023, well before the impact of AI Overview, we restrict our main analysis to the pre-May 2024 period.

 %We employ a multiple change-point detection approach \citep{killick2012optimal} and identify persistent breaks in traffic patterns, most prominently in November 2023 and August 2024, after which traffic levels shift to a lower level. %We estimate a synthetic difference-in-differences (SDID) \citep{arkhangelsky2021synthetic} model using log traffic in the six months before and six months after each detected breakpoint, with the top 100 retail websites as the control group. Following the August 2024 break, traffic to news publishing websites decreased by approximately 13.2\% relative to the control group. The point estimates for the November 2023 break are also negative but statistically insignificant. 

\paragraph{Many publishers block LLM crawlers, and blocking reduces total traffic; there is a negative, but imprecise effect on human traffic.} 
One lever that publishers have to respond to declining traffic is access control: they can block GenAI crawlers to prevent their content from being scraped for model training and query answering. We evaluate the effectiveness of this strategy using a staggered difference-in-differences analysis \citep{callaway2021difference}  of the period before May 2024, where identification is cleanest. Publishers could declare that crawling is not allowed in their \textit{robots.txt} file, which instructs web crawlers on what they are allowed and not allowed to access. We identify when each publisher first disallowed Generative AI-related crawlers using the HTTP Archive.\footnote{\url{https://httparchive.org/}} We find that news publishers choose to block LLM access more often than non-news websites. About 75\% of the top publishers blocked LLM crawlers at different times starting mid-2023.\footnote{The blocking was focused primarily on bot crawlers, and not on human traffic that was using a different HTTP user-agent} We use the staggered blocking pattern in a difference-in-differences analysis that compares blocking publishers to not-yet-blocking and never-blocking publishers to estimate the effect of blocking LLM crawlers on total traffic. We find a 7\% post-blocking decline in weekly visits measured by SimilarWeb  or Semrush within the 6 weeks after blocking. Using Comscore’s Web-Behavior panel of human browsing history, we find a similarly sized decline of approximately 7\% in weekly publisher visits, although the estimates are less precise, possibly because the panel is smaller. These results suggest that blocking LLM crawlers may have negative effects on publishers: it is followed by lower total traffic and potentially lower human traffic, not merely the mechanical removal of bot visits. Indeed, we observe that some publishers unblocked LLM crawlers in 2024. When we extend the analysis to smaller publishers with lower traffic in the Comscore and Semrush data, we find similarly negative but imprecisely estimated effects. Two channels can explain this decline: reduced brand exposure in LLM  responses as the source for information, which lowers direct visits, and lost referral clicks from LLM citations. The former is likely dominant given that LLM-driven referral traffic was small before May 2024.

\paragraph{Publishers do not scale up textual production; they shift toward richer pages and embedded components.} A recently emerging phenomenon associated with LLM usage is the production of content slop.\footnote{``Digital content of low quality that is produced usually in quantity by means of artificial intelligence'', Merriam-Webster.com Dictionary. Retrieved December 21, 2025, from \url{https://www.merriam-webster.com/dictionary/slop}.} We use page-structure metrics from the HTTP Archive\footnote{\url{https://httparchive.org/}} and URL histories from the Internet Archive\footnote{\url{https://web.archive.org/}} to estimate whether news publishers create more content or whether they alter their content after the introduction of LLMs. We find no evidence that publishers respond by expanding the number of sections or by accelerating growth in text and article-related URLs. Instead, pages exhibit substantial increases in interactive elements (68.1\%) and in advertising and targeting technology components (50.1\%) relative to retail websites, with growth concentrated in image-related URLs. This pattern is consistent with work linking monetization design to content incentives \citep{sun2013ad,chiou2013paywalls,lambrecht2017fee} and with evidence that multimedia and interactivity shape user engagement \citep{chung2008interactive,calder2009experimental,ghose2025tell}. Such a response is also consistent with a strategy of differentiation. As LLMs cannot easily replicate visual content without manipulation that triggers negative consumer response \citep{epstein2023art}, by choosing to emphasize visual content, news publishers are able to keep attracting consumers. 

\paragraph{There is no short-term contraction in newsroom hiring relative to other roles.} Another possible response by publishers to a decline in traffic and reduced cost of content production is to reduce their newsroom headcount. We use employer-linked job postings to track publishers’ hiring by occupation. We test whether publishers disproportionately reduce demand for editorial/content-production roles after the introduction of LLMs. We find no such pattern: editorial postings do not exhibit a discrete post-GenAI decrease. The share of postings not only does not decline but rather increases, suggesting that publishers do not respond to LLMs primarily by reducing newsroom headcount.

To summarize, we analyze how news publishers respond to GenAI. We find that many publishers block LLM crawlers, but that can adversely affect their traffic. We find evidence that content is becoming richer, but no evidence that more textual content is produced. We also find no evidence of reduced hiring in content creation and editing positions. These findings suggest that GenAI is not simply displacing traditional news production in the short run, but is prompting strategic adjustments in access, content format, and hiring. The impact of these adjustments, however, sometimes has surprising strategic effects that our analysis helps to identify. These findings offer guidance for decision makers considering how to  strategically respond to GenAI.

%The remainder of the paper describes the data construction and empirical strategy, documents traffic patterns, estimates the effects of LLM blocking on total and human traffic, characterizes changes in page structure, multimedia richness and advertisement density, and analyzes hiring composition.

\section{Contribution and Related Work}
\label{sec:literature}

Our central contribution is to show that the most widely adopted publisher response to generative AI, blocking LLM crawlers via \textit{robots.txt}, does not stop traffic declines and may accelerate it. Prior work establishes that news aggregators and discovery platforms can materially reshape publishers' traffic and strategic choices \citep{chiou2017content, calzada2020news, athey2021impact}, and recent evidence shows that LLM adoption specifically reduces downstream browsing and click-through to content sites \citep{padilla2025impact}. A parallel literature analyzes how GenAI changes the economics of content reuse, licensing, and copyright \citep{cage2020production, gans2024copyright, yang2024generative}, but this work does not study whether publishers' actual defensive actions are effective. We fill this gap by treating blocking as an observable strategic action and estimating its causal effect %using a staggered difference-in-differences design \citep{callaway2021difference} 
across three independently constructed traffic datasets. Our finding that blocking is followed by a decline in visits, including in a human-only browsing panel, documents a previously unmeasured managerial tradeoff: technical access control is easy to implement but can have adverse effects.
 
The finding that publishers emphasize rich content over creating more content is  consistent with a differentiation strategy: investing in formats that LLMs cannot easily replicate. Past research showed that multimedia and interactivity drive user engagement in news settings \citep{chung2008interactive, ghose2025tell}, and our research documents how publishers use it strategically. We also find no disproportionate contraction in editorial hiring, which adds publisher-specific evidence to a growing literature that finds limited near-term labor displacement from GenAI despite broad task exposure \citep{eloundou2024gpts, brynjolfsson2025generative, humlum2025large, demirci2025ai} and to work documenting reduced contributions in online knowledge communities after GenAI adoption \citep{burtch2024consequences, lyu2025wikipedia}. %Our findings suggest that publishers are adapting to generative AI not by producing more text but by differentiating through visual content that LLMs cannot easily replicate or substitute.

\section{Data}
\label{sec:data}
We used multiple data sources to construct a publisher panel by combining high-frequency measures of website traffic, page structure and content proxies, and hiring.

\subsection{Website traffic}
\paragraph{SimilarWeb} Our first traffic measure comes from SimilarWeb accessed via the Dewey Data Platform.\footnote{Similarweb. (2025). Website Traffic Visits [Dataset]. Dewey Data. \url{https://doi.org/10.82551/PRDY-D115}} The data provide daily, domain-level estimates of total visits (desktop and mobile) for each website in our sample from January 1, 2019 through February  28, 2026. For each domain–day, we observe the estimated number of daily worldwide visits. We use these data to characterize aggregate traffic patterns and to study traffic responses around changes in publisher GenAI bots policies.
\paragraph{Comscore Web-Behavior Panel} To measure human browsing, we use the Comscore Web-Behavior Panel available through WRDS for 2022–2024. This panel records desktop browsing behavior for a large sample of U.S. households. For each household–URL event, we observe the domain URL and timestamp. We aggregate these events to construct publisher-level measures of human consumer visits.

\paragraph{Semrush} We also collect daily traffic data from Semrush, which provides domain-level visit estimates by channel (direct, organic search, referral, social, etc) for desktop and mobile devices in the US and worldwide. We use the channel breakdown to examine traffic composition trends.

\subsection{Publisher characteristics and content quantity}

\paragraph{Robots.txt and page structure} For each publisher domain, we collected \textit{robots.txt} rules and page-level HTML metadata from the HTTP Archive,\footnote{\url{https://httparchive.org/}} which tracks how the web is built and how it changes over time. These data allow us to code whether and when a domain blocks major GenAI crawlers and to quantify changes in page composition, including the intensity of using images, videos, and interactive elements.
\paragraph{Content quantity proxies.}
To proxy for the scale and scope of published content, we use the Internet Archive’s Wayback Machine\footnote{\url{https://web.archive.org/details/}} to construct annual counts of unique URLs observed for each domain. We use these counts as a proxy for the number of distinct pages a publisher maintains over time and use them in our analysis of content production.

\subsection{Job postings and employment}

To study hiring patterns, we use job posting data from Revelio Labs via WRDS.\footnote{\url{https://wrds-www.wharton.upenn.edu/pages/about/data-vendors/revelio-labs/}}
Revelio aggregates job postings from multiple sources and provides employer identifiers, job titles, occupation codes, locations, and posting dates. We use these data to construct publisher-level monthly counts of new job postings by occupation category (e.g., editorial/content-production versus other roles) and to track changes in editorial/content postings and total postings over time.

\subsection{Sample construction and merging across sources}
We take all websites that appear in the Similarweb dataset that also have a corresponding Revelio record of employer-sponsored postings. This process yields 6,315 URLs from which we select those that belong to NAICS 513110 and are newspaper publishers.\footnote{A URL can have multiple NAICS codes. We count a URL in a sector if at least one of its NAICS codes belongs to that sector. Economic Census: NAICS Codes Understanding Industry Classification Systems,
\url{https://www.census.gov/programs-surveys/economic-census/year/2022/guidance/understanding-naics.html}.}  
This process yields 30 URLs; we present the list of URLs and their total 2023 traffic in Appendix Table \ref{tab:url_list}.

%We select websites classified under  as newspaper publishers. We then manually review these websites and filter out those that are not newspaper publishers, such as news aggregators. 

%We construct our sample by starting from website domains of the focal newspaper publishers that have employer-linked job postings in Revelio and that appear in at least one traffic source (SimilarWeb or Comscore). Domains are matched across sources using Revelio’s firm–URL mapping at the domain level. In total, Revelio and SimilarWeb share  common domains. Table \ref{tab:url_naics_distribution} presents the ratios of URLs in each industry sector (the first two digits of the NAICS code).
Semrush traffic data, SimilarWeb data, and Comscore panel browsing data are merged in when available. To construct stable Comscore-based measures of human traffic, we restrict the sample to active panelists with at least four browsing sessions in each month of a calendar year, and we aggregate their visits to the domain–week level. For analyses that rely on Semrush or Comscore traffic, we expand coverage beyond the 30 newspaper publishers by including the top 500 news-publisher domains with the highest Comscore traffic among those matched to Revelio.\footnote{Traffic to the 30 publishers accounts for 70\% of total Comscore traffic among the top 500 publishers.}

Finally, we retrieve robots.txt rules and HTML snapshots from the HTTP Archive, and historical URL coverage from the Internet Archive, for all of these domains. Table \ref{tab:combined_traffic_stats} presents summary statistics for the daily traffic of websites across different industries.

\begin{table}[h!]
\begin{center}
\caption{Descriptive Statistics of Website Traffic by Category}
% \resizebox{0.95\textwidth}{!}{
\begin{tabular}{lrrrrr}
\hline\hline
 & \textbf{All} & \textbf{Information} & \textbf{Newspapers} & \textbf{Retail (All)} & \textbf{Retail (Top)} \\
\hline
Min (k) & 0.02 & 0.04 & 448.32 & 0.02 & 236.43 \\
%25\% ($10^3$) & 0.51 & 12.82 & 2,269.37 & 0.43 & 360.92 \\
Median (k) & 1.95 & 490.69 & 3,589.32 & 1.76 & 528.15 \\
%75\% ($10^3$) & 14.95 & 3,178.72 & 4,768.40 & 14.98 & 1,178.15 \\
Max (M) & 2,770.46 & 2,770.46 & 20.17 & 23.18 & 23.18 \\
Mean (M) & 1.06 & 14.79 & 5.71 & 0.13 & 1.59 \\
Std Dev. (M) & 37.23 & 150.17 & 5.45 & 0.98 & 3.32 \\

\hline
N & 6,315 & 385 & 30 & 1,905 & 100 \\
\hline\hline

\end{tabular}
% }
\label{tab:combined_traffic_stats}

\end{center}
\footnotesize
This table summarizes the daily traffic of websites across different categories. For the retail trade category (NAICS 44 and 45), we exclude URLs that also appear in the information or technology sectors.
\end{table}

\section{Background: Publisher Traffic Trends in the GenAI Era}
\label{sec:results}

We begin by documenting aggregate traffic patterns to motivate the strategic responses analyzed in the remainder of the paper. We divide the discussion into two periods: before and after May 2024, which is when Google introduced AI Overviews in search results.

\begin{figure}[!ht]
\begin{center}
    \caption{Publishers' Daily Traffic}    \includegraphics[width=0.8\linewidth]{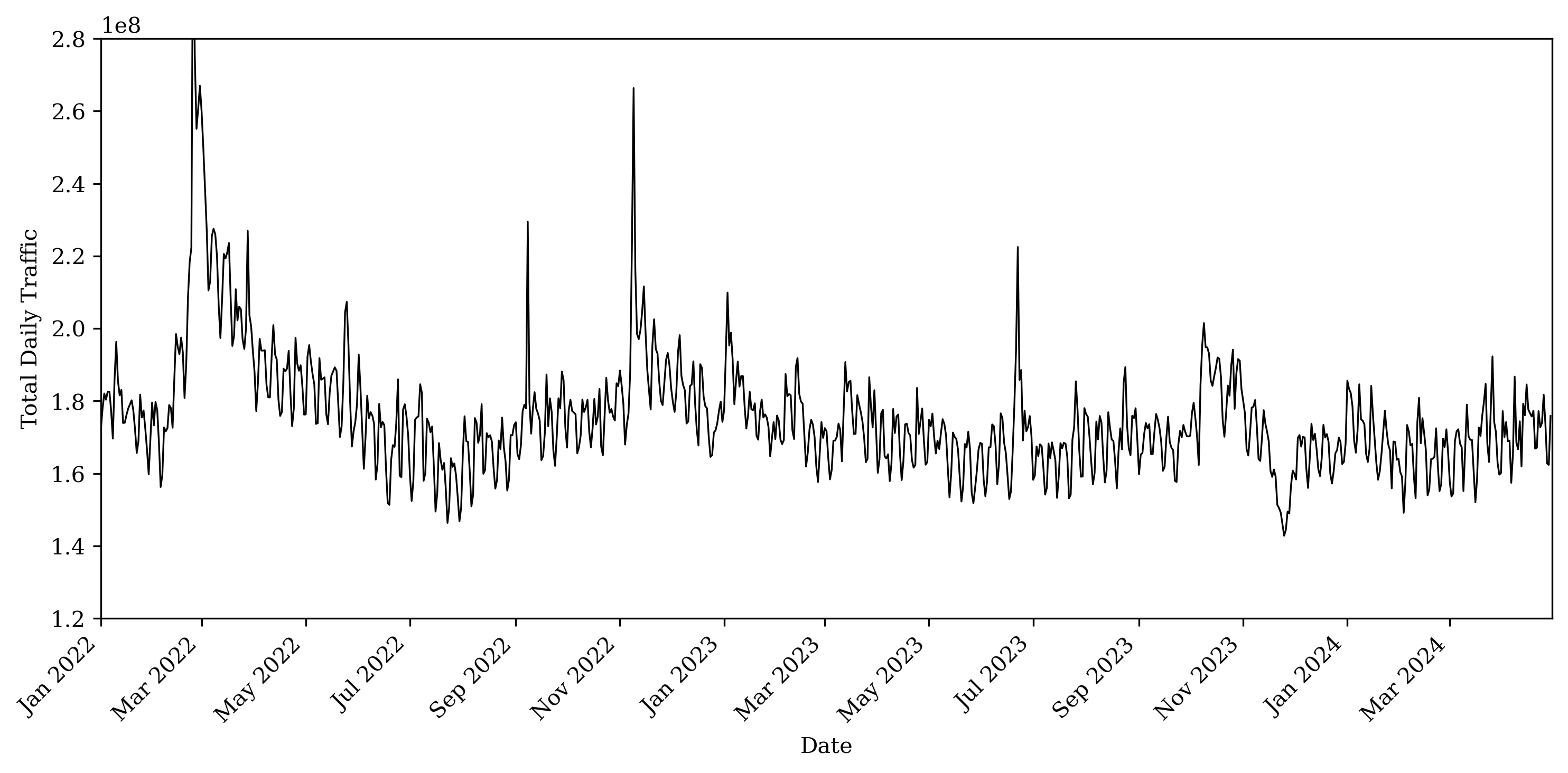}
    \label{fig:traffic}
\end{center}    
\footnotesize
This figure plots the total daily traffic ($\times 10^8$) across all news publishing websites using SimilarWeb.
\end{figure}

\noindent\emph{Before May 2024.}\quad
Figure~\ref{fig:traffic} plots the sum of daily visits across the newspaper domains in our sample from September 2022 through April 2024 using the SimilarWeb data. There appears to be a downward shift in traffic after early 2023, with visits remaining at lower levels.

\begin{figure}[!ht]
\begin{center}
    \caption{Change-Point Detection for Daily Traffic}
    \includegraphics[width=0.9\linewidth]{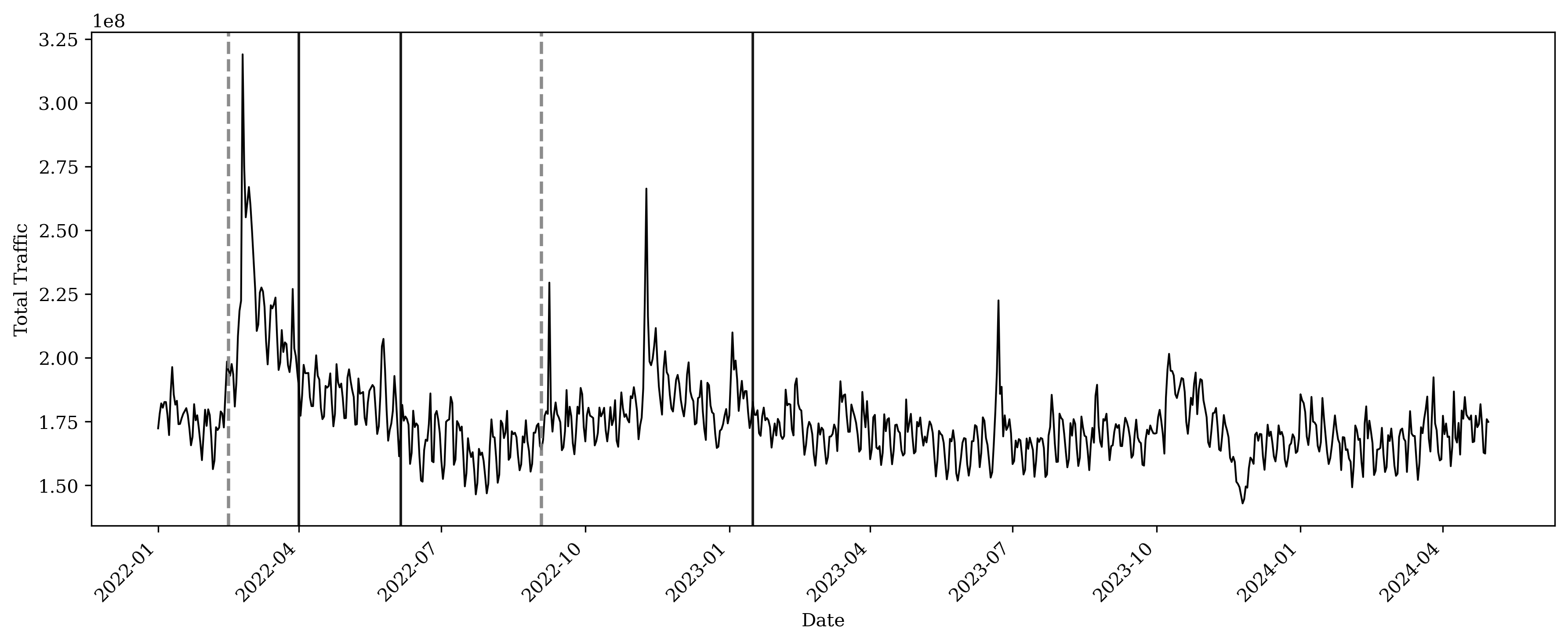}
    \label{fig:traffic_change_detection_similarweb_2022_2024}
\end{center}
\footnotesize
This plot visualizes change-point detection using the Pruned Exact Linear Time (PELT) algorithm applied to residualized log traffic after controlling for day-of-week, calendar-week, and month fixed effects. Vertical reference lines mark detected change points: dashed gray lines denote increases in the mean level of traffic following a detected change point, while solid black lines denote decreases in the mean levels.
\end{figure}

To test whether these visual patterns reflect structural breaks, we apply a multiple change-point detection procedure \citep{killick2012optimal} to the logarithm of daily total visits. Specifically, we residualize log-traffic with day-of-week fixed effects, calendar-week fixed effects, and month fixed effects to absorb systematic seasonality and within-week variation. We then apply the Pruned Exact Linear Time (PELT) algorithm to the residual series, assuming a piecewise-constant mean and selecting the number of breaks via a penalized least-squares criterion. The procedure identifies major breaks in early 2023, after which the traffic declines; we visualize the detected change points in Figure \ref{fig:traffic_change_detection_similarweb_2022_2024}. Although the magnitude of this early decline is modest, it coincides with growing industry concern about LLM-mediated substitution. Table \ref{tab:lag_traffic_regression} and \ref{tab:lag_traffic_regression_placebo} in the appendix present an interrupted time series analysis around the detected change points, and placebo analysis to rule out spurious results.

%We present additional parametric checks and placebo tests for the change-point analysis in the Appendix.

\begin{figure}[!ht]
      \centering
      \caption{Weekly Traffic by Channel (Semrush, 30 News Publishers,
  Worldwide, All Devices)}
      \label{fig:semrush_channel_breakdown}

      % --- First Plot (Left) ---
      \begin{subfigure}[t]{0.49\textwidth}
          \centering
          \caption{Jan 2023 -- Apr 2024}
          \includegraphics[width=\linewidth]{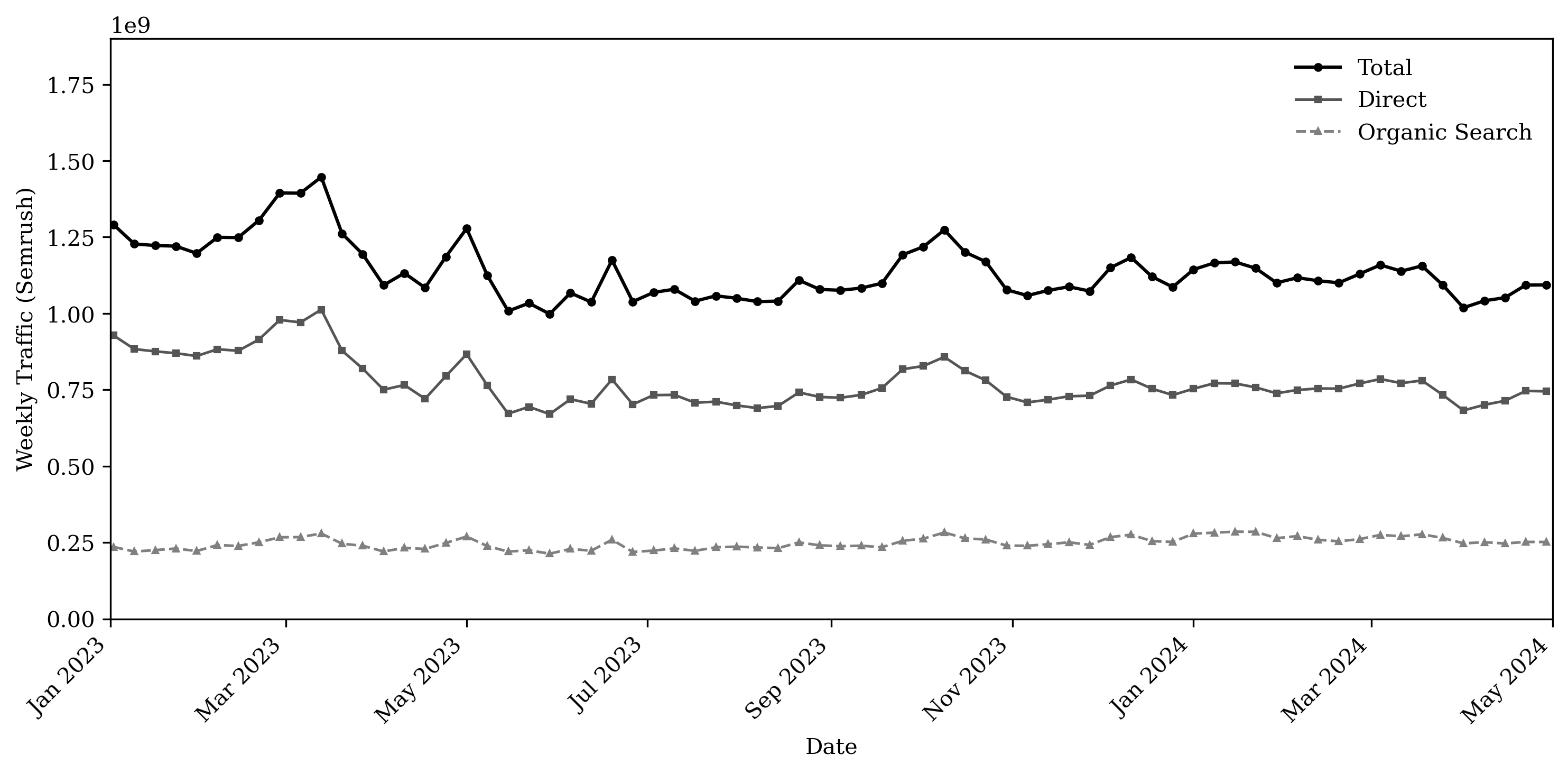}
          \label{fig:channel_2023_2024}
      \end{subfigure}
      \hfill
      % --- Second Plot (Right) ---
      \begin{subfigure}[t]{0.49\textwidth}
          \centering
          \caption{May 2024 onward}
          \includegraphics[width=\linewidth]{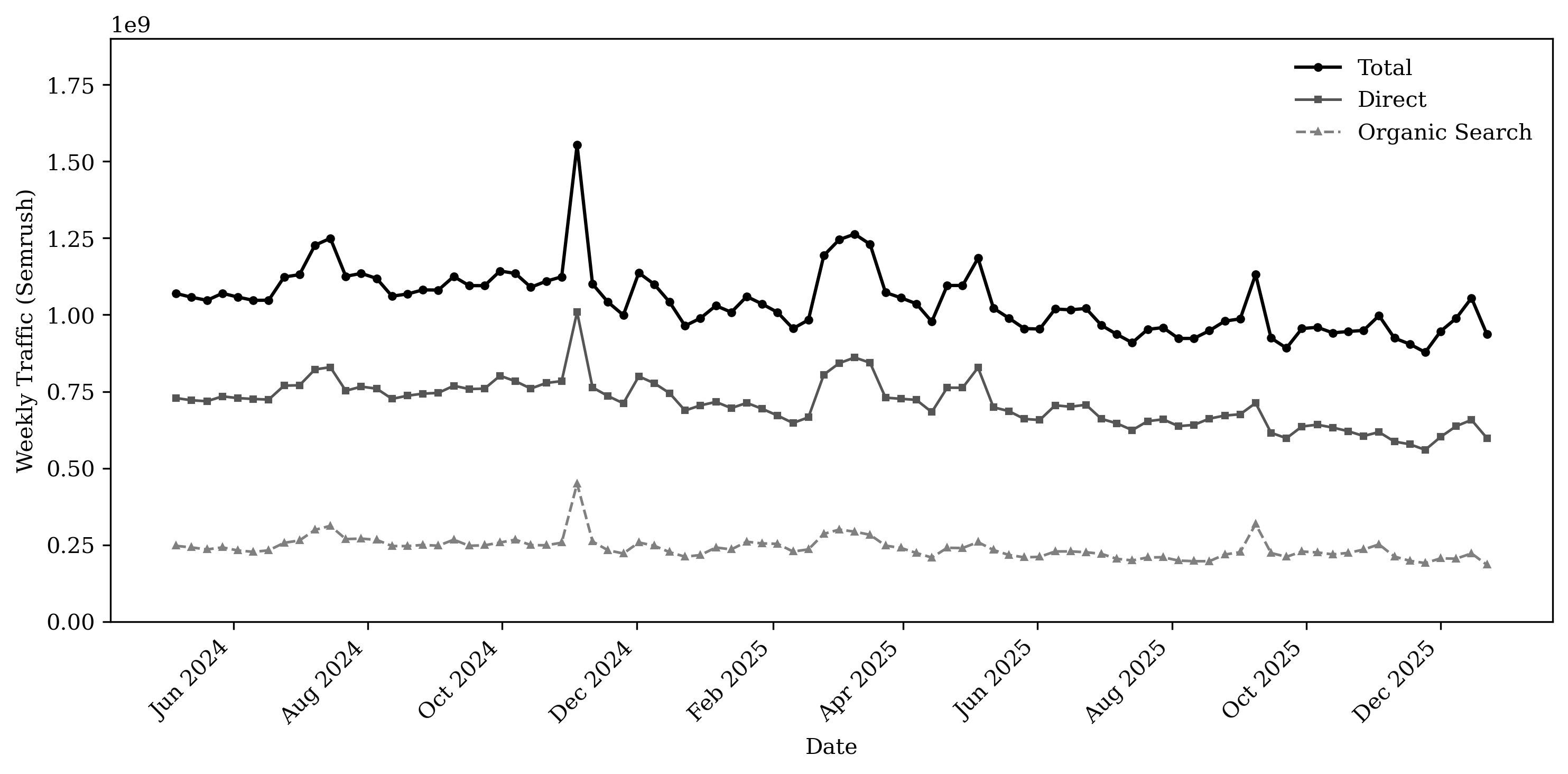}
          \label{fig:channel_2024_onward}
      \end{subfigure}

      \par\bigskip
      \footnotesize
      Weekly traffic for the top 30  news publisher websites, decomposed into
  Total, Direct, and Organic Search channels. Data from Semrush.
  \end{figure}

\noindent\emph{Traffic After May 2024.}\quad
Google introduced AI Overviews in May 2024 and rolled out a core algorithm update in August 2024, both of which could mechanically affect publishers' search-referral traffic. To examine this period, we decompose publisher visits into direct and organic-search channels using Semrush  data (mobile and desktop, worldwide). Figure~\ref{fig:semrush_channel_breakdown} presents the weekly breakdown. Before May 2024, the traffic decline is concentrated in direct visits, while organic search referrals remain broadly stable. After May 2024, both channels decline, with organic search referrals falling alongside continued decreases in direct traffic.

These patterns motivate our empirical design. We restrict our main causal analysis of the blocking effect to the pre-May 2024 period,  because the blocking decisions were concentrated in mid-to-late 2023 before AI Overview was introduced, providing the cleanest identification.

% Figure \ref{fig:rank_total} presents trends in website rankings using Tranco data.\footnote{\url{https://tranco-list.eu/}} For each week, we compute each publisher’s average Tranco rank across days and then take the mean across publishers. The resulting series shows a gradual deterioration in publishers’ relative rank over time (i.e., higher numerical ranks). This pattern is consistent with newspaper domains losing prominence relative to other sites, which could reflect both the entry and growth of GenAI-related destinations and changes in ranking relative to other non-GenAI sites.

% Figure \ref{} presents the trends in website rankings using Tranco data.\footnote{\url{https://tranco-list.eu/}} For each week, we first compute each publisher’s daily average Tranco rank, and then take the mean across all publishers. The resulting series shows a gradual decline in the relative ranking of news websites over time. This pattern can reflect both downward re-ranking driven by the entry and growth of GenAI-related websites and a decline in rankings relative to existing non-GenAI sites.
 
\section{The Effect of Blocking GenAI Crawlers on Publisher Traffic}

Restricting automated access through \textit{robots.txt} is one of the most immediate strategic responses available to publishers facing a threat from LLM engines. Blocking such traffic may protect content from being scraped for model training and query answering, but it may also reduce a publisher’s visibility in LLM-mediated interfaces and yield lower subsequent visits. We first document the staggered pattern of adoption of blocking and then use that staggered timing in a difference-in-differences design to estimate the effect of blocking on publisher traffic. We further examine heterogeneity across publisher size and consider robustness with placebo tests, selection-on-trends checks, and  concurrent changes audits.

\subsection{Blocking adoption patterns: Staggered adoption started in mid-2023.} 

\begin{figure}[!ht]
     \centering
          \caption{Fraction of websites that disallow GenAI bots}
\includegraphics[width=0.9\linewidth]{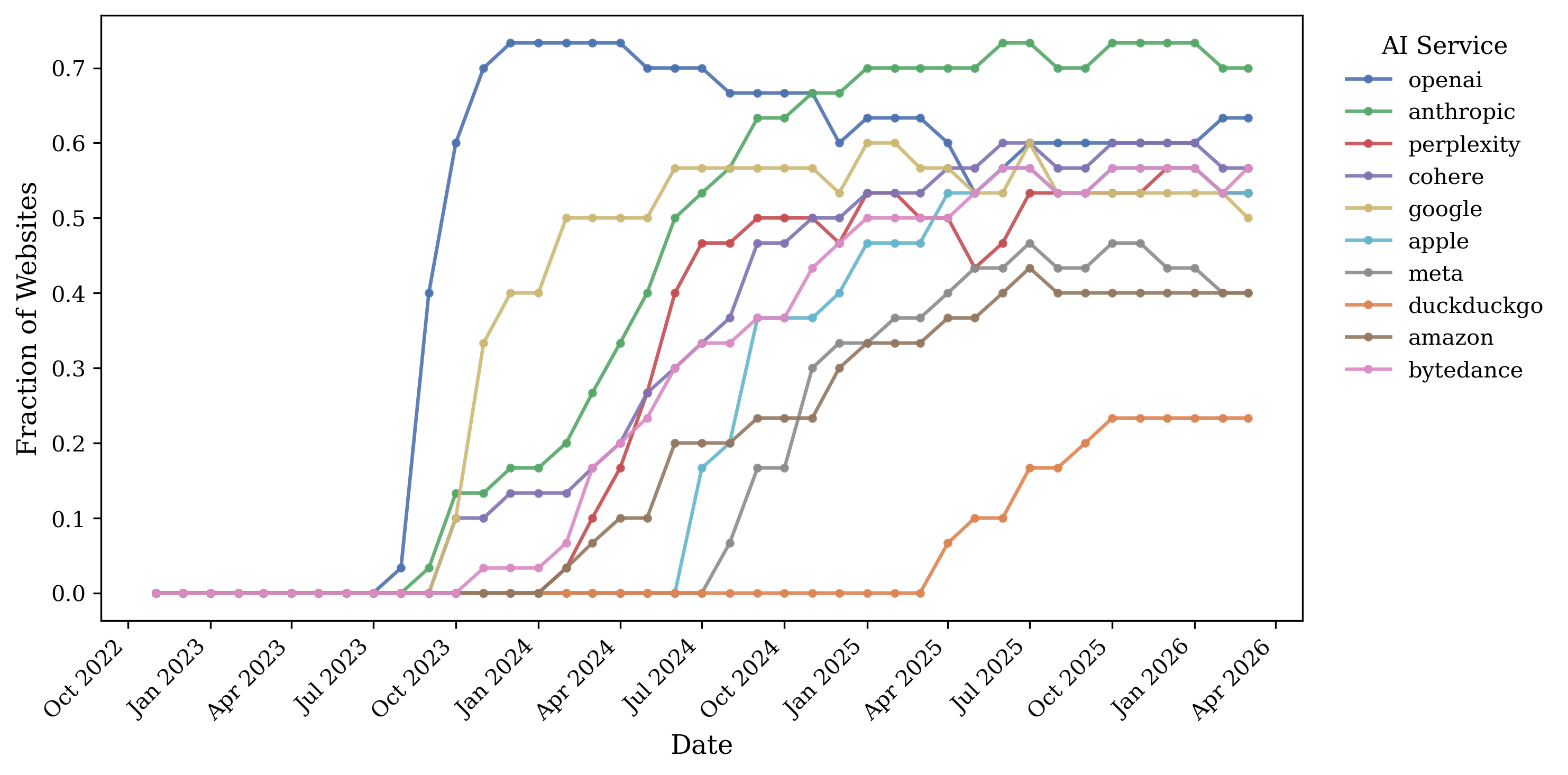}
     \label{fig:genai_bot_blocking}
     
     \footnotesize
     This figure plots the fraction of websites that disallow GenAI bots over time.
 \end{figure}

We use historical \textit{robots.txt} snapshots from the HTTP Archive for the analysis. We code a publisher as blocking a given GenAI bot if its \textit{robots.txt} contains an explicit \texttt{Disallow} directive for that bot’s user agent (Table \ref{tab:genai_bot}). Most user agents correspond to official crawlers documented by the vendors; we supplement these with a small set of widely used user agents observed frequently in \textit{robots.txt} files. Figure~\ref{fig:genai_bot_blocking} plots the fraction of websites that disallow GenAI bots over time.

Publishers begin blocking GenAI crawlers as early as mid-2023, with staggered adoption thereafter. OpenAI-related crawlers are blocked first, followed by Anthropic, Perplexity, and others. Roughly 75\% of publishers in our sample block an OpenAI-related crawler at some point. News publishers also exhibit substantially higher GenAI bot blocking rates compared with other domains in the information sector, and compared with domains in other sectors such as retail. Figure \ref{fig:blocking_groups} in the Appendix shows blocking ratios for lower-traffic publishers and retailers. The blocking ratio among the top 500 news publishing websites is below 60\%, whereas it is below 10\% among the top retailers.

\subsection{Empirical strategy: Staggered Difference-in-Differences (DiD)}

We quantify the effect of blocking GenAI crawlers on publisher traffic. The outcome $Y_{it}$ is the log of weekly visits to publisher $i$ in week $t$. We denote the week in which publisher $i$ first introduces a \texttt{Disallow} rule for any GenAI crawler in its \textit{robots.txt} by $G_i$, and $G_{i,g} = \mathbf{1}\{G_i = g\}$ indicate membership in adoption cohort $g$. The treatment indicator is $D_{i,t} = \mathbf{1}\{t \ge G_i\}$, so $D_{i,t}=1$ from the first blocking week onward. We focus on the ATT for eventual adopters. Never-blocking publishers are in the control group together with not-yet-treated publishers. For publishers that later remove the \texttt{Disallow} rule, we drop observations after the first unblocking week.

% Following \citet{callaway2021difference}, the parameter of interest is
% \begin{align}
%    \nonumber \text{ATT}_{g,t}
%     &= \mathbb{E}\big[ Y_{i,t}(g) - Y_{i,t}(0) \,\big|\, G_{i,g} = 1 \big] \\
%     &= \mathbb{E}\big[ Y_{i,t} - Y_{i,g-1} \,\big|\, G_{i,g} = 1 \big]
%      \;-\;
%        \mathbb{E}\big[ Y_{i,t} - Y_{i,g-1} \,\big|\, D_{i,t} = 0,\, G_{i,g} = 0 \big].
% \end{align}
% Here, $Y_{i,t}(g)$ denotes the potential outcome for publisher $i$ in week
% $t$ if it starts blocking in week $g$, and $Y_{i,t}(0)$ denotes the potential
% outcome if it has not yet blocked by week $t$. In the second line, the first
% term is the observed change in outcomes for cohort $g$ between the pre-period
% $g-1$ and week $t$, while the second term is the corresponding change for
% publishers that are not yet treated in week $t$ ($D_{i,t}=0$) and belong to
% other cohorts ($G_{i,g}=0$). The controls are all future adopters (not-yet-treated) and never-treated units.

% \medskip
% \noindent\emph{Identification.}
% Identification  relies on a conditional parallel trends
% assumption. For all $t \ge g$ and for all
% cohorts $g$ and $g' \neq g$,
% \[
% \mathbb{E}\big[ Y_{i,t}(0) - Y_{i,g-1}(0) \,\big|\, G_{i,g}=1, X_i \big]
% =
% \mathbb{E}\big[ Y_{i,t}(0) - Y_{i,g-1}(0) \,\big|\, D_{i,t}=0, G_{i,g}=0, X_i \big],
% \]
% This requires that the evolution of log-weekly traffic that cohort $g$
% would have experienced in the absence of blocking is the same as that of
% publishers that have not yet blocked by week $t$ or never blocked. 

Following \citet{callaway2021difference}, the ATT of blocking vs not blocking for unit $i$ who blocked in cohort $g$ in time $t$ is 
\begin{equation}
\label{eq:att-def}
\mathrm{ATT}(g,t)
=
\mathbb{E}\!\left[ Y_{i,t}(g)-Y_{i,t}(0)\mid G_i=g \right],
\end{equation}
where $Y_{i,t}(g)$ is publisher $i$'s potential weekly traffic if blocking
first begins in week $g$, and $Y_{i,t}(0)$ is the potential weekly traffic if
publisher $i$ never blocks. We denote $G_i=\infty$ for never-adopters/blockers.

\medskip
\noindent\emph{Identification.}
Identification relies on parallel trends and no anticipation. For
all $t\ge g$,
\[
\mathbb{E}\!\left[ Y_{i,t}(0)-Y_{i,g-1}(0)\mid G_i=g \right]
=
\mathbb{E}\!\left[ Y_{i,t}(0)-Y_{i,g-1}(0)\mid G_i>t \right],
\]
This requires that the untreated traffic trajectory for cohort $g$ would have evolved in
parallel to that of publishers untreated by week $t$. Under these assumptions, $\mathrm{ATT}(g,t)$ is identified by comparing cohort
$g$ to publishers untreated in week $t$:
\begin{equation}
\label{eq:att-id}
\mathrm{ATT}(g,t)
=
\mathbb{E}\!\left[ Y_{i,t}-Y_{i,g-1}\mid G_i=g \right]
-
\mathbb{E}\!\left[ Y_{i,t}-Y_{i,g-1}\mid G_i>t \right].
\end{equation}
The first term is the observed change for cohort $g$ from the pre-period
$g-1$ to week $t$, and the second is the corresponding change among publishers
untreated in week $t$, including both future adopters and never-adopters.

\subsection{Results: Blocking reduces total traffic}

\begin{figure}[!ht]
\begin{center}
       \caption{Staggered DiD of blocking GenAI bots on  publisher traffic.}

    \begin{subfigure}[t]{0.33\linewidth}
        \centering
                \caption{log(SimilarWeb Traffic)}
\includegraphics[width=\linewidth]{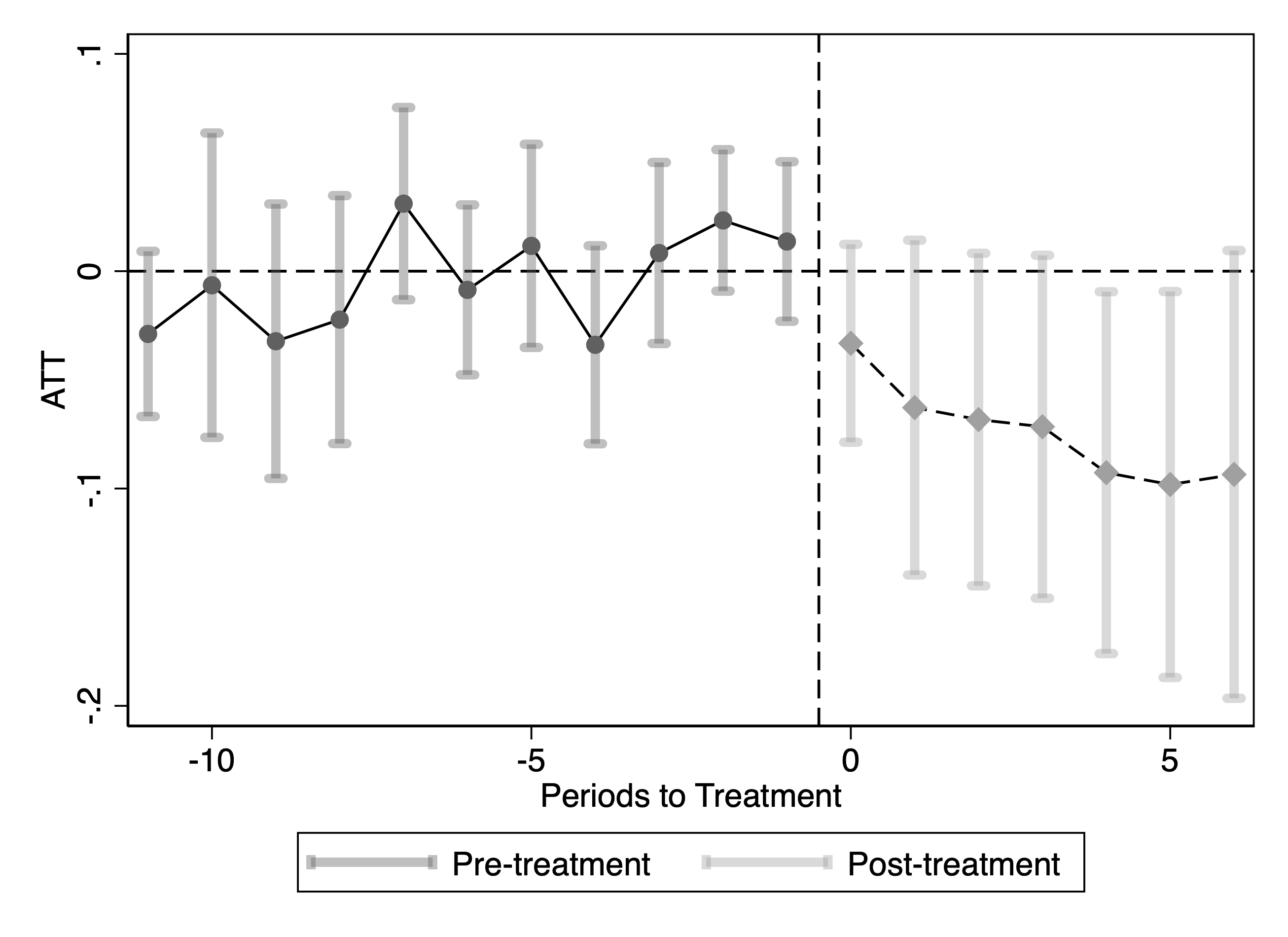}
        \label{fig:csdid_similarweb}
    \end{subfigure}\hfill
    \begin{subfigure}[t]{0.33\linewidth}
        \centering
                \caption{log(Semrush Traffic)}
\includegraphics[width=\linewidth]{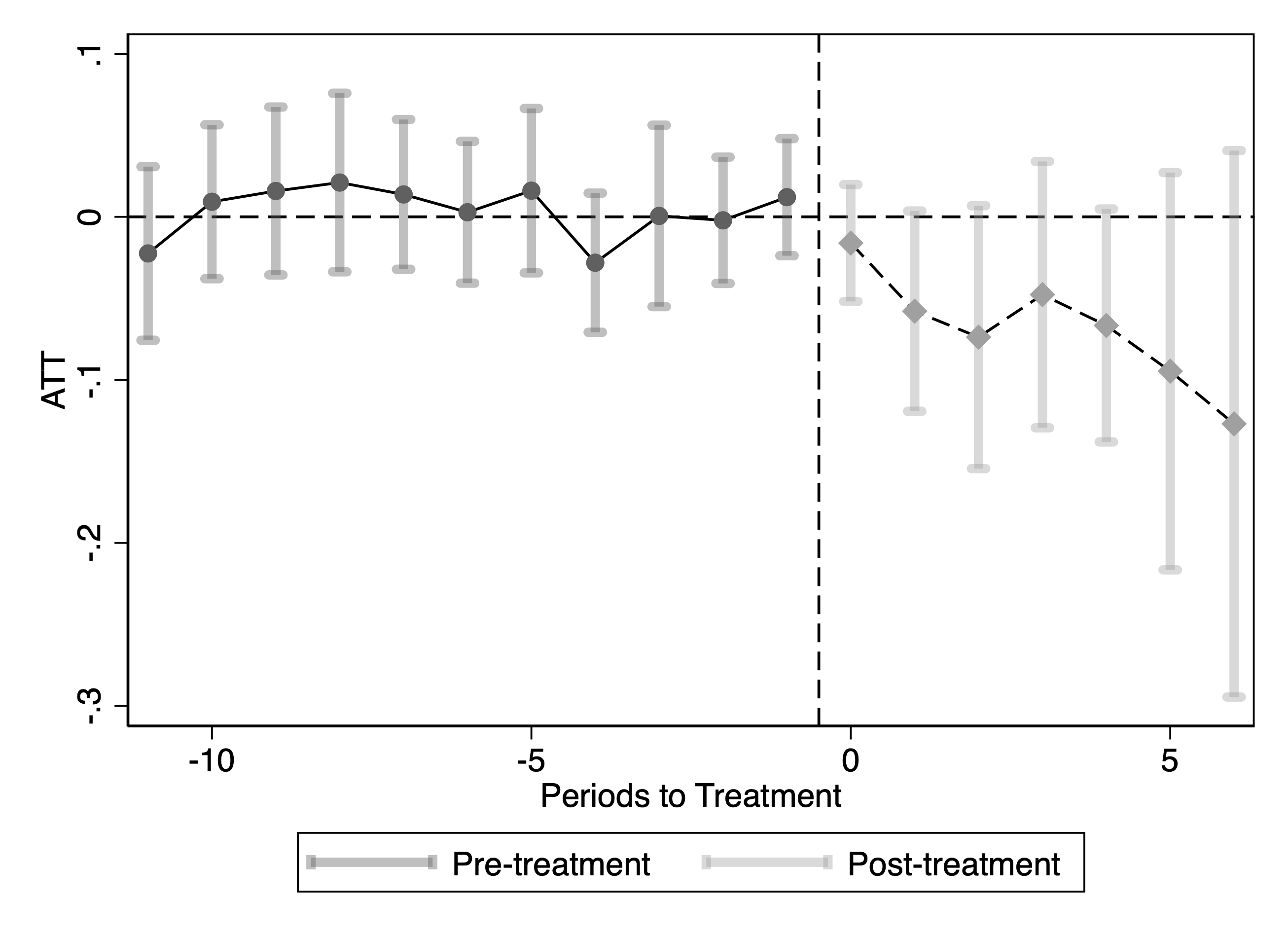}
        \label{fig:csdid_weekly_sr}
    \end{subfigure}
        \begin{subfigure}[t]{0.33\linewidth}
        \centering
                \caption{log(Comscore Traffic)}
\includegraphics[width=\linewidth]{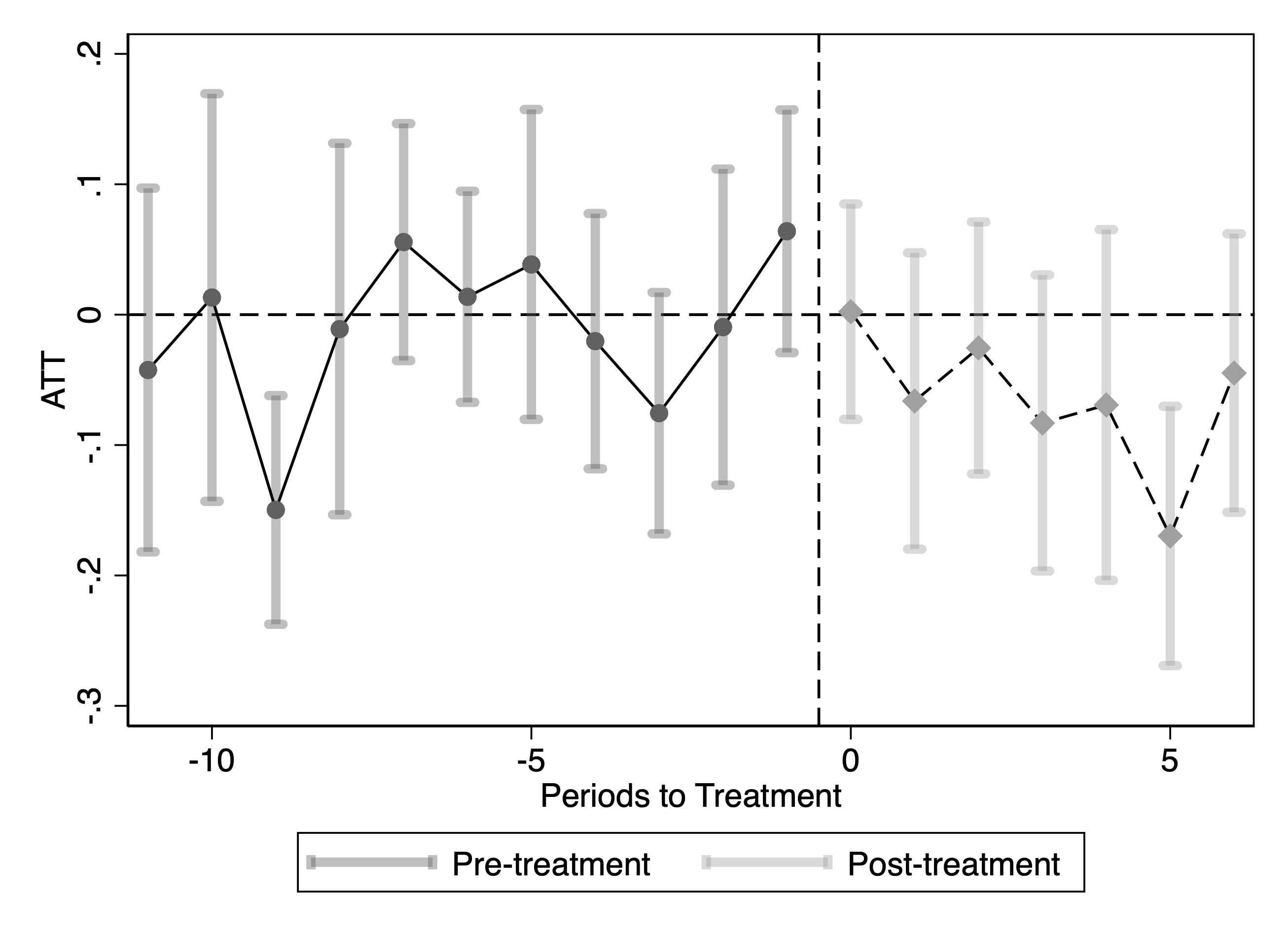}
        \label{fig:csdid_weekly_cs}
    \end{subfigure}
\end{center} 
    \label{fig:csdid_combined}
    \footnotesize
This figure reports staggered DiD event-study estimates of the effect of blocking GenAI web crawlers on publisher traffic, using a 12-week before and 6-week after blocking window. Panel (a) uses SimilarWeb traffic; Panel (b) uses Semrush traffic; Panel (c) uses Comscore traffic. The outcome in each panel is the logarithm of weekly visits. Confidence intervals are based on 50 bootstrap replications.
\end{figure}

Figure~\ref{fig:csdid_similarweb} displays event-time aggregate estimates using log weekly traffic from SimilarWeb as the outcome.\footnote{At each relative week, the plotted coefficient is a weighted average of the cohort-specific effects for publisher cohorts observed at that point in event time, with weights proportional to each cohort’s share among treated publishers observed at that relative time.} The event-time coefficients are close to zero and statistically insignificant in the pre-blocking weeks, providing evidence of no systematic pre-trends. After blocking, the coefficients became negative and remain below zero, indicating that blocking GenAI bots is associated with a decline in total traffic for blocking publishers in the short term.

To alleviate the concern that the reduction in traffic might reflect a drop in automated/bot visits rather than human visits, and to examine the robustness of the traffic decline results, we replicate the analysis using two other datasets: Comscore human panelist records and Semrush traffic data. Figures \ref{fig:csdid_weekly_sr} and \ref{fig:csdid_weekly_cs} show similar negative post-blocking patterns in Semrush and Comscore. Because these datasets aim to filter automated traffic, and because Comscore is based on human-panel browsing history, these results suggest that the estimated decline is unlikely to be driven solely by bot traffic. We report the aggregated ATT estimates in Table \ref{tab:att_estimates}.  The point estimates are similar in magnitude across all three data sources: publishers experience approximately a 7\% decline in traffic within 6 weeks after blocking GenAI crawlers, whether measured by SimilarWeb, Semrush, or Comscore, relative to the pre-blocking period. The consistency across independently constructed datasets, including Comscore's human-only panel, strengthens the interpretation that blocking leads to a reduction in traffic, though the Comscore estimate is imprecise, which might be due to the smaller panel size. We also report ATT estimates from two other methods: Synthetic DiD \citep{arkhangelsky2021synthetic} and Two-Way Fixed Effects in Table \ref{tab:att_sdid_estimates}, which yield similarly negative effects.\footnote{The Synthetic DiD method requires a balanced panel, yielding SimilarWeb estimates that differ slightly from those in Table~\ref{tab:att_estimates}.}%, because the restricted sample includes only the first few months after blocking.} 

\normalsize
\begin{table}[!ht]
\begin{center}
    \caption{Staggered DiD  estimates of blocking GenAI crawlers on publisher traffic}
\label{tab:att_estimates}
\begin{tabular}{l*{3}{c}}
\hline\hline
    &\multicolumn{1}{c}{SimilarWeb}&\multicolumn{1}{c}{Semrush}&\multicolumn{1}{c}{Comscore}\\
\hline
ATT                 &      -0.074\sym{***}  &      -0.069\sym{*}         &      -0.065         \\
                    &[-0.141,-0.007]         &[-0.145,0.007]         &[-0.150,0.021]         \\
\hline\hline
\end{tabular}
\end{center}
\footnotesize
This table reports staggered DiD ATT estimates of the effect of blocking GenAI web crawlers on publisher traffic. The dependent variables in the columns are the logarithm of weekly visits from  SimilarWeb, Semrush, and Comscore data, respectively. Confidence intervals are based on 50 bootstrap replications. *$p<$0.1; **$p<$0.05; ***$p<$0.01.
\end{table}
\normalsize

\begin{table}[!ht]
\begin{center}
    \caption{Synthetic and TWFE DiD  estimates of blocking GenAI crawlers on publisher traffic}
\label{tab:att_sdid_estimates}
\resizebox{\textwidth}{!}{
\begin{tabular}{l*{6}{c}}
\hline\hline
        &\multicolumn{2}{c}{SimilarWeb}&\multicolumn{2}{c}{Semrush}&\multicolumn{2}{c}{Comscore}\\
        Method      &               SDID  &  TWFE &               SDID  &               TWFE & SDID  &               TWFE\\
\hline
ATT       &               -0.0912*  &               -0.0727***&               -0.0734   &               -0.0404** &               -0.0185   &               -0.0647** \\
                    &      [-0.189,0.00626]   &      [-0.111,-0.0342]   &       [-0.174,0.0274]   &   [-0.0799,-0.000895]   &       [-0.116,0.0790]   &      [-0.116,-0.0136]   \\
\hline
\hline
\end{tabular}
}
\end{center}
\footnotesize
This table reports Synthetic and TWFE DiD ATT estimates of the effect of blocking GenAI web crawlers on publisher traffic. The dependent variables in columns (1)–(2), (3)–(4) and (5)–(6) are the logarithm of weekly visits from SimilarWeb, Semrush, and Comscore, respectively. Confidence intervals are based on 50 bootstrap repetitions. *p$<$0.1; **p$<$0.05; ***p$<$0.01.
\end{table}

The negative effects on human traffic could arise through two channels, which we summarize in Figure~\ref{fig:two_channels}. A plausible mechanism is that blocking lowers the number of times the publisher is mentioned as a source of the query results, reducing brand exposure. This is consistent with our earlier evidence that, before May 2024, the traffic decline is concentrated in direct visits while organic search referrals remain broadly stable. When LLMs mention blocked publishers less often in query responses or AI-powered search summaries, subsequent direct visits may decline, an effect that would not appear in referral metrics. Blocking may also prevent LLM-mediated interfaces from generating referral traffic through citations and links, though this channel is less likely to be the primary driver given that AI-referral traffic was low before May 2024.

\begin{figure}[!ht]                  
  \centering 
  \caption{Possible mechanisms linking blocking to reduced publisher traffic}
  \label{fig:two_channels}    
  \begin{tikzpicture}[  
      box/.style={rectangle, draw=black!70, fill=black!5, rounded corners,      
                  minimum height=0.7cm, minimum width=3.2cm, align=center,      
  font=\small},   
      mainbox/.style={rectangle, draw=orange!80!black, fill=orange!10, rounded  
  corners,     minimum height=0.7cm, minimum width=3.2cm, align=center,
  font=\small},   
      minorbox/.style={rectangle, draw=gray, fill=gray!5, rounded corners,
                  minimum height=0.7cm, minimum width=3.2cm, align=center,      
  font=\small},                     
      arrow/.style={->, thick, >=stealth}                
  ]                                     
  % Main channel (left)
  \node[mainbox]                              (block)   at (0,0)    {Publisher
  blocks LLM};   
  \node[box,    below=0.5cm of block]         (mention) {LLM mentions publisher
  less};         
  \node[box,    below=0.5cm of mention]       (recall)  {Consumers recall brand
  less};    
  \node[mainbox,below=0.5cm of recall]        (direct)  {Direct
  visits decrease};        
  \draw[arrow] (block)   -- (mention);                                          
  \draw[arrow] (mention) -- (recall);
  \draw[arrow] (recall)  -- (direct);       
  \node[font=\footnotesize\itshape, text=orange!80!black, below=0.15cm of       
  direct]
      {Likely primary channel};  
  % Minor channel (right)
  \node[minorbox, right=2.2cm of mention]  (nolink)   {No citations / links};
  \node[minorbox, below=0.5cm of nolink]   (referral) {LLM         
  referrals decrease} ;  
  \draw[arrow, gray] (block.east) -- ++(0.5,0) |- (nolink.west);                
  \draw[arrow, gray] (nolink) -- (referral);    
  \node[font=\footnotesize\itshape, text=gray, below=0.15cm of referral]
      {Possible secondary channel};  
  \end{tikzpicture} 
  \footnotesize
  
  % The main channel proposes that blocking reduces LLM mentions of the publisher, which in turn weakens consumer brand recall and lowers direct visits. The    
  % minor channel proposes that blocking removes citations and links in LLM answers, reducing referral clicks; we expect this to be small during the      
  % pre--May 2024 period.
  \end{figure}

\subsubsection{Longer time window and heterogeneity by publisher size}

When we extend the post-treatment period of analysis, Figure~\ref{fig:csdid_weekly_longer} shows that the point estimates remain negative but become statistically insignificant after approximately 20 weeks.

We also extend the analysis to the broader set of 500 largest news publishing websites, stratified by Semrush traffic rank. Table~\ref{tab:att_size_heterogeneity} Panel A shows the estimated effects are negative and significant for the top 50 publishers, consistent with our previous findings.  For mid-ranked publishers (ranks 51–100), the point estimates remain negative but are no longer statistically significant. For lower-ranked publishers (ranks 101–500), the estimate is no longer negative, which may reflect less accurate traffic estimates for lower-traffic websites and a lower impact of blocking for smaller publishers.

% \begin{table}[!ht]
% \begin{center}
%     \caption{Staggered DiD  estimates of blocking GenAI crawlers on publisher traffic}
% \label{tab:att_semrush_top}
% \begin{tabular}{l*{3}{c}}
% \hline\hline
% &\multicolumn{3}{c}{Log(Semrush traffic) }\\
% \hline
% ATT                 &               -0.0691*  &               -0.0343   &                0.0164   \\
%                     &      [-0.145,0.00686]   &       [-0.141,0.0725]   &       [-0.0971,0.130]   \\
% \hline   
% Group (rank in Semrush)  & Top 50 & 51-100 & 100-500\\
% \hline\hline
% \end{tabular}
% \end{center}
% \footnotesize
% This table reports staggered DiD ATT estimates of the effect of blocking GenAI web crawlers on publisher traffic. The dependent variables in the columns are the logarithm of weekly visits from  Semrush data. Confidence intervals are based on 50 bootstrap repetitions. *$p<$0.1; **$p<$0.05; ***$p<$0.01.
% \end{table}
% \normalsize

\begin{figure}[!ht]
\begin{center}
        \caption{Staggered DiD estimates for Comscore traffic by publisher size group}
    \begin{subfigure}[t]{0.33\textwidth}
        \centering
                \caption{Group 1}
\includegraphics[width=\textwidth]{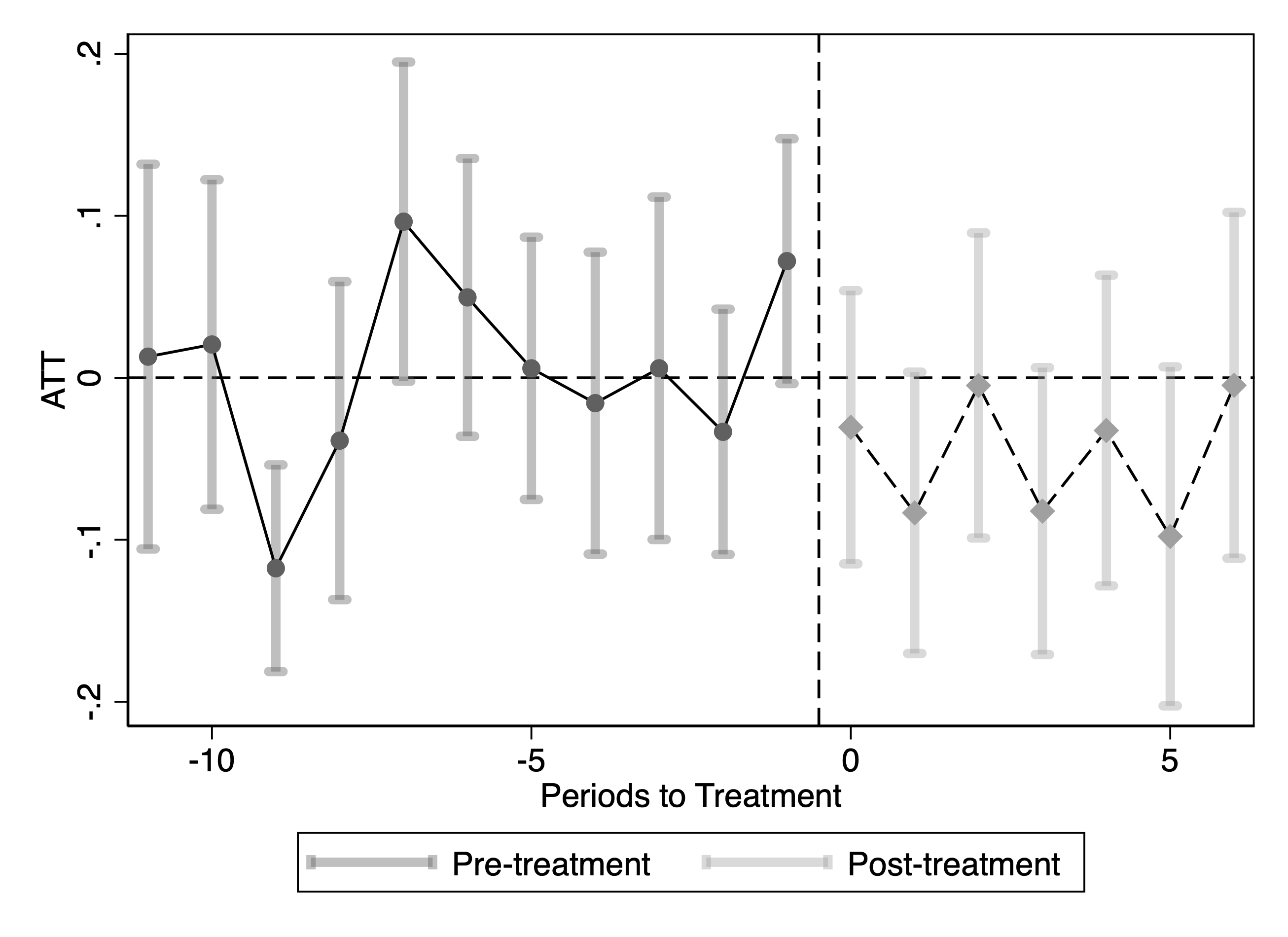}
        \label{fig:csdid_top_1}
    \end{subfigure}\hfill
    \begin{subfigure}[t]{0.33\textwidth}
        \centering
                \caption{Group 2}
\includegraphics[width=\textwidth]{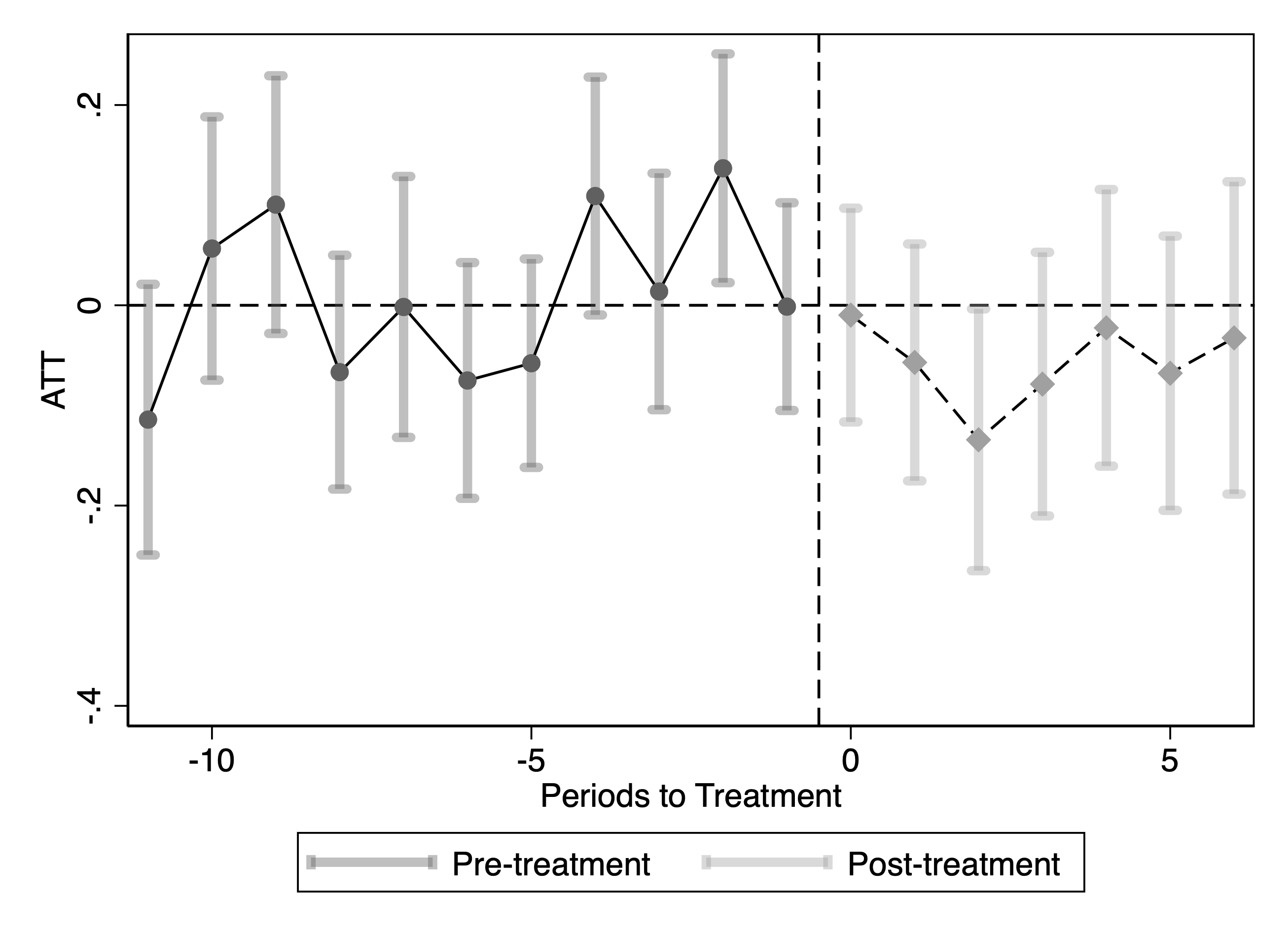}
        \label{fig:csdid_top_2}
    \end{subfigure}\hfill
    \begin{subfigure}[t]{0.33\textwidth}
        \centering
                \caption{Group 3}
\includegraphics[width=\textwidth]{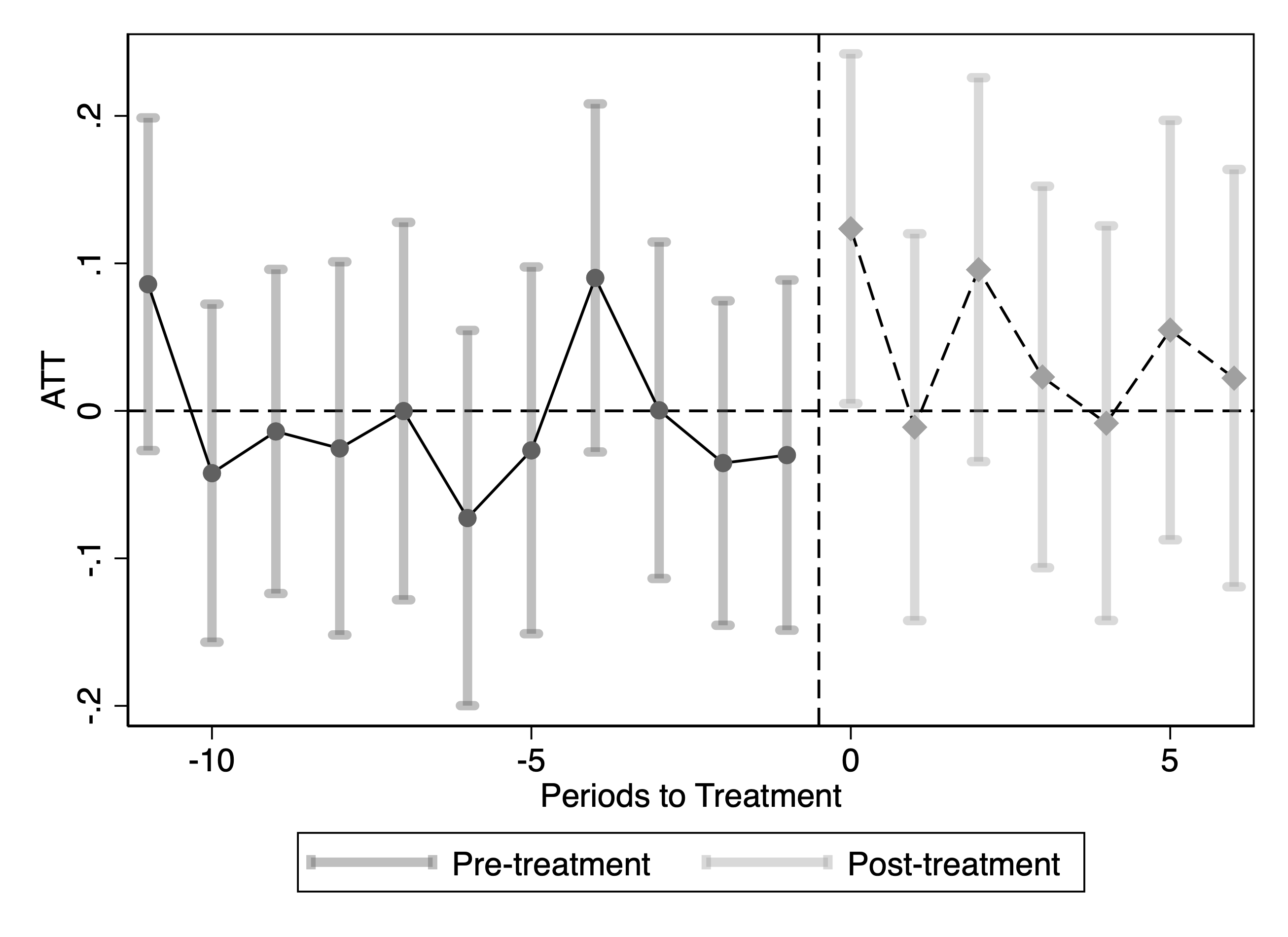}
        \label{fig:csdid_top_3}
    \end{subfigure}

    \label{fig:csdid_comscore_all}
\end{center}
\footnotesize
This figure reports staggered DiD event-study estimates of the effect of blocking GenAI crawlers on publisher traffic measured using Comscore data. Panel (a), Group~1, includes the largest publishers (more than 10 visits per day on average); Panel (b), Group~2, includes mid-sized publishers (1--10 visits per day; ranked 34th--164th by average daily visits); and Panel (c), Group~3, includes the smallest publishers (fewer than 1 visit per day on average). The outcome variable is the logarithm of Comscore weekly visits. Confidence intervals are based on 50 bootstrap replications.
\end{figure}

We find a similar pattern using Comscore data, stratified by average daily visits in Table~\ref{tab:att_size_heterogeneity} Panel B. The negative blocking effect is concentrated among higher-traffic publishers (more than 10 visits per day) and mid-tier websites (1–10 visits per day; Figure~\ref{fig:csdid_comscore_all}). For the lowest tier (fewer than 1 visit per day), we do not find such a negative effect, likely due to less accurate traffic measurement at low visit levels and a smaller impact of blocking for these publishers.

% \begin{table}[!ht]
% \begin{center}
%     \caption{Staggered DiD  estimates of blocking GenAI crawlers on publisher traffic}
% \label{tab:att_comscore_top}
% \begin{tabular}{l*{3}{c}}
% \hline\hline
% &\multicolumn{3}{c}{Log(Comscore traffic) }\\
% \hline
% ATT                 &               -0.0480   &               -0.0575   &                0.0425   \\
%                     &       [-0.121,0.0249]   &       [-0.157,0.0417]   &       [-0.0579,0.143]   \\  \hline   
% Group (websites daily traffic in Comscore)  & $\ge$ 10& 1-10& $\le$1\\
% \hline\hline
% \end{tabular}
% \end{center}
% \footnotesize
% This table reports staggered DiD ATT estimates of the effect of blocking GenAI web crawlers on publisher traffic. The dependent variables in the columns are the logarithm of weekly visits from  Comscore data. Confidence intervals are based on 50 bootstrap repetitions. *$p<$0.1; **$p<$0.05; ***$p<$0.01.
% \end{table}
% \normalsize

\begin{table}[!ht]
\begin{center}
\caption{Heterogeneity in the effect of blocking GenAI crawlers on publisher traffic}
\label{tab:att_size_heterogeneity}
\begin{tabular}{lccc}
\hline\hline
& Group 1 & Group 2 & Group 3 \\
\hline
\multicolumn{4}{l}{\textit{Panel A: Semrush (rank in Semrush)}} \\\hline
Group definition & Top 50 & 51--100 & 101--500 \\\hline
ATT & -0.0691* & -0.0343 & 0.0164 \\
    & [-0.145,0.00686] & [-0.141,0.0725] & [-0.0971,0.130] \\
\hline
\\\hline
\multicolumn{4}{l}{\textit{Panel B: Comscore (websites' daily traffic in Comscore)}} \\\hline
Group definition & $\geq 10$ & 1--10 & $\leq 1$ \\\hline
ATT & -0.0480 & -0.0575 & 0.0425 \\
    & [-0.121,0.0249] & [-0.157,0.0417] & [-0.0579,0.143] \\
\hline\hline
\end{tabular}
\end{center}
\footnotesize
This table reports staggered DiD ATT estimates of the effect of blocking GenAI web crawlers on publisher traffic across publisher size groups. Panel A uses the logarithm of weekly visits from Semrush data, with groups defined by Semrush traffic rank. Panel B uses the logarithm of weekly visits from Comscore data, with groups defined by average daily Comscore visits. Confidence intervals are based on 50 bootstrap repetitions. *$p<$0.1; **$p<$0.05; ***$p<$0.01.
\end{table}
\normalsize

\normalsize
\subsection{Robustness: Placebo tests, selection checks, and concurrent \textit{robots.txt} changes}

We assess the robustness of the main estimates with three checks. First, selection-on-trends tests examine whether publishers block in anticipation of traffic changes. Second, three placebo exercises break the link between blocking timing and traffic outcomes, helping rule out the possibility that the estimates reflect coincident traffic shocks or generic time-series patterns. Third, we verify that concurrent \textit{robots.txt} changes unrelated to GenAI, such as SEO- or paywall-related modifications, are not driving the results.

\subsubsection{Selection-on-trends checks}
One concern is that publishers may choose to block in anticipation of traffic declines or other strategic changes. The event-study coefficients are statistically indistinguishable from zero in the pre-blocking period, providing no evidence of differential pre-trends. Figure~\ref{fig:csdid_weekly_longer} extends the event-study window further back (24 weeks prior to blocking) and shows similarly flat pre-trends. 

As an additional check on endogenous adoption, we estimate a discrete-time adoption regression in the pre-adoption sample, where the dependent variable is an indicator for first-time blocking in week $t$ and the covariate is lagged traffic growth:
\[
Block_{it}=\alpha_i + \gamma_t + \beta\, \Delta \log(\text{Traffic}_{i,t-1})+ \varepsilon_{it}, \quad \text{for } t \le T_i^{adopt}.
\]
Table~\ref{tab:adoption_traffic} reports the results; the coefficient on lagged traffic growth is not statistically distinguishable from zero, suggesting that recent traffic declines do not predict blocking adoption.\footnote{If publishers block in response to traffic declines, we would expect the blocking indicator to be negatively correlated with lagged traffic changes. Instead, our estimate is close to zero and not statistically distinguishable from zero.}

\begin{table}[!ht]
\begin{center}
    
\caption{Lagged traffic and blocking adoption}
\label{tab:adoption_traffic}
\setlength{\tabcolsep}{8pt}
\renewcommand{\arraystretch}{1.15}
\begin{tabular}{l*{3}{c}}
\hline\hline
            &\multicolumn{3}{c}{Block}\\
\hline
$\Delta \log(\text{Traffic}_{i,t-1})$
&     -0.0129         &   -0.00757    &     0.0106                                 \\
            &     (0.285)         &          (0.322)         &    (0.658)                     \\\hline
Traffic data & SimilarWeb & Semrush & Comscore\\
\hline\hline
\end{tabular}
\end{center}
\vspace{0.5em}
\footnotesize 
The dependent variable is an indicator for first-time blocking adoption in  week $t$ (at-risk sample). The regressors are lagged weekly traffic growth, $\Delta \log(\text{Traffic})_{t-1}$, measured using SimilarWeb, Semrush, and Comscore traffic, respectively. Week and url fixed effects are included.
\textit{p}-values are in parentheses. *p$<$0.1; **p$<$0.05; ***p$<$0.01. Robust standard errors clustered at the URL level. \end{table}

\subsubsection{Placebo tests for treatment-timing} Another concern is that our negative post-blocking estimates could reflect coincident traffic shocks or generic time-series patterns rather than a causal effect of blocking. We conduct three placebo exercises that preserve the panel structure but break the link between the \emph{timing} of blocking and traffic outcomes. In each exercise, we repeat the procedure for $R = 50$ random draws, re-estimate the same staggered DiD event study, and compare the realized estimate to the placebo distribution.

\noindent\emph{Placebo 1: Re-randomize treatment timing for treated publishers.}
For each eventual blocker $i$, we draw a placebo adoption week $\widetilde{G}_i$ uniformly over the weeks in which publisher $i$ \emph{not yet blocked} in the actual data.
We then define a placebo treatment indicator $\widetilde{D}_{i,t}=\mathbf{1}\{t\ge \widetilde{G}_i\}$ for treated publishers, while leaving never-blockers untreated.
We re-estimate $\text{ATT}^{\text{pl}}_{g,t}$ using the same \cite{callaway2021difference} procedure. Because the placebo timing is independent of traffic shocks by construction, the absence of any systematic post-event decline under this design would support that our main estimates reflect the effect of blocking rather than spurious patterns induced by staggered timing.

\smallskip
\noindent\emph{Placebo 2: ``Control-only'' placebo treatments.}
We next restrict attention to publishers that never block AI crawlers during our sample window. We randomly assign each such publisher a placebo adoption week $\widetilde{G}_i$ (e.g., drawn from the empirical distribution of adoption weeks among actual blockers), define $\widetilde{D}_{i,t}=\mathbf{1}\{t\ge \widetilde{G}_i\}$, and re-run the same staggered DiD estimator under this placebo treatment timing. Since these publishers never change blocking behavior, the null of no treatment effect should hold, and estimated post-event effects should be centered near zero.

\smallskip
\noindent\emph{Placebo 3: Placebo-treated never-blockers with an expanded pre-treatment control pool.}
Finally, we again assign placebo adoption weeks to never-blockers as in Placebo 2, but we also include the pre-blocking observations of eventual blockers (i.e., we keep treated publishers only for $t < G_i$ and drop their observations thereafter). This design uses a richer set of untreated outcomes while ensuring that no post-blocking observations from actual treated publishers enter the estimation. We would expect placebo estimates to remain close to zero and exhibit no systematic post-event pattern.

\begin{figure}[!ht]
    \centering
    % First Image
        \caption{Distribution of placebo estimates.}
    \begin{subfigure}[b]{0.3\textwidth}
        \centering
                \caption{Placebo 1}
\includegraphics[width=\textwidth]{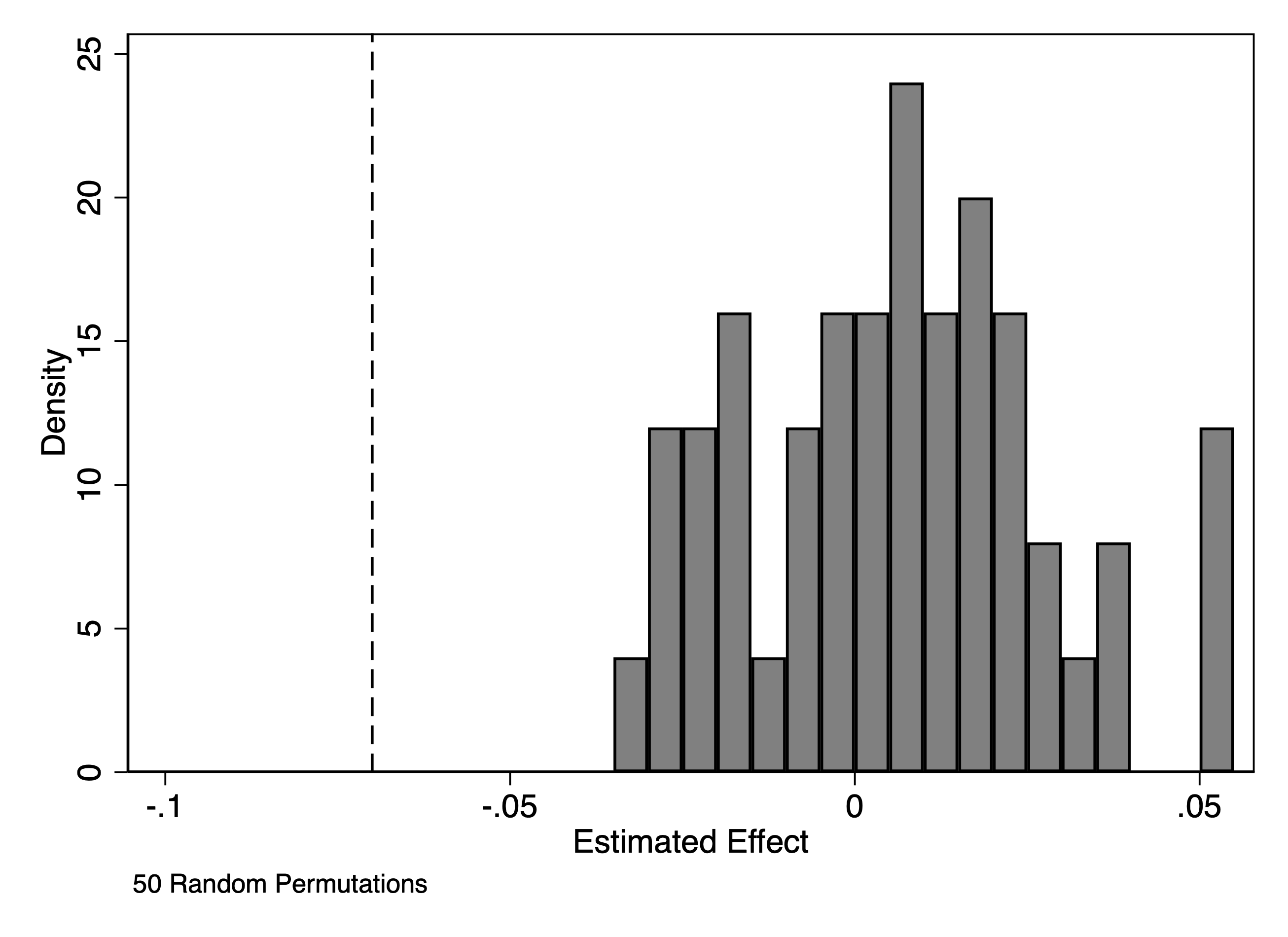}
        \label{fig:placebo1}
    \end{subfigure}
    \hfill % Adds space between images
    % Second Image
    \begin{subfigure}[b]{0.3\textwidth}
        \centering
                \caption{Placebo 2}
\includegraphics[width=\textwidth]{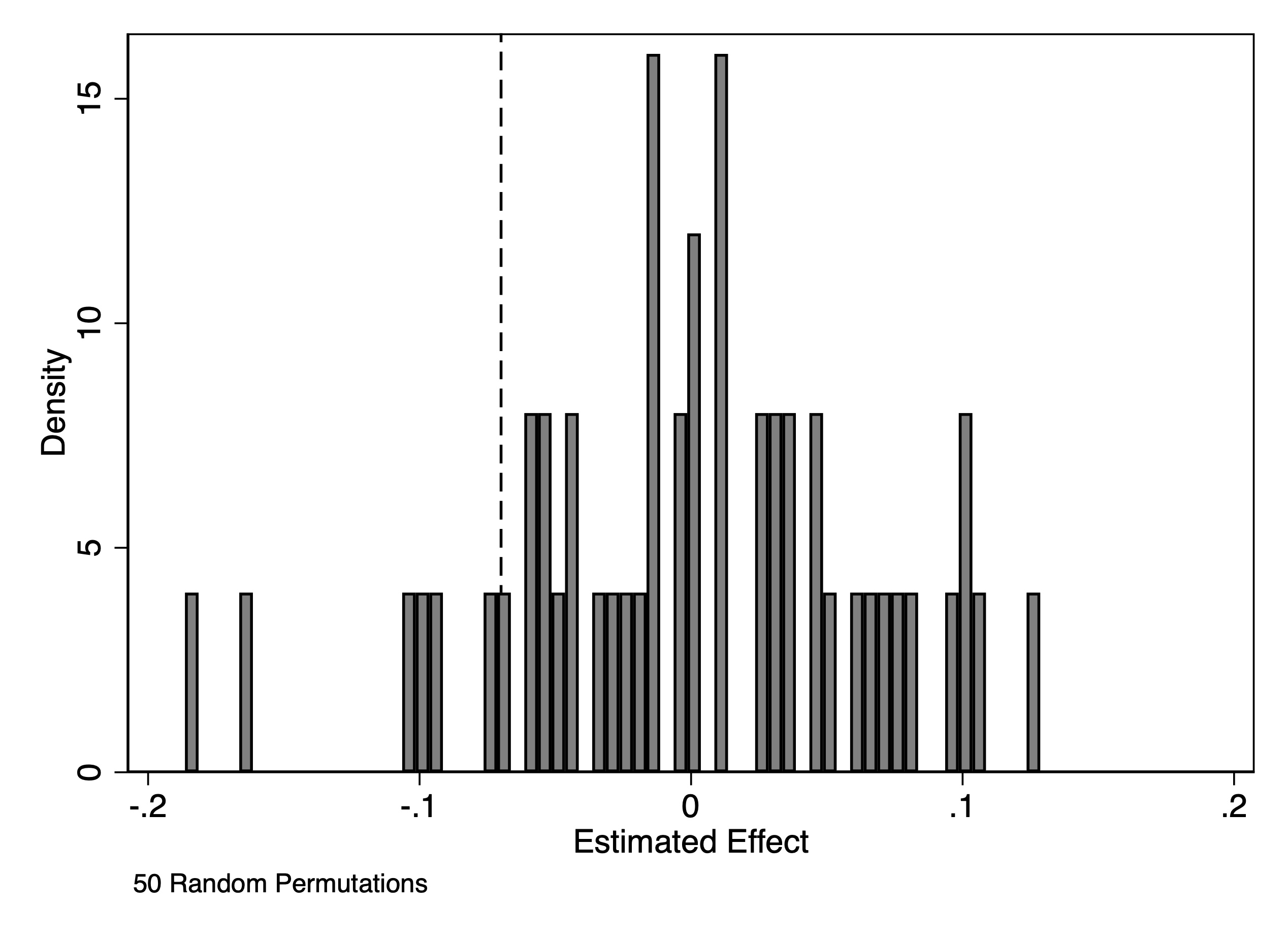}
        \label{fig:placebo2}
    \end{subfigure}
    \hfill % Adds space between images
    % Third Image
    \begin{subfigure}[b]{0.3\textwidth}
        \centering
                \caption{Placebo 3}
\includegraphics[width=\textwidth]{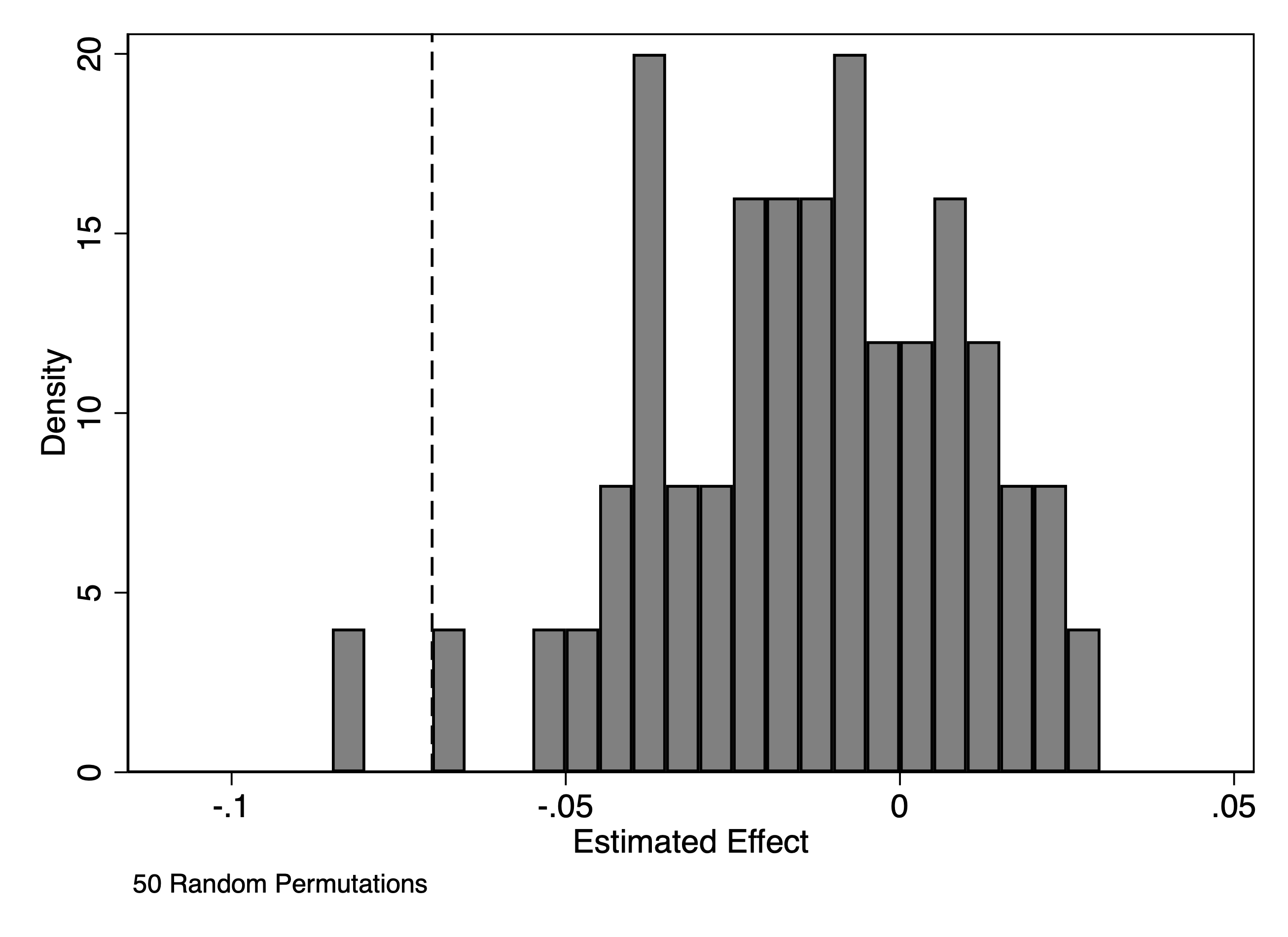}
        \label{fig:placebo3}
    \end{subfigure}
    \label{fig:three_placebos}
    \footnotesize
    
    This figure reports results from placebo analyses. The analysis mimics the main specification but randomizes the treatment assignment. The reported coefficients represent the average effect over 50 random permutations.
\end{figure}

\smallskip
\noindent\emph{Inference and interpretation.} For each placebo design, Figure~\ref{fig:three_placebos} plots the distribution of placebo ATT point estimates, $\{\widetilde{\text{ATT}}^{(r)}\}_{r=1}^R$. Across designs, the placebo distributions are centered near zero and show no systematic negative post-event pattern, while the observed post-blocking decline in total traffic lies in the left tail of these distributions. In Placebos 1 and 3, none of the random draws produces an aggregated ATT more negative than our baseline estimate from Table \ref{tab:att_estimates}, and the placebo estimates are uniformly statistically indistinguishable from zero. In Placebo 2, the sample is limited to never-blockers, so the placebo estimates are particularly noisy; none is statistically significant, although some point estimates are negative. These placebo tests suggest that the estimated decline following blocking is unlikely to be an artifact of treatment timing or generic time-series fluctuations.

\subsubsection{Ruling out concurrent robots.txt changes: SEO, paywall, etc.} 

Another threat to identification is other concurrent changes to the robots.txt file beyond blocking LLM crawling. %Our analysis of the files from the month before blocking compared to the month of blocking shows that 19 out of 24 made only LLM focused changes, while the others made only minor changes. Importantly none of these changes are related to major search engine traffic.
To confirm that our estimates are not driven by concurrent strategic or technical changes, for example, adjustments to search-indexing and crawling policies, paywalls, site redesigns, or other anti-bot measures that could coincide with blocking, we compare each site’s \textit{robots.txt} file in the month before and the month after the blocking event and parse directive changes beyond the LLM-crawler rules using the HTTP Archive’s parsed \textit{robots.txt} metadata. Of the 24 sites in our main sample that eventually blocked GenAI bots, 19 made no changes beyond adding GenAI-specific user-agent directives (e.g., GPTBot, CCBot, Google-Extended). The remaining five exhibit only minor concurrent changes.\footnote{Specifically, three News Corp properties (marketwatch.com, nypost.com, the-sun.com) added entries for archival crawlers (mj12bot, ia\_archiver, omgili), Reuters renamed a user-agent label (pipl to piplbot), and ESPN added an analytics-related directive (claritybot).} Critically, none of these changes involve directives for major search-engine crawlers (e.g., Googlebot, Bingbot), user-facing access restrictions, or other traffic-generating referral channels, the primary mechanisms through which \textit{robots.txt} changes could confound our blocking estimates. Thus, our results are unlikely to be driven by broader contemporaneous shifts in crawler restrictions.

\section{Publisher Responses Beyond Blocking}

\subsection{Publishers Shift Toward Media-Rich Content Rather Than Scaling Text}

Beyond access control, publishers can adjust what they produce. One possibility is scaling up textual output, potentially aided by AI writing tools.  Publishers could differentiate by shifting toward formats that LLMs cannot easily replicate, such as images, video, and interactive features. 

\begin{figure}[htbp]
    \centering
    \caption{Aggregate element counts}
 \begin{subfigure}{0.48\textwidth}    \centering
    \caption{DOM count}
\includegraphics[width=0.95\linewidth]{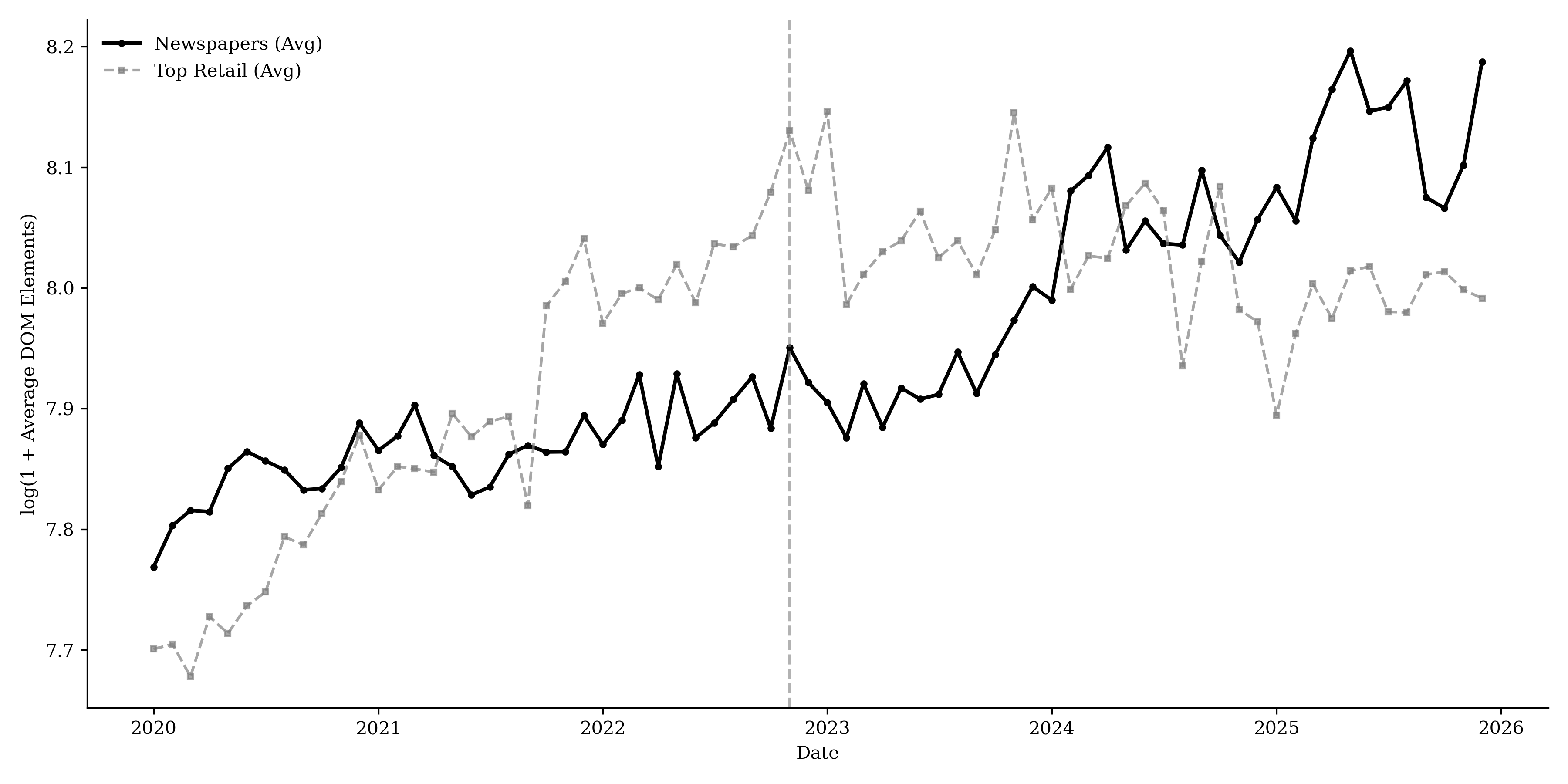}
    \label{fig:avg_dom_elements}
\end{subfigure}
    \begin{subfigure}{0.45\textwidth}
        \centering
                \caption{Site Framework \& Layout}
\includegraphics[width=\linewidth]{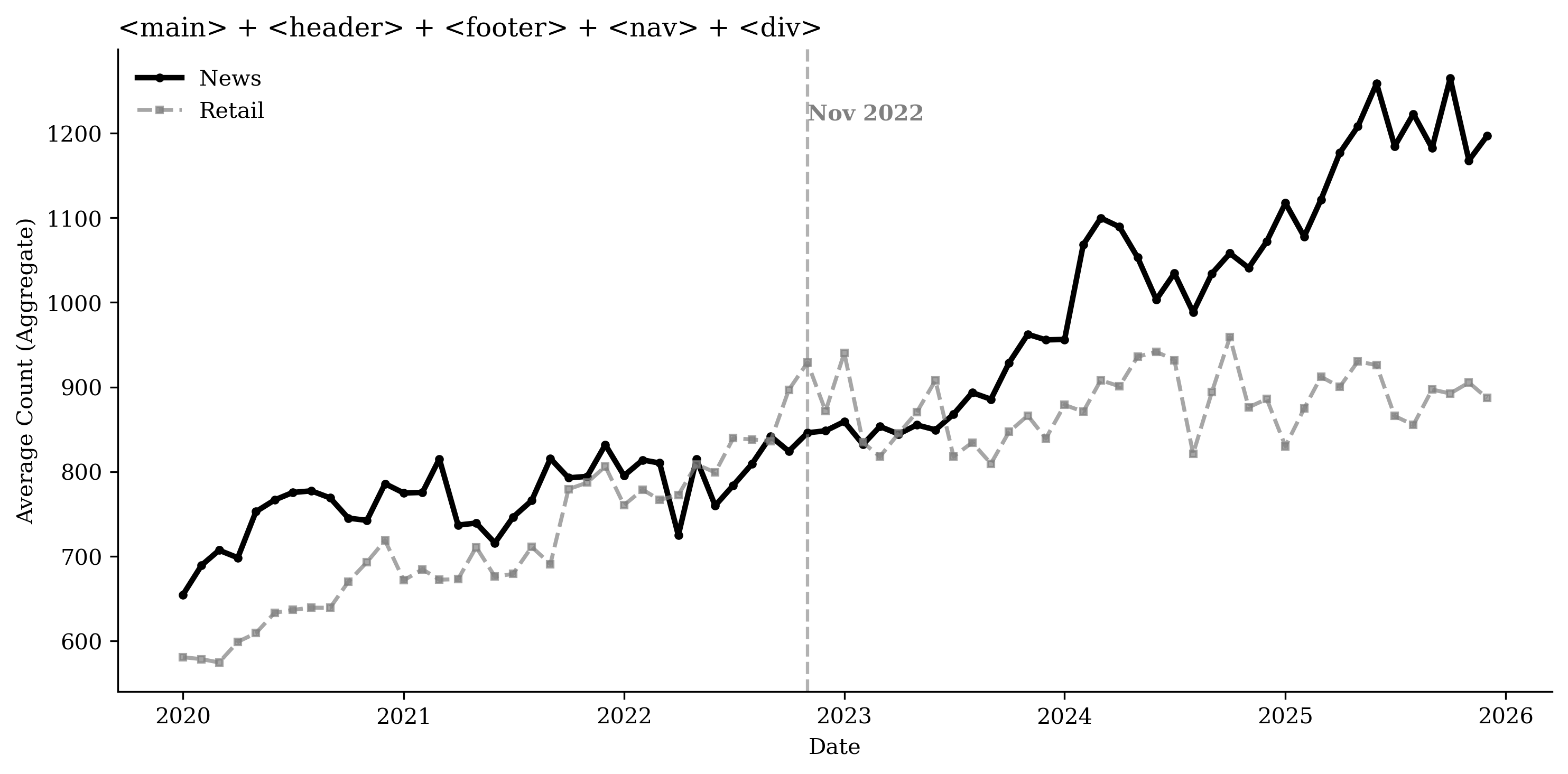}
        \label{fig:site_framework}
    \end{subfigure}
    \begin{subfigure}{0.45\textwidth}
        \centering
                \caption{Article Volume}
\includegraphics[width=\linewidth]{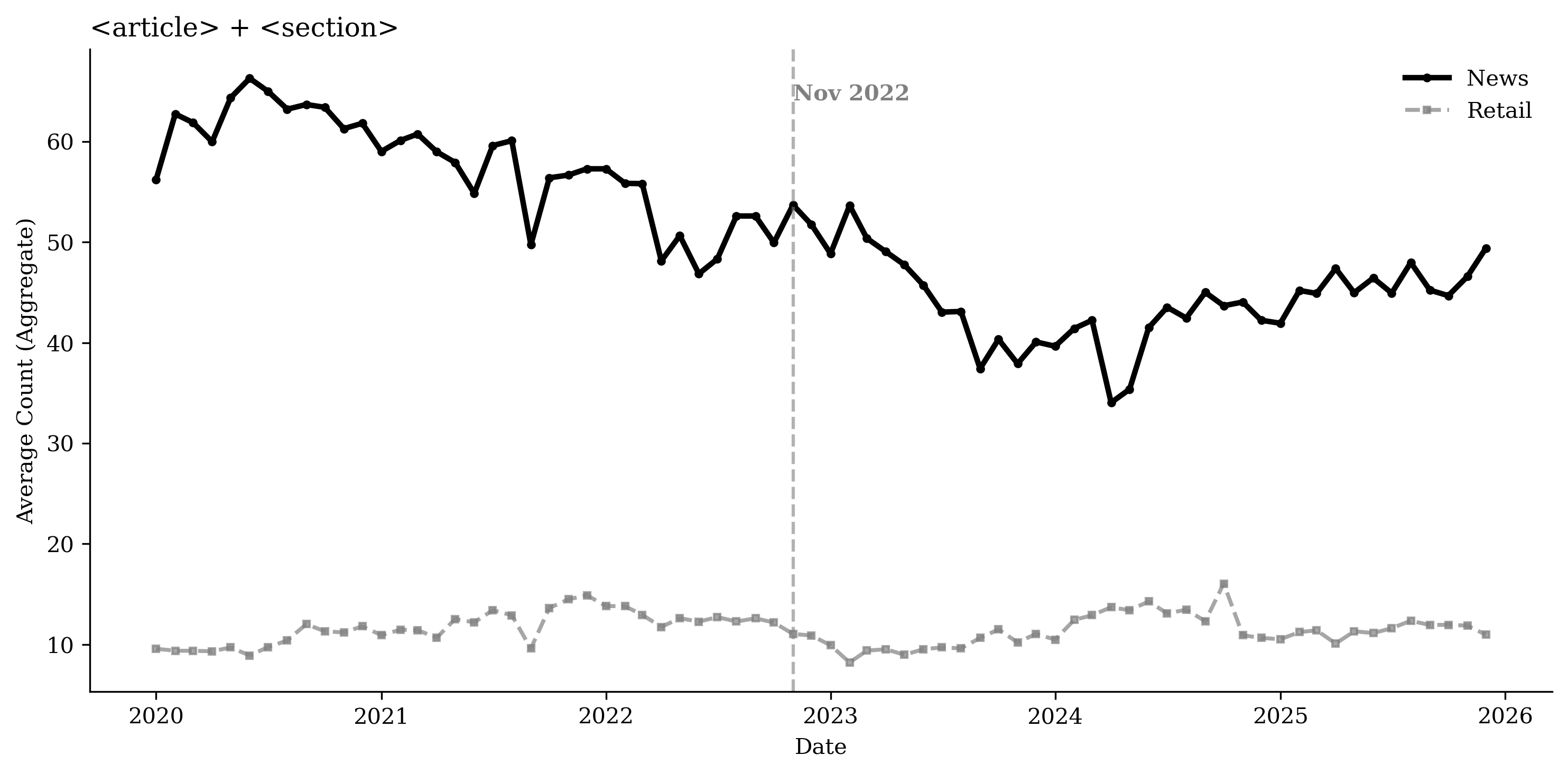}
        \label{fig:semantic_density}
    \end{subfigure}
    \begin{subfigure}{0.45\textwidth}
        \centering
                \caption{Visual \& Multimedia}
\includegraphics[width=\linewidth]{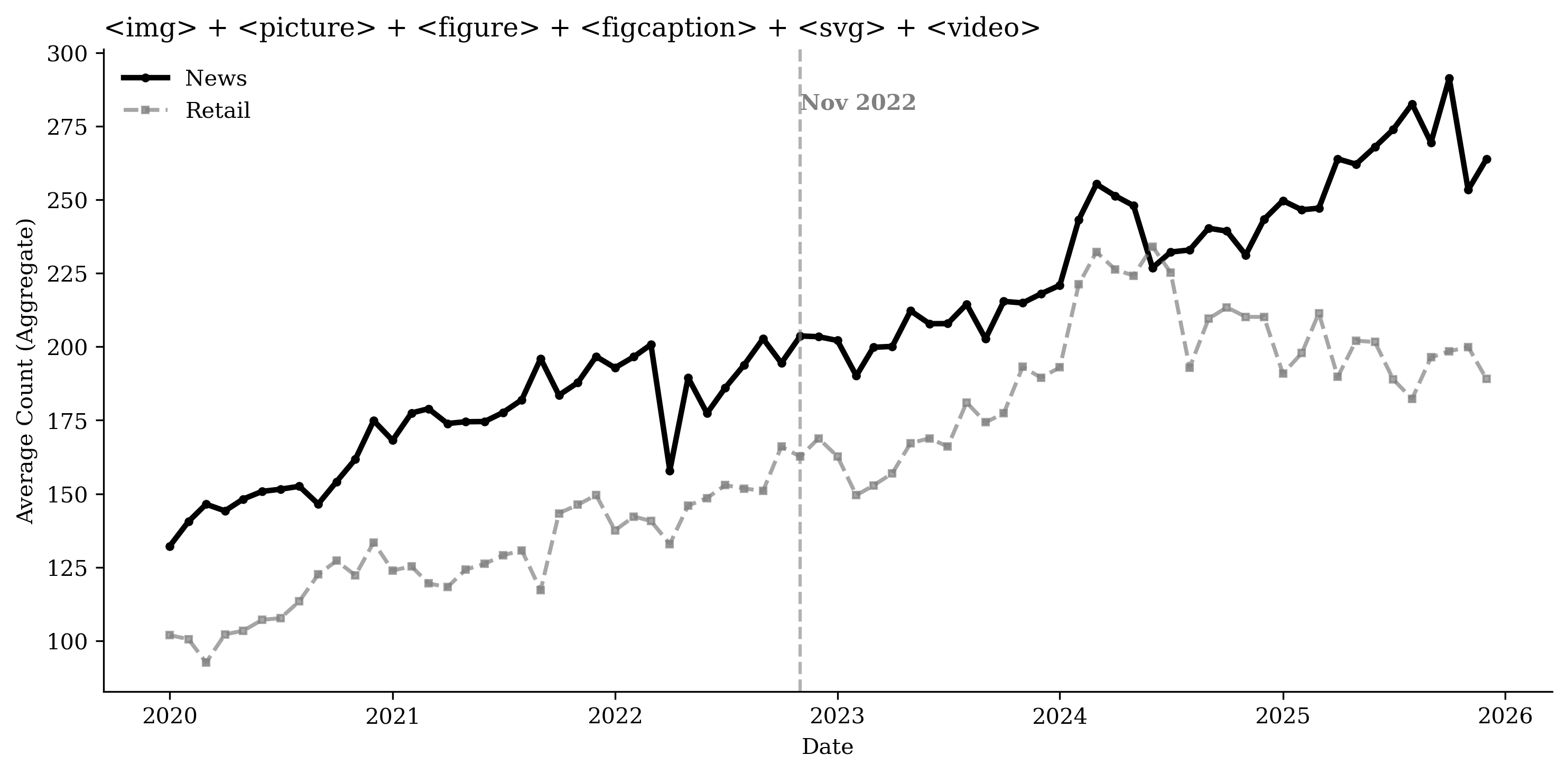}
        \label{fig:visual_multimedia}
    \end{subfigure}
    \begin{subfigure}{0.44\textwidth}
        \centering
                \caption{Interactive \& Engagement}
\includegraphics[width=\linewidth]{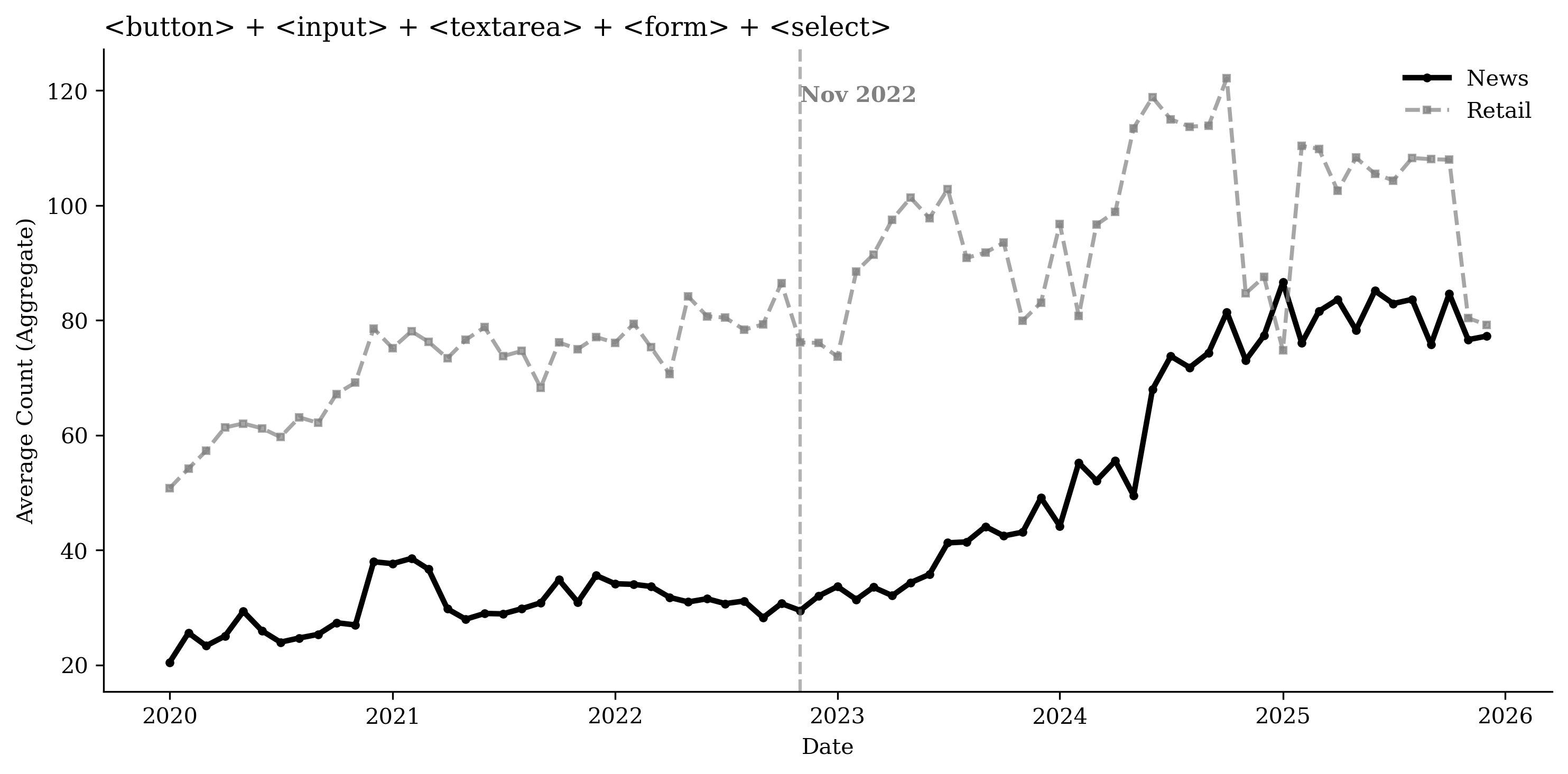}
        \label{fig:interactive_engagement}
    \end{subfigure}
\begin{subfigure}{0.44\textwidth}
        \centering
                \caption{Ads \& Targeting}
\includegraphics[width=\linewidth]{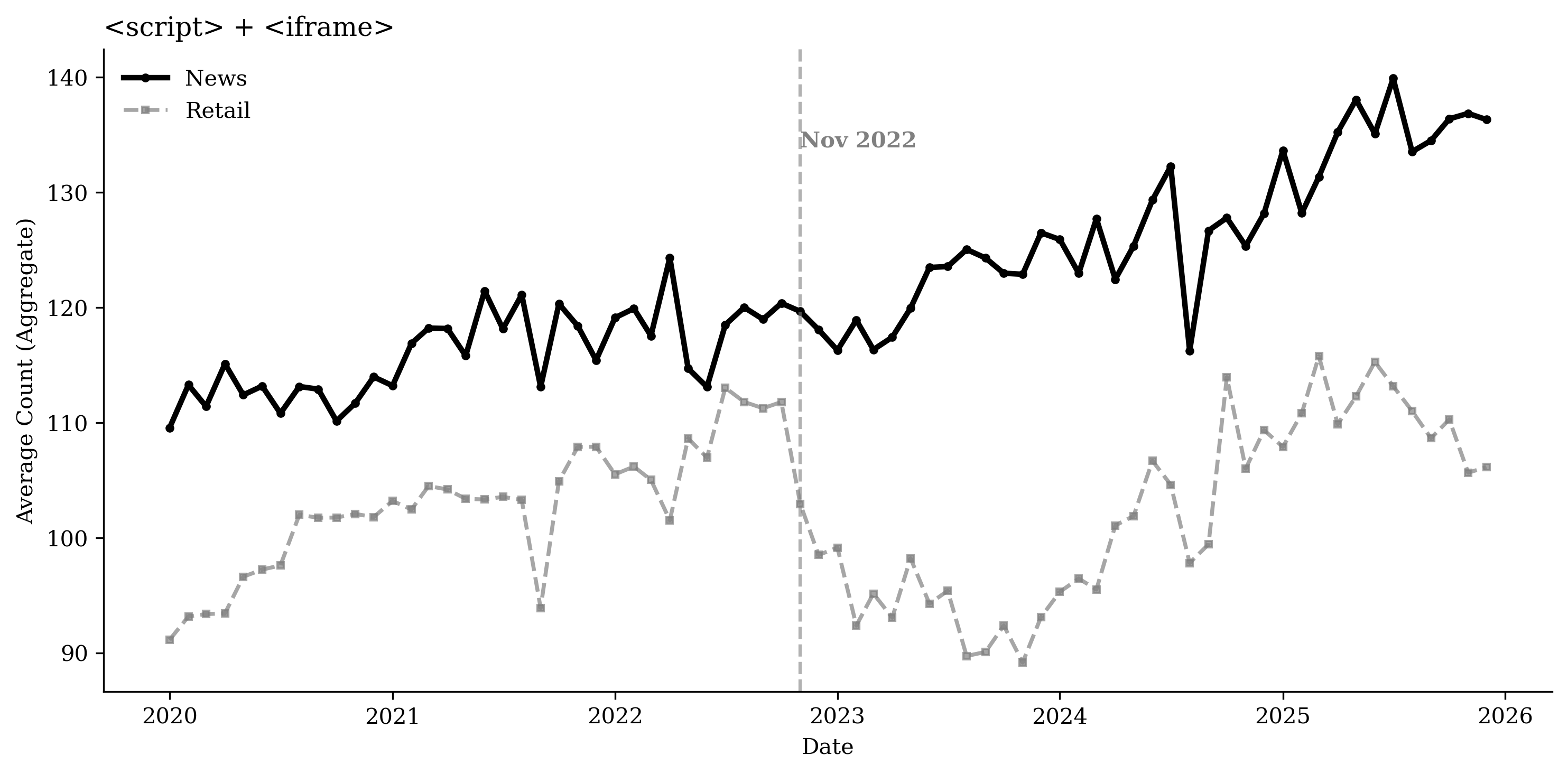}
        \label{fig:commercial_technical}
    \end{subfigure}
    \label{fig:aggregate_dom_counts}
\end{figure}

We test which pattern emerges using webpage composition data from the HTTP Archive. We count total DOM elements\footnote{A webpage's Document Object Model (DOM) is the tree of elements---headings, images,  buttons, scripts---that the browser renders.} as an aggregate measure of page complexity, as well as specific HTML elements grouped into functional categories that map to distinct strategic margins: textual article elements (\texttt{<article>}, \texttt{<section>}) proxy for the volume of editorial output; visual and multimedia elements (\texttt{<img>}, \texttt{<video>}, \texttt{<figure>}) capture investment in rich media; interactive elements (\texttt{<button>}, \texttt{<input>}) reflect efforts to deepen on-site engagement; and advertising technologies (\texttt{<script>}, \texttt{<iframe>}) indicate monetization intensity, which publishers may increase to compensate for declining traffic. We benchmark against the top 100 retail domains, which have similar total  traffic levels to our publisher sample, to separate publisher-specific  shifts from broader web development trends.
% Using the HTTP Archive, we measure webpage composition by counting all HTML elements (called DOM elements), as well as counting specific HTML elements across functional categories: number of textual article elements (e.g., \texttt{<article>}, \texttt{<section>}) which would indicate more content, visual and multimedia richness (e.g., \texttt{<img>}, \texttt{<video>}, \texttt{<figure>}) which would indicate richness in content presentation, interactive mechanisms (e.g., \texttt{<button>}, \texttt{<input>}) which would indicate attempts to engage consumers, and advertising and targeting technologies (e.g., \texttt{<script>}, \texttt{<iframe>}) which would indicate monetization opportunities. To control for broader web development trends, we compare these patterns with a control group of the top 100 retail domains. 
Figure \ref{fig:aggregate_dom_counts} plots a time series of average element counts for each category for both the news publisher and retail sectors, while Figures \ref{fig:Site_Framework_Layout_RAW} through \ref{fig:Commercial_Technical_Weight_RAW} in the Appendix present the corresponding disaggregated element counts.

While overall page complexity increased, as shown by the rise in average DOM elements (Figure \ref{fig:avg_dom_elements}, with a lagged increase relative to retailers) and general layout containers (Figure \ref{fig:site_framework}), the growth is not driven by expanded content volume. In fact, article volume declines, reflected in a reduction in the core \texttt{<article>} and \texttt{<section>} tags (Figure \ref{fig:semantic_density}). News sites are becoming more multimedia-heavy: visual assets increase at a similar rate to the retail sector (Figure \ref{fig:visual_multimedia}), while interactive engagement mechanisms, such as buttons and forms, increase at a significantly faster rate than the retail control group (Figure \ref{fig:interactive_engagement}). Growth in advertising and targeting technologies (Figure \ref{fig:commercial_technical}) indicates greater reliance on third-party scripts and embedded \texttt{<iframe>} modules, likely as publishers expand advertising and targeting to recoup revenue in a shifting traffic landscape.

Table \ref{tab:content_structure} presents two-way fixed effects regression results for the different content and structure measures, controlling for URL and month fixed effects, consistent with the patterns in Figure \ref{fig:aggregate_dom_counts}. The coefficient of interest is the interaction between an indicator for news-publisher websites and an indicator for the post--November 2022 period. Relative to the top retail websites, publishers’ advertising and targeting technologies increase by about 50\% in the post–November 2022 period. Interactive elements increase by 68.1\%, and site framework and layout increase by about 70.2\%, while article volume decreases by 31.2\%. We do not find evidence that newspapers’ overall DOM elements or multimedia elements increase faster than those of retailers.

\begin{table}[!ht]
\begin{center}
\caption{Effects on Website Content and Structure}
\label{tab:content_structure}
\resizebox{\textwidth}{!}{%
\begin{tabular}{lcccccc}
\hline\hline
& $\log(\text{DOM})$ 
& $\log(\text{Site Framework})$ 
& $\log(\text{Articles})$ 
& $\log(\text{Multimedia})$ 
& $\log(\text{Interactive})$ 
& $\log(\text{Ads})$ \\
\hline
\addlinespace
ATT & -0.259 & 0.532** & -0.374* & 0.342 & 0.519** & 0.405** \\
$p$-value & (0.266) & (0.0161) & (0.0487) & (0.0873) & (0.0123) & (0.00847) \\
BH-adjusted $p$-value
& 0.266  & 0.032  & 0.073 
& 0.105  & 0.032  & 0.032 \\\hline
$\exp(\text{ATT})-1$ (\%)
& -22.8 
& 70.2 
& -31.2 
& 40.8 
& 68.1 
& 50.0 \\
\hline
\end{tabular}%
}
\end{center}
\footnotesize
This table reports estimates from TWFE regressions with publisher and month fixed effects. The coefficient of interest is the interaction between an indicator for news-publisher websites and an indicator for the post--November 2022 period. $p$-values are reported in parentheses, and Benjamini–Hochberg adjusted $p$-values \citep{benjamini1995controlling} are reported to account for multiple hypothesis testing. Robust standard errors are clustered at the publisher level. *$p<$0.1; **$p<$0.05; ***$p<$0.01 based on Benjamini--Hochberg adjusted $p$-values.
\end{table}

Using the Internet Archive’s Wayback Machine, we collect annual counts of newly observed URLs by content type. Figure \ref{fig:content_evolution_comparison} plots the number of unique URLs of different types, which is consistent with findings from the HTTP Archive's element counts. We do not find the number of new text/article URLs to be increasing over the years; rather, there is a moderate decrease, suggesting that publishers are not responding by producing more articles. Instead, growth is concentrated in non-text assets: the number of unique image URLs increases substantially. These patterns imply that publishers’ primary adjustment margin is to enrich existing content with additional media and embedded components rather than scaling up text output.

To examine whether richer media is associated with better traffic outcomes, we correlate per-publisher linear time trends in log traffic and log media element counts across the 30 publishers. Publishers that increase their use of responsive image elements (\texttt{<picture>}) exhibit modestly better traffic  trajectories ($r = 0.37$, $p = 0.045$), though the association for video or aggregate multimedia elements  is positive but noisier ($r = 0.27$, $p = 0.14$).\footnote{Aggregate multimedia elements is the sum of \texttt{<img>}, \texttt{<picture>}, \texttt{<figure>}, \texttt{<figcaption>}, \texttt{<svg>}, \texttt{<video>}, and \texttt{<audio>} counts per page.}

% \begin{figure}[htbp]
%     \centering
%     \caption{Stock price trends of major publicly listed U.S. newspaper publishers.}
%     \begin{subfigure}[b]{0.45\textwidth}
%         \centering
%                 \caption{NYT}
% \includegraphics[width=\textwidth]{Figures/nyt.png}
%         \label{fig:nyt}
%     \end{subfigure}
%     \hfill
%     \begin{subfigure}[b]{0.45\textwidth}
%         \centering
%                 \caption{LEE}
% \includegraphics[width=\textwidth]{Figures/lee.png}
%         \label{fig:lee}
%     \end{subfigure}
%     \hfill
%     \begin{subfigure}[b]{0.45\textwidth}
%         \centering
%                 \caption{GCI}
% \includegraphics[width=\textwidth]{Figures/gci.png}
%         \label{fig:gci}
%     \end{subfigure}
%     \hfill
%     \begin{subfigure}[b]{0.45\textwidth}
%         \centering
%                 \caption{NWSA}
% \includegraphics[width=\textwidth]{Figures/nwsa.png}
%         \label{fig:nwsa}
%     \end{subfigure}
%     %------------------------%
%     \label{fig:news_stock_panel}
% \end{figure}

\subsection{No Near-Term Contraction in Newsroom Hiring Relative to Other Roles}

We next examine whether GenAI adoption is associated with a contraction in publishers’ demand for editorial labor. We measure editorial hiring using monthly job postings from Revelio and classify postings into (i) producer/editorial roles (e.g., writer, editor, content specialist, technical writer) and (ii) all other roles.

\begin{figure}[!ht]
    \begin{center}
    \caption{Job Posting Trends for Editorial and Non-Editorial Roles}
    % Panel A: Raw Counts
    \begin{subfigure}[t]{0.48\linewidth}
        \centering
                \caption{Number of Postings}
\includegraphics[width=\linewidth]{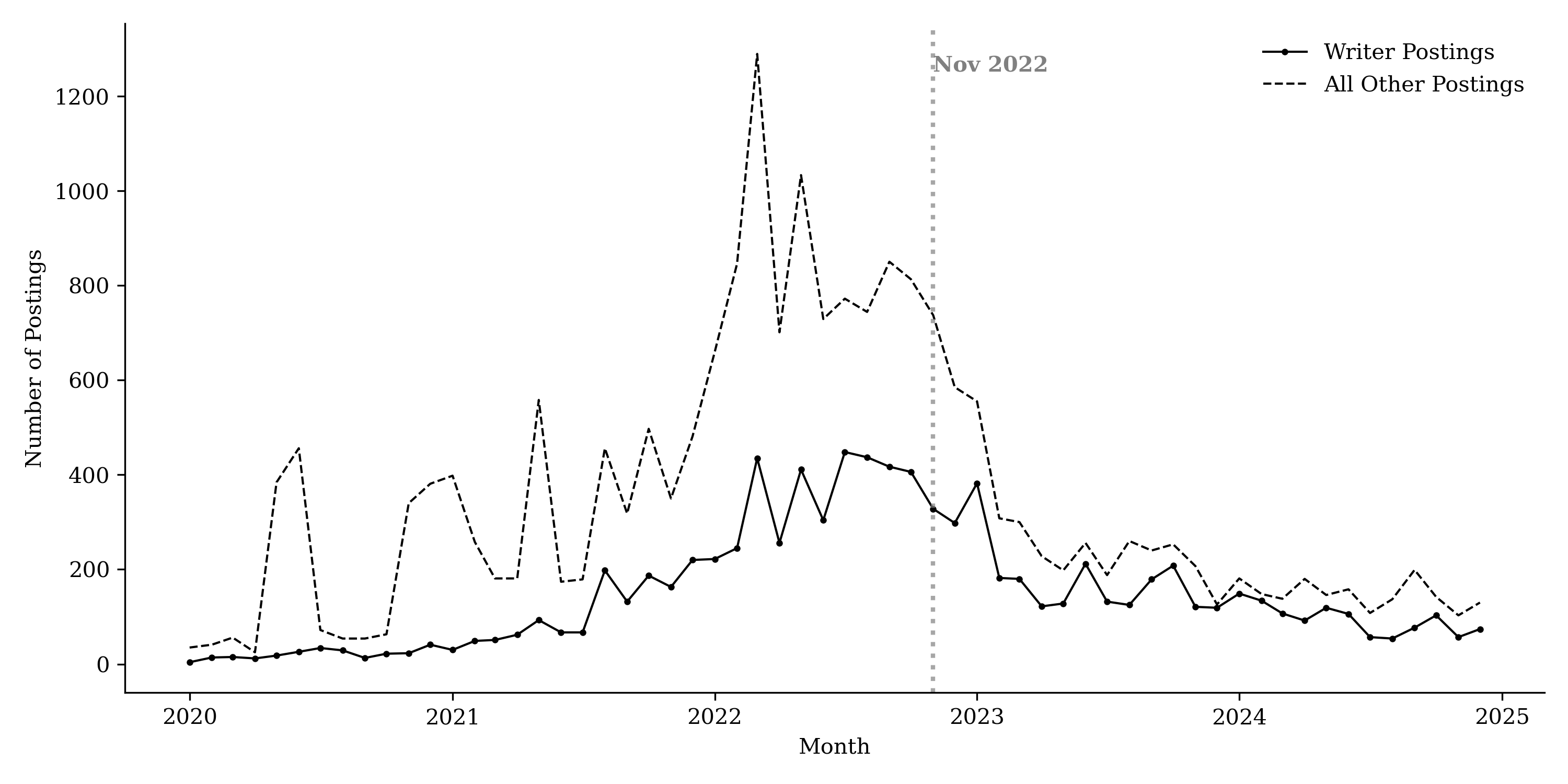}
        \label{fig:postings_counts}
    \end{subfigure}
    \hfill
    % Panel B: Ratio
    \begin{subfigure}[t]{0.48\linewidth}
        \centering
        \caption{Editorial Job Postings Ratio}
\includegraphics[width=\linewidth]{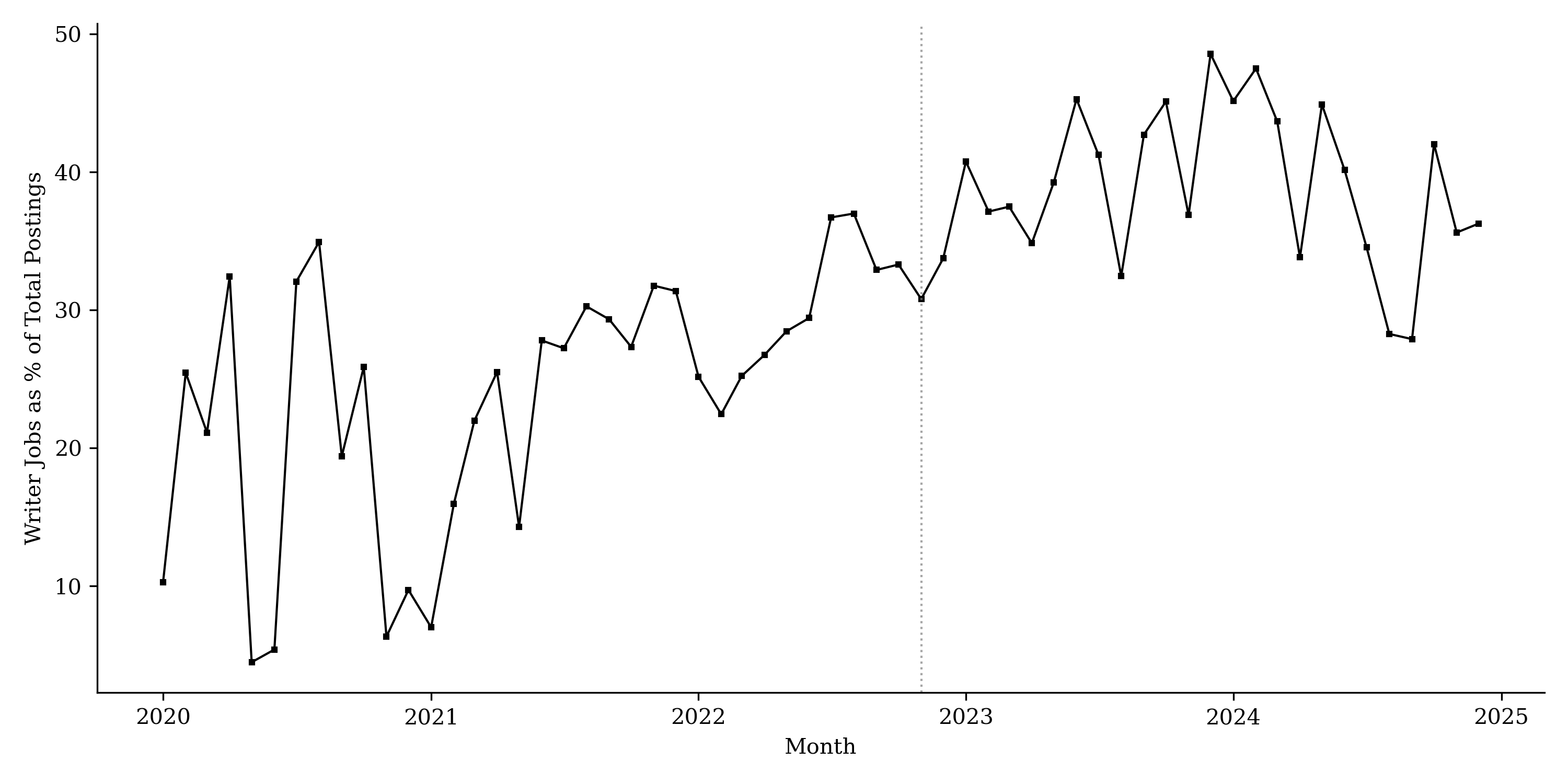}
        \label{fig:postings_ratio}
    \end{subfigure}
        \end{center}
        \footnotesize
  This figure plots the timeline of job postings for newspapers in the sample. Panel (a) displays the counts of writer-related postings (solid line) versus all other roles (dashed line). Panel (b) shows the percentage share of writer roles relative to the total number of postings. The vertical gray line indicates the release of ChatGPT in November 2022. The writer category includes Producers, Writers, Content Specialists, and Technical Writers.
    \label{fig:number_of_job_posting}
\end{figure}

Figure \ref{fig:postings_counts} plots the monthly number of producer-role postings over time. While job postings fluctuate month to month and exhibit a secular post-COVID decline in overall hiring, we do not observe a discrete collapse in producer-role postings coinciding with the expansion of GenAI use. Instead, producer-role postings appear to decline more gradually than other roles. Figure \ref{fig:postings_ratio} plots the share of producer/editorial postings relative to total postings among newspaper publishers. The plot shows that the editorial share does not fall in the post-GenAI period and, in several periods, it increases. This pattern indicates that publishers do not disproportionately reduce demand for editorial labor relative to other job categories.

\normalsize

We construct a publisher--month--category panel where non-editorial postings are considered as never treated, and editorial postings are considered treated after November 2022. Table \ref{tab:sdid_employment} reports the results of a TWFE analysis  controlling for publisher and month fixed effects.\footnote{The reason we do not implement a synthetic DiD is that the same company can appear in both the treatment and control groups for editorial and non-editorial jobs, which makes inference at the company level challenging.} Consistent with the plots, the estimated coefficient on editorial hiring after November 2022 is positive and statistically significant, providing no clear evidence of a disproportionate contraction in editorial roles relative to non-editorial hiring. This evidence aligns with our broader interpretation that GenAI has not yet functioned as a direct substitute for core newsroom labor demand during our sample period, even though job postings measure labor demand rather than realized employment and may also reflect changes in recruiting intensity or posting practices.

\begin{table}[!ht]
\begin{center}
    \def\sym#1{\ifmmode^{#1}\else\(^{#1}\)\fi}
\caption{Effect on Editorial Job Posting}
\begin{tabular}{l*{1}{c}}
\hline\hline
&   \multicolumn{1}{c}{Number of job postings} \\
        &\multicolumn{1}{c}{TWFE}  \\
\hline
ATT  &       7.578\sym{***}             \\
            &[2.897,12.260]                   \\
\hline\hline
\end{tabular}
\label{tab:sdid_employment}
\end{center}
\footnotesize
This table reports the TWFE estimate for the interaction between an indicator for editorial job postings and an indicator for the post–November 2022 period, with non-editorial postings as the comparison group. The outcome is the number of job postings. 95\% confidence intervals are reported in brackets and are based on robust standard errors clustered at the publisher level.*$p<$0.1; **$p<$0.05; ***$p<$0.01.
\end{table}

\section{Discussion and Conclusion}
\label{sec:conclusion}

This paper investigates how news publishers adjust their strategies in response to the emergence of generative AI, examining three dimensions: crawler access policies, content format choices, and hiring patterns. A key finding is that the most common protective step, restricting AI bot access through \textit{robots.txt} is associated with lower, not higher, traffic. Staggered difference-in-differences estimates across three independently constructed traffic panels indicate an approximately 7\% reduction in weekly visits after publishers begin blocking, a pattern that also appears in household-level browsing data. This suggests that limiting crawler access, while straightforward to implement, may reduce a publisher's presence in AI-generated responses and summaries, diminishing brand awareness and audience reach it was intended to preserve.

Regarding content, publishers do not appear to increase the volume of text they produce. Instead, they gradually shift their pages toward greater use of multimedia, interactive elements, formats that are less easily reproduced by language models. This movement toward richer, more experiential content is consistent with a differentiation approach rather than an effort to compete on quantity. Demand for editorial and content-production roles remains stable relative to other job categories, suggesting that early-stage GenAI has not meaningfully displaced newsroom hiring.

Our analysis also has some limitations. The traffic data we have access to do not capture consumption that takes place entirely within AI interfaces, and the study period precedes the wider adoption of AI-integrated search products. Future research incorporating direct measures of LLM-driven referral traffic, and licensing arrangements would help clarify how these dynamics evolve as the technology becomes more widespread.

In terms of practical considerations for publishers and other content providers, we can offer the following: technical access restrictions carry understated costs, while investing in distinctive content formats may offer a steadier path forward. At the same time, some publishers may view blocking as part of a longer-run strategic posture, accepting short-run traffic losses to strengthen their position in future negotiations over licensing, attribution, or platform access.

% Several limitations qualify interpretation. Our traffic measures combine modeled aggregates (SimilarWeb) with a U.S. desktop panel (Comscore), and neither directly observes LLM-mediated consumption paths (e.g., in-chat answers, referrals, or citation links). The study covers an early phase of the technology, before newer interfaces and integration (e.g., more prominent AI summaries and search-chat products) may fully diffuse. 

% Future work can sharpen mechanisms by incorporating direct measures of LLM discovery and referral, richer firm-level data on AI access/licensing negotiations and enforcement. More broadly, continued monitoring will help determine whether the patterns documented here intensify, attenuate, or shift toward new adjustments as AI capabilities and adoption change.

\section*{Funding and Competing Interests}
Partial financial support was received from (removed for blind review). All authors certify that they have no affiliations with or involvement in any organization or entity with any financial interest or non-financial interest in the subject matter or materials discussed in this manuscript.

\singlespacing
\bibliography{references}

% ==================================================
%   Appendix counters and formatting
% ==================================================

% --------------------------------------------------
%   Regular Appendices (A.1, A.2, A.1.1, etc.)
% --------------------------------------------------
\newcounter{appsection}
\newcounter{appsubsection}[appsection]
\renewcommand{\theappsection}{A.\arabic{appsection}}
\renewcommand{\theappsubsection}{A.\arabic{appsection}.\arabic{appsubsection}}

\newcommand{\appendicesstart}{%
  \clearpage
  \setcounter{table}{0}%
  \setcounter{figure}{0}%
  \setcounter{equation}{0}%
  \renewcommand{\thetable}{A.\arabic{table}}%
  \renewcommand{\thefigure}{A.\arabic{figure}}%
  \renewcommand{\theequation}{A.\arabic{equation}}%
}

\newcommand{\appsection}[1]{%
  \refstepcounter{appsection}%
  \section*{Appendix~\theappsection.~#1}%
  \addcontentsline{app}{section}{Appendix~\theappsection.~#1}%
}

\newcommand{\appsubsection}[1]{%
  \refstepcounter{appsubsection}%
  \subsection*{\theappsubsection.~#1}%
  \addcontentsline{app}{subsection}{\theappsubsection.~#1}%
}

% --------------------------------------------------
%   Web Appendices (WA.1, WA.2, WA.1.1, etc.)
% --------------------------------------------------
\newcounter{webappsection}
\newcounter{webappsubsection}[webappsection]
\renewcommand{\thewebappsection}{WA.\arabic{webappsection}}
\renewcommand{\thewebappsubsection}{WA.\arabic{webappsection}.\arabic{webappsubsection}}

\newcommand{\webappendicesstart}{%
  \clearpage
  \setcounter{page}{1}%
  \renewcommand{\thepage}{WA-\arabic{page}}%
  \setcounter{table}{0}%
  \setcounter{figure}{0}%
  \setcounter{equation}{0}%
  \renewcommand{\thetable}{WA.\arabic{table}}%
  \renewcommand{\thefigure}{WA.\arabic{figure}}%
  \renewcommand{\theequation}{WA.\arabic{equation}}%
}

\newcommand{\webappsection}[1]{%
  \refstepcounter{webappsection}%
  \section*{Web Appendix~\thewebappsection.~#1}%
  \addcontentsline{app}{section}{Web Appendix~\thewebappsection.~#1}%
}

\newcommand{\webappsubsection}[1]{%
  \refstepcounter{webappsubsection}%
  \subsection*{\thewebappsubsection.~#1}%
  \addcontentsline{app}{subsection}{\thewebappsubsection.~#1}%
}

\onehalfspacing
\webappendicesstart

\section*{Web Appendix}
Table   \ref{tab:url_naics_distribution} summarizes the number of unique URLs and their proportions in each  NAICS sector. 

\begin{table}[!ht]
  \begin{center}
  \caption{Distribution of URLs by NAICS Sector}
   \resizebox{0.98\textwidth}{!}{
  \begin{tabular}{llll}
  \hline
  \textbf{Sector} & \textbf{URL} & \textbf{Ratio (\%)} & \textbf{Industry Name}
  \\ \hline
  44-45 & 2,423 & 38.37 & Retail Trade \\
  72 & 1,599 & 25.32 & Accommodation and Food Services \\
  31-33 & 1,015 & 16.07 & Manufacturing \\
  52 & 650 & 10.29 & Finance and Insurance \\
  62 & 589 & 9.33 & Health Care and Social Assistance \\
  81 & 531 & 8.41 & Other Services (except Public Administration) \\
  42 & 500 & 7.92 & Wholesale Trade \\
  54 & 485 & 7.68 & Professional, Scientific, and Technical Services \\
  51 & 387 & 6.13 & Information \\
  92 & 319 & 5.05 & Public Administration \\
  71 & 318 & 5.04 & Arts, Entertainment, and Recreation \\
  53 & 267 & 4.23 & Real Estate and Rental and Leasing \\
  61 & 244 & 3.86 & Educational Services \\
  48-49 & 238 & 3.77 & Transportation and Warehousing \\
  56 & 233 & 3.69 & Administrative, Support, Waste Management and Remediation
  Services \\
  11 & 105 & 1.66 & Agriculture, Forestry, Fishing and Hunting \\
  23 & 102 & 1.62 & Construction \\
  99 & 88 & 1.39 & Unclassified \\
  22 & 51 & 0.81 & Utilities \\
  55 & 45 & 0.71 & Management of Companies and Enterprises \\ \hline
  \end{tabular}}
  \label{tab:url_naics_distribution}
  \end{center}
  \footnotesize
A single URL may be associated with multiple NAICS codes.
  \end{table}

\normalsize

Table \ref{tab:url_list} presents the total website traffic for 2023 from SimilarWeb data.

\begin{table}[!ht]
  \begin{center}
      \caption{Website Traffic in 2023 SimilarWeb Data}
  \label{tab:url_list}
  \begin{tabular}{lr | lr}
  \hline
  \textbf{URL} & \textbf{Traffic ($10^9$)} & \textbf{URL} & \textbf{Traffic
  ($10^9$)} \\ \hline
  cnn.com & 6.75 & reuters.com & 1.11 \\
  nytimes.com & 6.73 & npr.org & 1.08 \\
  bbc.com & 5.83 & businessinsider.com & 1.08 \\
  espn.com & 5.48 & screenrant.com & 1.06 \\
  dailymail.co.uk & 4.26 & the-sun.com & 1.01 \\
  theguardian.com & 3.99 & nbcnews.com & 0.99 \\
  foxnews.com & 3.95 & wsj.com & 0.94 \\
  nypost.com & 2.06 & si.com & 0.93 \\
  usatoday.com & 1.74 & apnews.com & 0.85 \\
  people.com & 1.68 & cbssports.com & 0.83 \\
  cnbc.com & 1.60 & tmz.com & 0.79 \\
  forbes.com & 1.52 & insider.com & 0.74 \\
  washingtonpost.com & 1.51 & marketwatch.com & 0.65 \\
  ign.com & 1.37 & 247sports.com & 0.47 \\
  buzzfeed.com & 1.21 & foxsports.com & 0.19 \\ \hline
  \end{tabular}
  \end{center}
  \end{table}

As a complementary parametric check, we estimate a simple autoregressive specification around each detected break date. For a given change point $\tau_k$ (with the next change point denoted $\tau_{k+1}$), we estimate the following regression:
\begin{equation}
    y_{t}
    = \alpha
    + \rho_0\, y_{t-1}
    + \rho_1\, y_{t-1} \cdot \mathbf{1}\{ \tau_k \le t \le \tau_{k+1}\}
    + \rho_2\, \mathbf{1}\{\tau_k \le t \le \tau_{k+1}\}
    + \epsilon_{t},
\end{equation}
where $y_{t}$ denotes residualized log traffic (after controlling for calendar-week, day-of-week, and month fixed effects) on day $t$.
The indicator $\mathbf{1}\{\tau_k \le t \le \tau_{k+1}\}$ identifies the post-break segment (from $\tau_k$ up to $\tau_{k+1}$).
In this specification, $\rho_2$ captures a level shift in traffic at the break, while $\rho_1$ captures a change in persistence of traffic dynamics after the break.
We report the results in Table \ref{tab:lag_traffic_regression}. The significantly negative estimates of $\rho_2$ after the January  2023 change points suggest a decrease in the mean level of residualized traffic. Table \ref{tab:lag_traffic_regression_placebo} presents results from the same regression model using randomly selected placebo dates. Across these specifications, we find no significant effect of the post-placebo indicators, confirming that the main results at the actual detection points capture systematic changes rather than random fluctuations.

\begin{table}[!ht] 
\begin{center}
  \caption{AR(1) Break Regressions at Detected Cutoff Dates}
      % \resizebox{\textwidth}{!}{%
\begin{tabular}{lccccc}
\hline\hline
 & \multicolumn{5}{c}{\textit{Dependent variable: log traffic residuals}} \\
\cline{2-6}
  & 2022-02-15 & 2022-04-01 & 2022-06-05 & 2022-09-03 & 2023-01-16 \\
\hline
(lag $\times$ post) $\rho_1$ & 0.037 & -0.149 & 0.314** & 0.047 & 0.098 \\
  & (0.151) & (0.159) & (0.152) & (0.127) & (0.063) \\
Lagged residuals $\rho_0$  & 0.810*** & 0.756*** & 0.301** & 0.586*** & 0.622*** \\
  & (0.135) & (0.070) & (0.128) & (0.112) & (0.052) \\
Post indicator  $\rho_2$ & 0.011 & -0.010 & -0.011* & 0.032*** & -0.022*** \\
  & (0.013) & (0.010) & (0.006) & (0.008) & (0.005) \\
\hline
Observations & 89 & 109 & 154 & 224 & 605 \\
Adjusted $R^2$ & 0.724 & 0.560 & 0.316 & 0.568 & 0.565 \\
F Statistic & 77.863*** & 46.909*** & 24.537*** & 98.921*** & 262.873*** \\
\hline\hline
\end{tabular}
\label{tab:lag_traffic_regression}
\end{center}
\footnotesize
Each column reports an AR(1) regression estimated around the indicated cutoff date. The post indicator equals 1 for observations in the post-break segment (from the cutoff date up to the next detected change point)  *p$<$0.1; **p$<$0.05; ***p$<$0.01.
\end{table}

\begin{table}[!ht] 
\begin{center}
  \caption{AR(1) Break Regressions at Placebo Cutoff Dates}
      % \resizebox{\textwidth}{!}{%
\begin{tabular}{lcccc}
\hline\hline
 & \multicolumn{4}{c}{\textit{Dependent variable: log traffic residuals}} \\
\cline{2-5}
  & 2019-08-09 & 2020-07-16 & 2021-06-22 & 2023-10-30 \\
\hline
(lag $\times$ post) $\rho_1$ & -0.055 & -0.038 & -0.299*** & -0.070 \\
  & (0.087) & (0.051) & (0.085) & (0.074) \\
Lagged residuals $\rho_0$ & 0.599*** & 0.909*** & 0.839*** & 0.744*** \\
  & (0.068) & (0.041) & (0.035) & (0.048) \\
Post indicator $\rho_2$ & 0.007 & 0.004 & 0.000 & -0.005 \\
  & (0.005) & (0.007) & (0.005) & (0.004) \\
\hline
Observations & 365 & 366 & 365 & 366 \\
Adjusted $R^2$ & 0.340 & 0.786 & 0.635 & 0.521 \\
F Statistic & 63.578*** & 446.892*** & 212.325*** & 133.562*** \\
\hline\hline
\multicolumn{5}{l}{\small *p$<$0.1; **p$<$0.05; ***p$<$0.01} \\
\end{tabular}
\label{tab:lag_traffic_regression_placebo}
% }
\end{center}
\footnotesize
Each column reports the AR(1) regression estimates within a six-month window before and after four random placebo cutoff dates. The post indicator equals 1 for observations in the post-break segment (from the cutoff date up to the next detected change point). *p$<$0.1; **p$<$0.05; ***p$<$0.01.
\end{table}

% \begin{figure}[!ht]
%     \centering
%     \caption{Total, Direct, and Google Referral weekly traffic trends.}
%     \label{fig:traffic_aggregated_sources}
%     \includegraphics[width=0.75\textwidth]{Figures/Traffic_Weekly_Aggregated_Sources.png}
    
%     \par\smallskip % Adds a small vertical space
%         \footnotesize
%         The plot compares Total Traffic against specific referral sources (Google.com and Direct) from Comscore data. The vertical dashed line indicates July 1, 2024.
% \end{figure}

\clearpage

\begin{table}[htbp]
\centering
\caption{Major GenAI / LLM-related crawlers and user agents}
\label{tab:genai_bot}
\small
    \resizebox{0.9\textwidth}{!}{%
\begin{tabularx}{\textwidth}{@{}l p{0.15\textwidth}  p{0.45\textwidth} p{0.27\textwidth}@{}}
\toprule
Vendor & User agent & Description & Reference URL \\
\midrule
\multirow{3}{*}{OpenAI} 
  & gptbot        & OpenAI crawler that collects web content to train and improve models like ChatGPT. 
                  & \url{https://platform.openai.com/docs/gptbot} \\
  & chatgpt-user  & User-triggered retrieval agent that fetches pages when a ChatGPT user opens links or uses tools. 
                  & \\ % URL omitted (same as above vendor docs)
  & oai-searchbot & OpenAI browsing/search bot used to fetch content for ChatGPT and related products. 
                  & \\ % URL omitted
\midrule
\multirow{4}{*}{Anthropic} 
  & claudebot       & Anthropic training crawler that gathers public web data for Claude models. 
                    & \url{https://privacy.claude.com/en/articles/8896518-does-anthropic-crawl-data-from-the-web-and-how-can-site-owners-block-the-crawler} \\
  & claude-user     & User-initiated Claude crawler that fetches specific URLs during a conversation. 
                    & \\ % same official article
  & claude-searchbot& Search/browsing agent attributed to Anthropic; behavior inferred from industry reports, not separately documented. 
                    & \\ % only main URL kept above
  & anthropic-ai    & Undocumented Anthropic-related agent string occasionally seen in logs. 
                    & \\ % only main URL kept above
\midrule
\multirow{2}{*}{Perplexity} 
  & perplexitybot   & Perplexity crawler used to index and collect data for its answer engine and LLMs. 
                    & \url{https://perplexity.mintlify.app/guides/bots} \\
  & perplexity-user & User-initiated fetches from Perplexity sessions to retrieve specific web pages. 
                    & \\ % same docs as above
\midrule
\multirow{2}{*}{Cohere} 
  & cohere-ai                & General Cohere agent string used by their services and tooling. 
                             & \url{https://darkvisitors.com/agents/cohere-training-data-crawler} \\
  & cohere-training-data-crawler & Reported Cohere crawler for collecting training data; not separately documented. 
                             & \\ % only main site URL kept above
\midrule
Google 
  & google-extended & Robots.txt control user agent that governs whether content is used to train and ground Gemini and Vertex~AI models. 
                    & \url{https://developers.google.com/crawling/docs/crawlers-fetchers/google-common-crawlers} \\
\midrule
Apple 
  & applebot-extended & Control agent that lets sites opt out of having Applebot data used to train Apple foundation models and Apple Intelligence. 
                      & \url{https://support.apple.com/en-us/119829} \\
\midrule
\multirow{2}{*}{Meta} 
  & meta-externalagent  & Meta crawler that collects web data for AI model training and content indexing across Meta products. 
                         & \url{https://developers.facebook.com/docs/sharing/webmasters/web-crawlers/} \\
  & meta-externalfetcher & Meta crawler that performs user-initiated fetches of individual links for specific product features. 
                         & \\ % same crawler docs as above
\midrule
\multirow{2}{*}{DuckDuckGo} 
  & duckassistbot & DuckDuckGo crawler that powers the DuckAssist generative answers feature. 
                  & \url{https://duckduckgo.com/duckduckgo-help-pages/results/duckduckbot} \\
  & duckassist    & Alternative user-agent token for the same DuckAssist AI answer system. 
                  & \\ % same DuckAssist page as above
\midrule
Amazon 
  & amazonbot   & Amazon crawler used to improve products like Alexa and may be used to train Amazon AI models. 
                & \url{https://developer.amazon.com/amazonbot} \\
\midrule
ByteDance 
  & bytespider  & ByteDance crawler widely reported as collecting data for search, recommendation, and LLM training. 
                & \url{https://datadome.co/bots/bytedance-crawler/} \\
\bottomrule
\end{tabularx}
}
\end{table}

\normalsize

\begin{figure}[!ht]
     \begin{center}
          \caption{GenAI Bot Blocking Fractions by News Publishers Traffic Rank Group and Retailers}
     % --- Group 1: Top 1-33 ---
     \begin{subfigure}[b]{0.49\textwidth}
         \centering
                  \caption{Group 1}
\includegraphics[width=\textwidth]{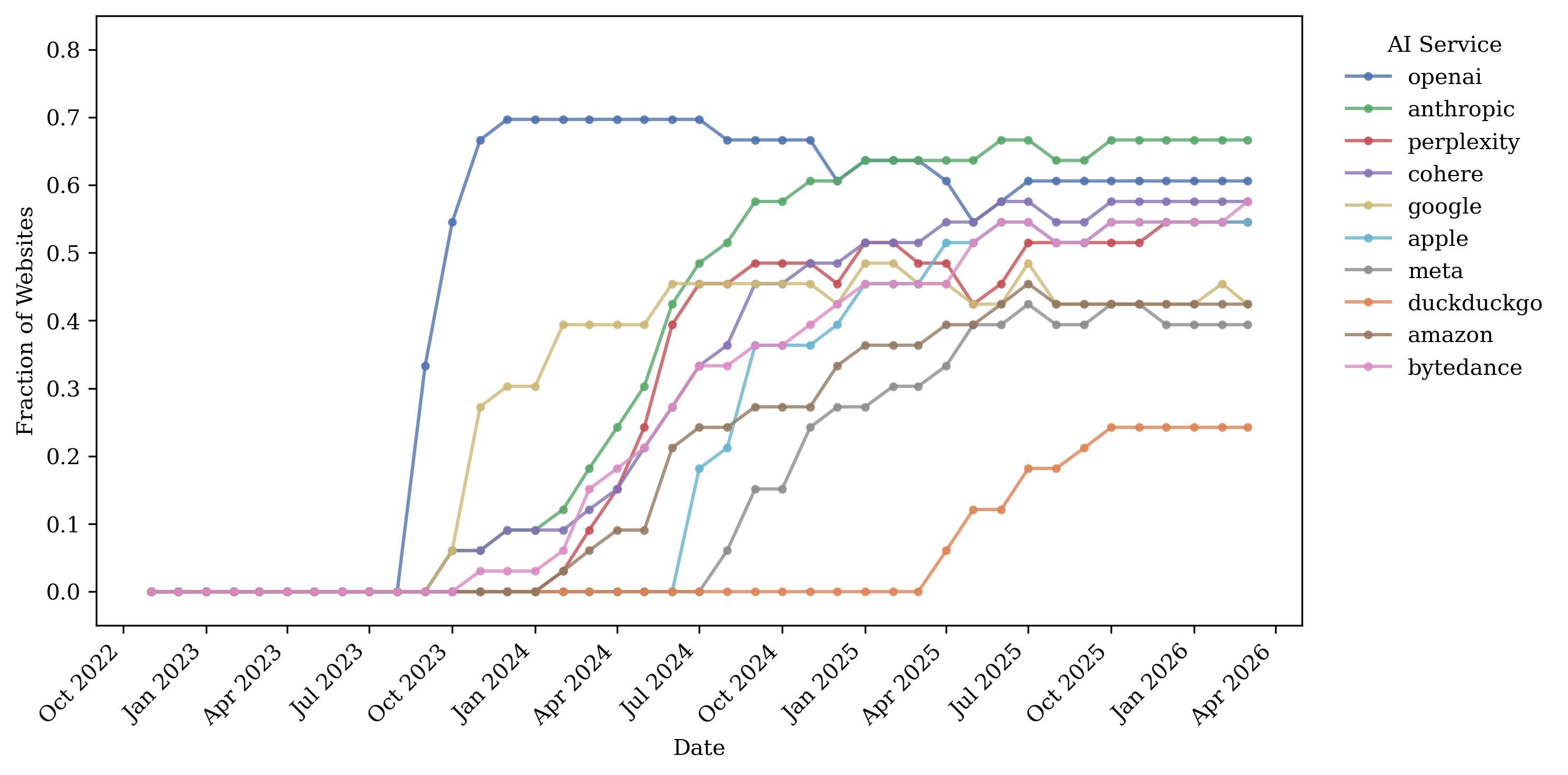}
         \label{fig:block_top}
     \end{subfigure}
     \hfill % Fill space between figures
     % --- Group 2: Rank 34-163 ---
     \begin{subfigure}[b]{0.49\textwidth}
         \centering
        \caption{Group 2}
\includegraphics[width=\textwidth]{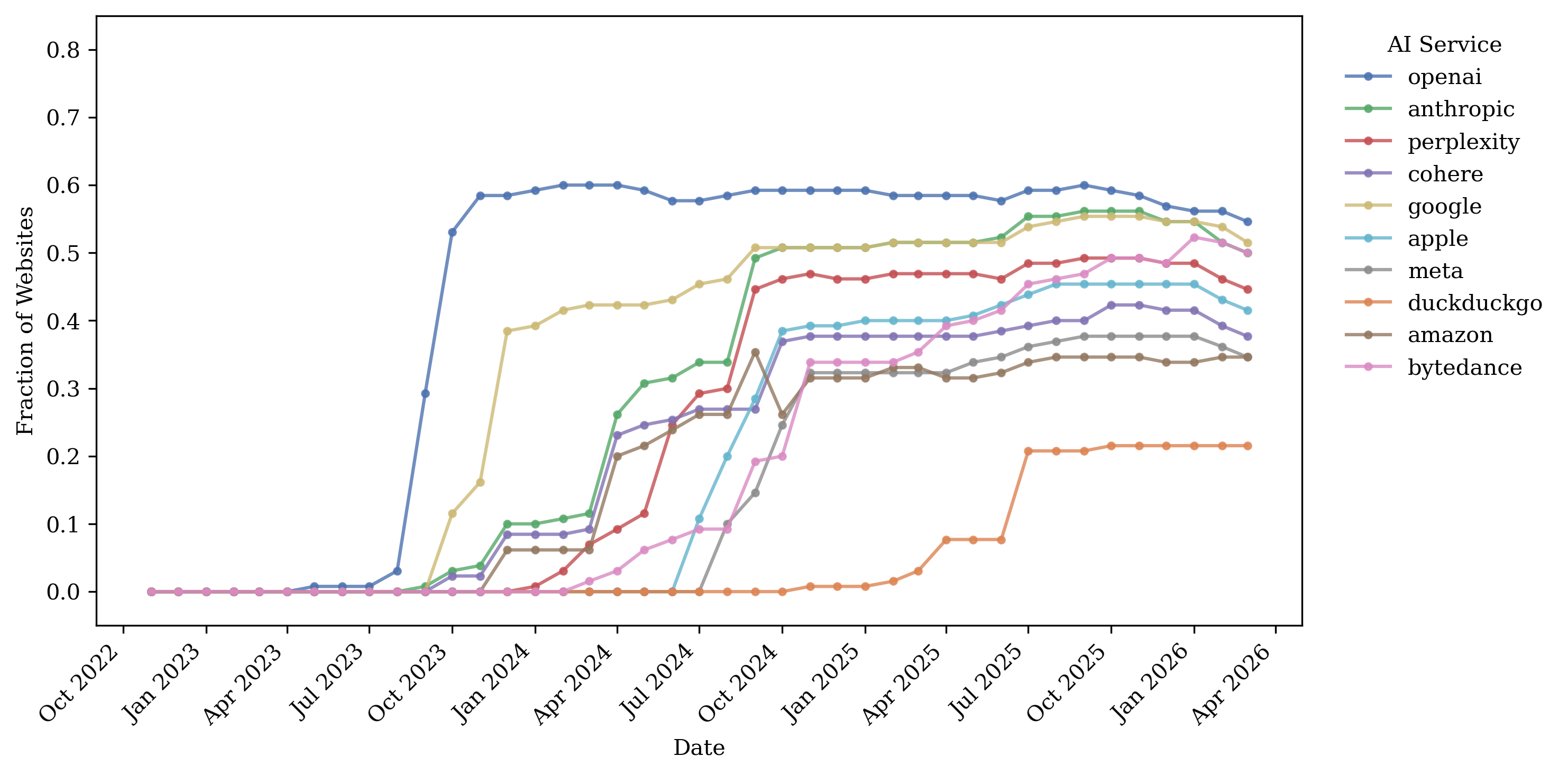}
         \label{fig:block_mid}
     \end{subfigure}
     \hfill
     % --- Group 3: Rank 164-501 ---
     \begin{subfigure}[b]{0.49\textwidth}
         \centering
        \caption{Group 3}
\includegraphics[width=\textwidth]{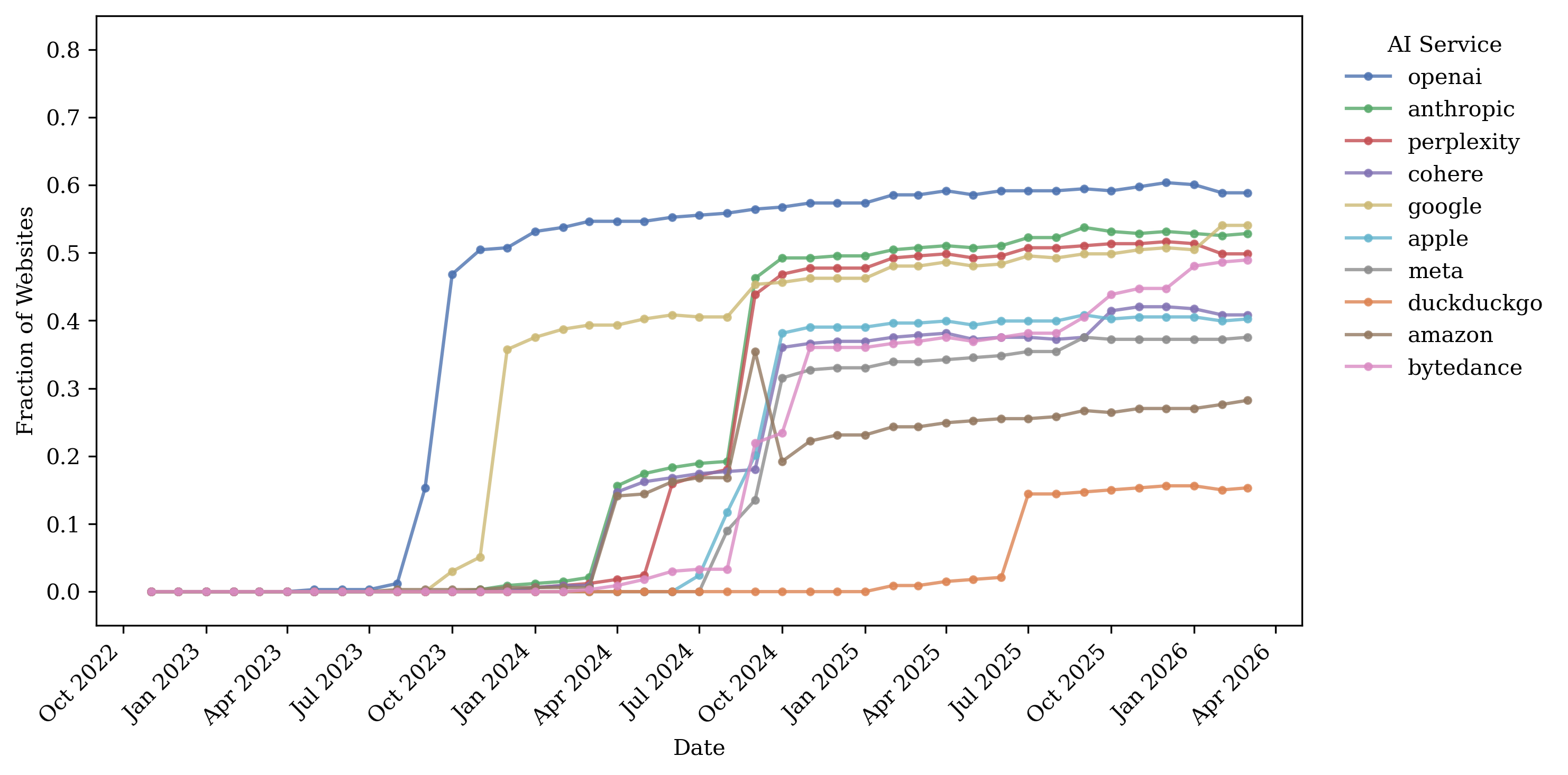}         \label{fig:block_low}
     \end{subfigure}
\begin{subfigure}[b]{0.49\textwidth}
     \centering
     \caption{Top 100 Retailers}
\includegraphics[width=\textwidth]{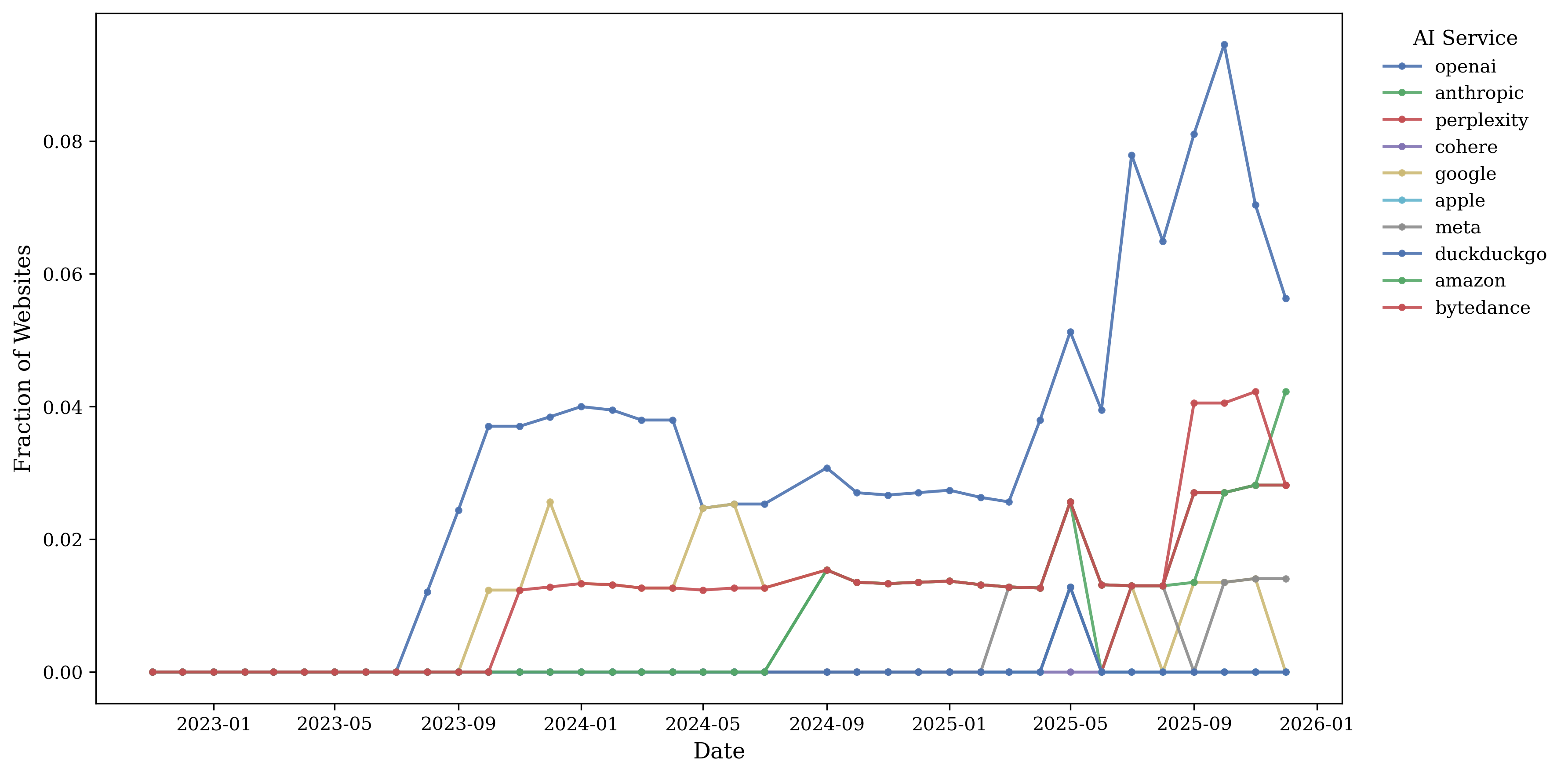}
     \label{fig:genai_bot_blocking_retail}
 \end{subfigure}
\label{fig:blocking_groups}
          \end{center}
          \footnotesize
This figure plots the fraction of websites that disallow GenAI bots over time. Panel (a), Group 1, includes the largest publishers (more than 10 visits per day on average); Panel (b), Group 2, includes mid-sized publishers (1–10 visits per day; ranked 34th–164th by average daily visits); Panel (c), Group 3, includes the smallest publishers (fewer than 1 visit per day on average); and Panel (d) includes the top 100 retailers.
\end{figure}

 \begin{figure}[!ht]
\begin{center}
        \caption{Staggered DiD of blocking GenAI bots on publisher traffic
  (extended window)}
      \label{fig:csdid_weekly_longer}

      \begin{subfigure}[t]{0.49\textwidth}
          \centering
          \caption{SimilarWeb}
          \includegraphics[width=\linewidth]{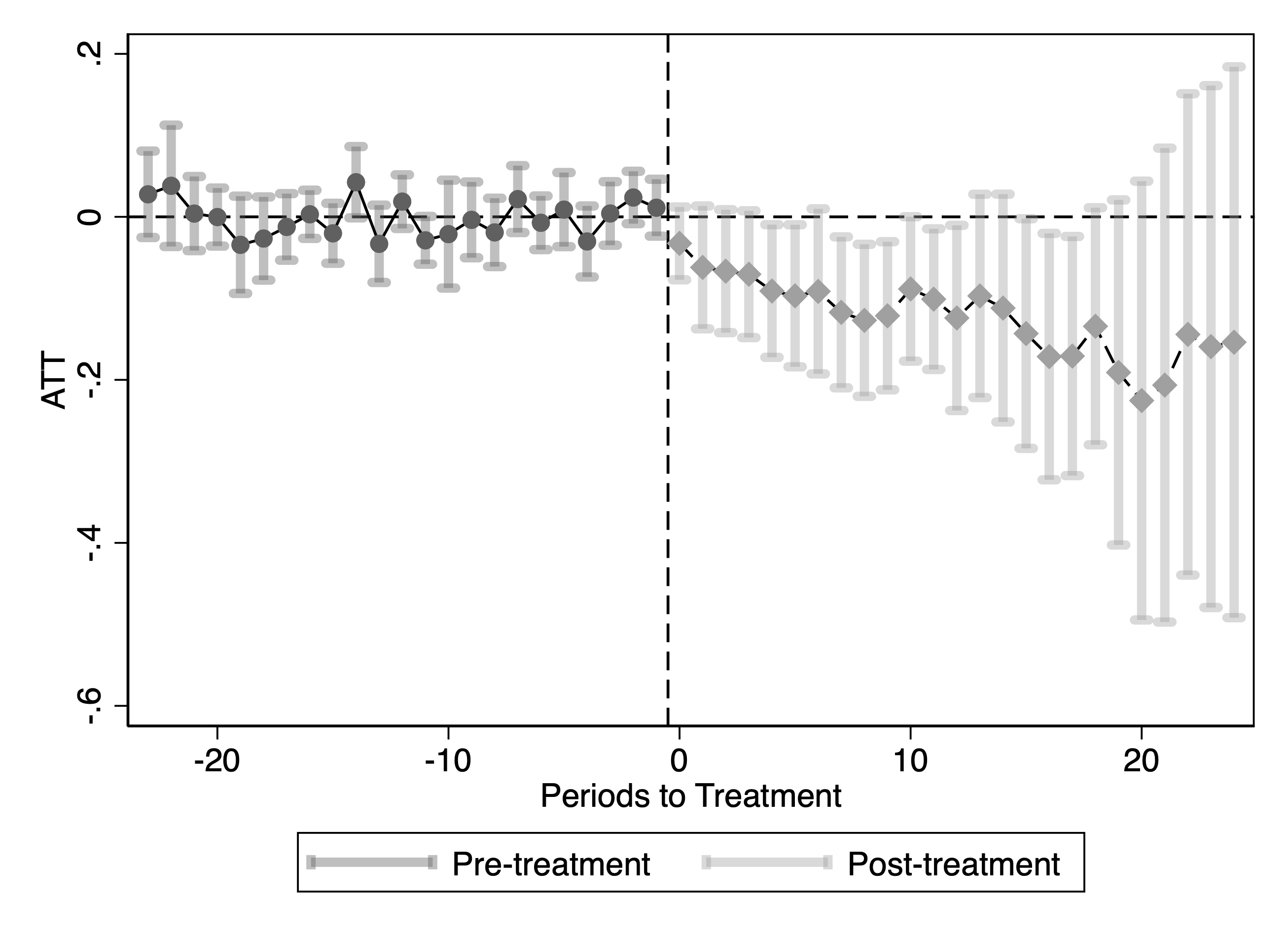}
          \label{fig:csdid_weekly_sw_longer}
      \end{subfigure}
      \hfill
      \begin{subfigure}[t]{0.49\textwidth}
          \centering
          \caption{Semrush}
          \includegraphics[width=\linewidth]{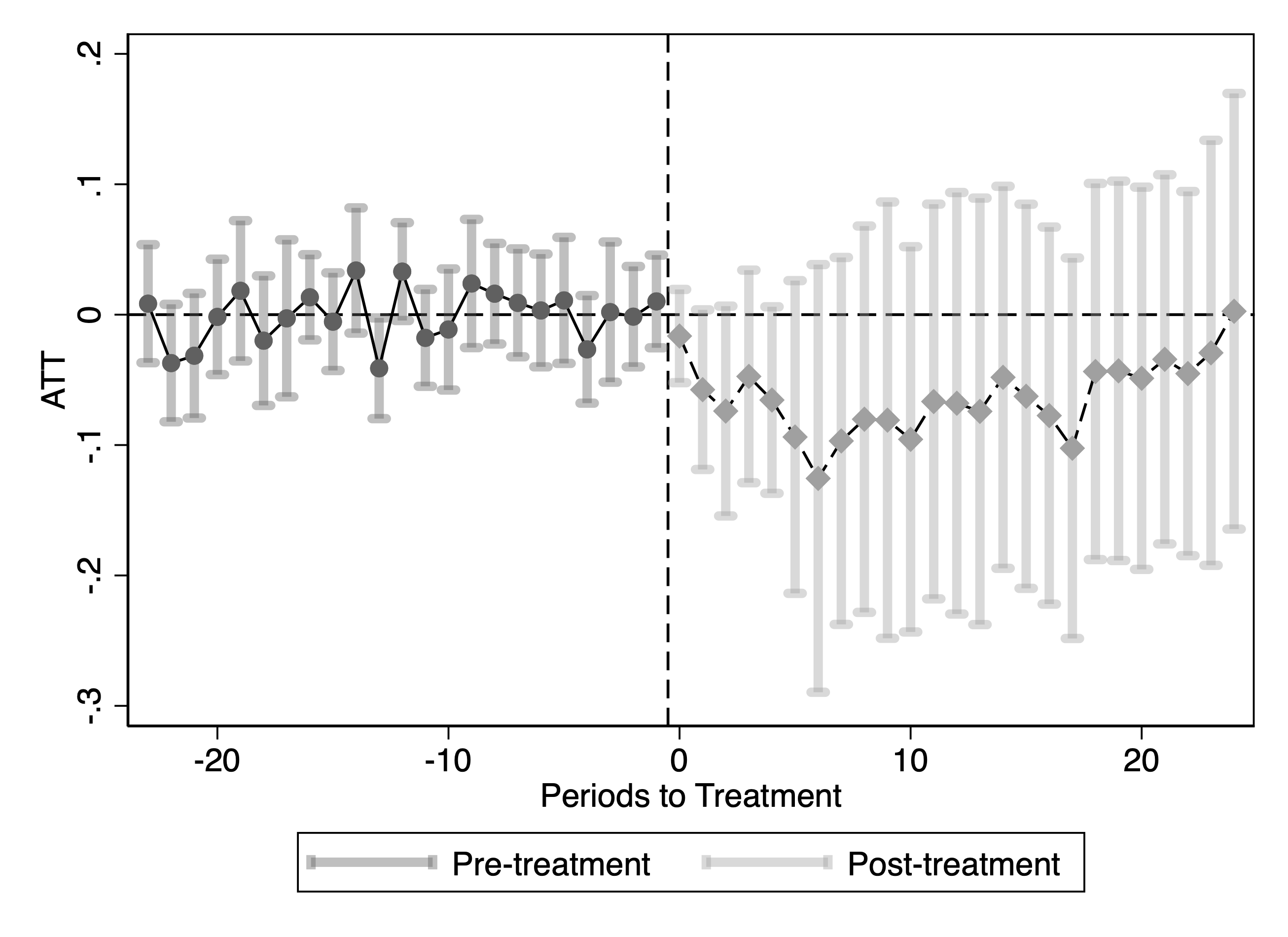}
          \label{fig:csdid_weekly_sr_longer}
      \end{subfigure}
\end{center}  
      \par\bigskip
      \footnotesize
      Staggered DiD event-study estimates of the effect of blocking GenAI web
  crawlers on publisher traffic, using a 24-week window before and after
  blocking. The outcome is the logarithm of weekly visits from SimilarWeb (a)
  and Semrush (b). Confidence intervals are based on 50 bootstrap replications.
  \end{figure}

% \begin{table}[!ht]
% \begin{center}
%     \caption{Staggered DiD  estimates of blocking GenAI crawlers on publisher traffic}
% \label{tab:att_estimates_longer}
% \begin{tabular}{l*{4}{c}}
% \hline\hline
%                     &\multicolumn{2}{c}{SimilarWeb}&\multicolumn{2}{c}{Semrush}\\
% \hline
% ATT                 &               -0.0926*** &                -0.124** &               -0.0746   &               -0.0630   \\
%                     &      [-0.163,-0.0219]   &     [-0.240,-0.00749]   &       [-0.184,0.0345]   &       [-0.184,0.0579]   \\\hline
% Window                 &               12-week post &                24-week post &               12-week post   &               24-week post   \\
% \hline
% \hline
% \end{tabular}
% \end{center}
% \footnotesize
% This table reports staggered DiD ATT estimates of the effect of blocking GenAI web crawlers on publisher traffic, using a 12-week or 24-week window before and after blocking. The dependent variables in the columns (1)-(2), (3)-(4) are the logarithm of weekly visits from  SimilarWeb and Semrush data, respectively. Confidence intervals are based on 50 bootstrap replications. *$p<$0.1; **$p<$0.05; ***$p<$0.01.
% \end{table}
% \normalsize

  \begin{figure}[!ht]
     \centering
          \caption{Site Framework and Layout}
\includegraphics[width=\linewidth]{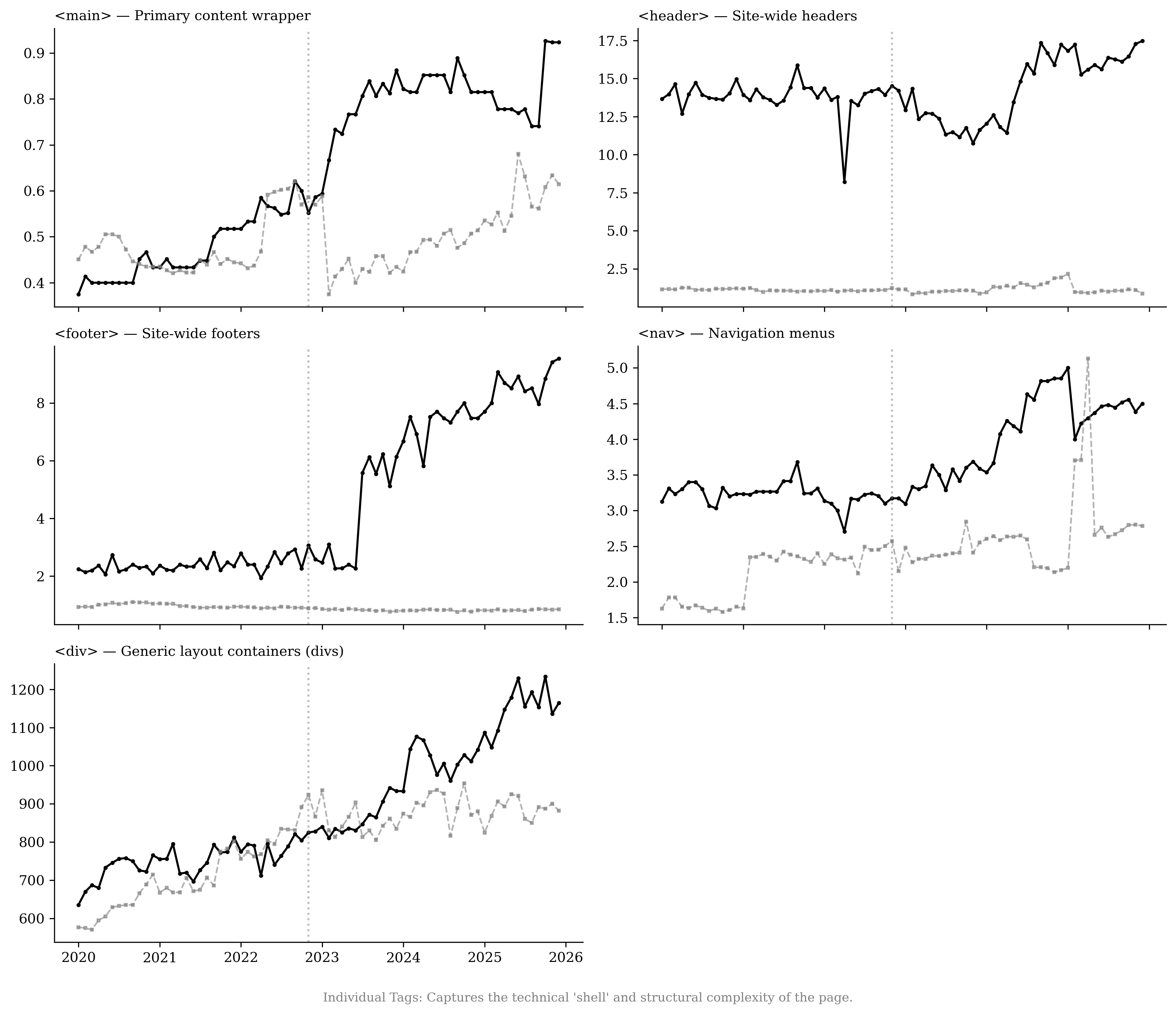}
\label{fig:Site_Framework_Layout_RAW}
 \end{figure}

 \begin{figure}[!ht]
     \centering
          \caption{Article Volume}
\includegraphics[width=0.95\linewidth]{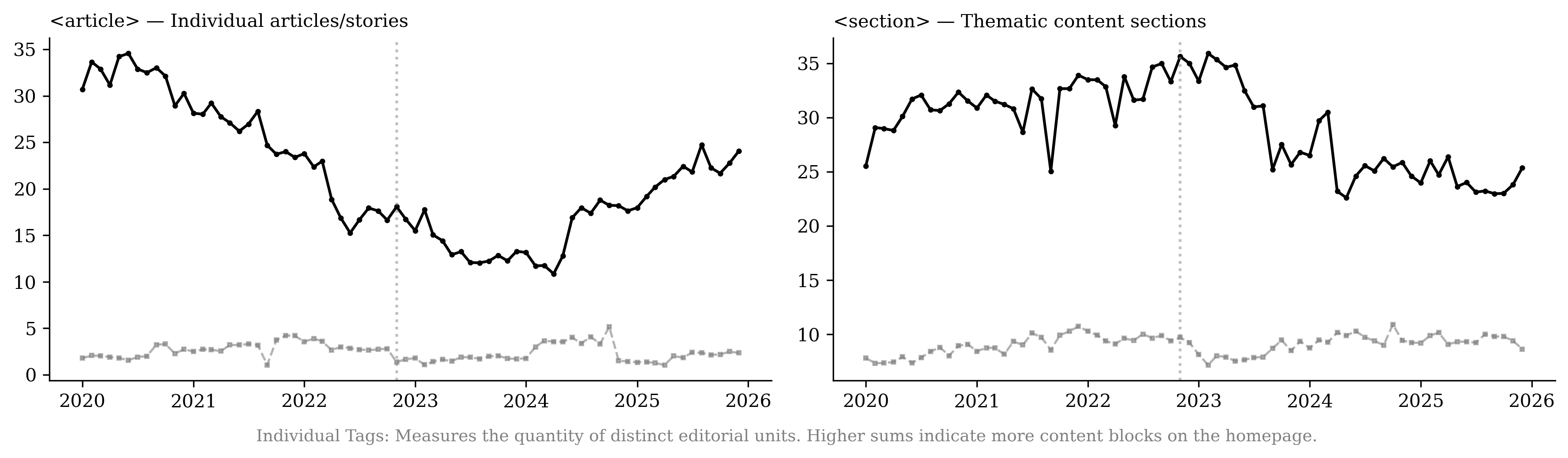}
     \label{fig:Semantic_Editorial_Density_RAW}
 \end{figure}

\begin{figure}[!ht]
    \centering
    \caption{Interactive Features}
\includegraphics[width=0.95\linewidth]{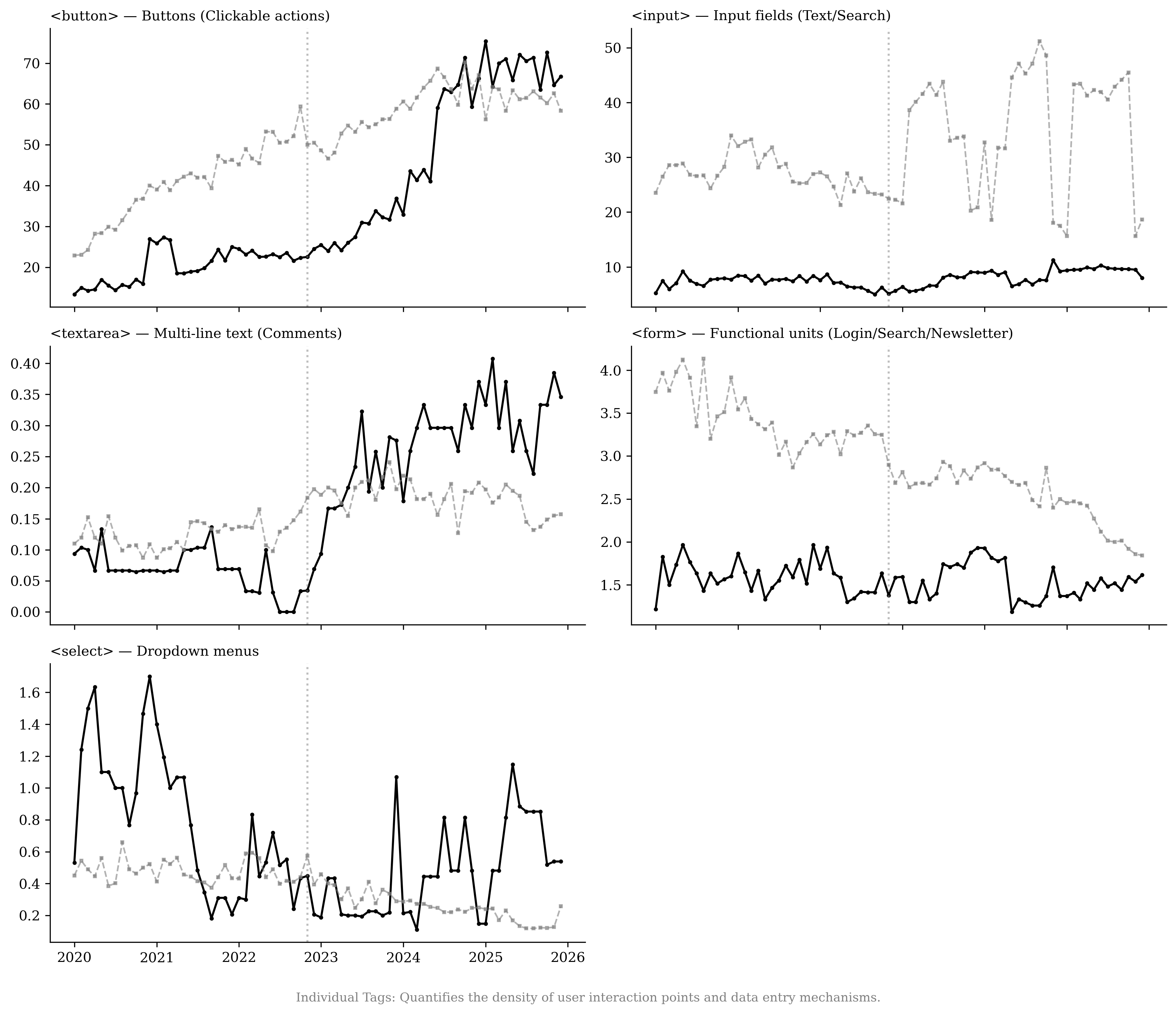}
    \label{fig:Interactive_Features}
\end{figure}

 \begin{figure}[!ht]
     \centering
          \caption{ Visual \& Media Richness  }
\includegraphics[width=0.95\linewidth]{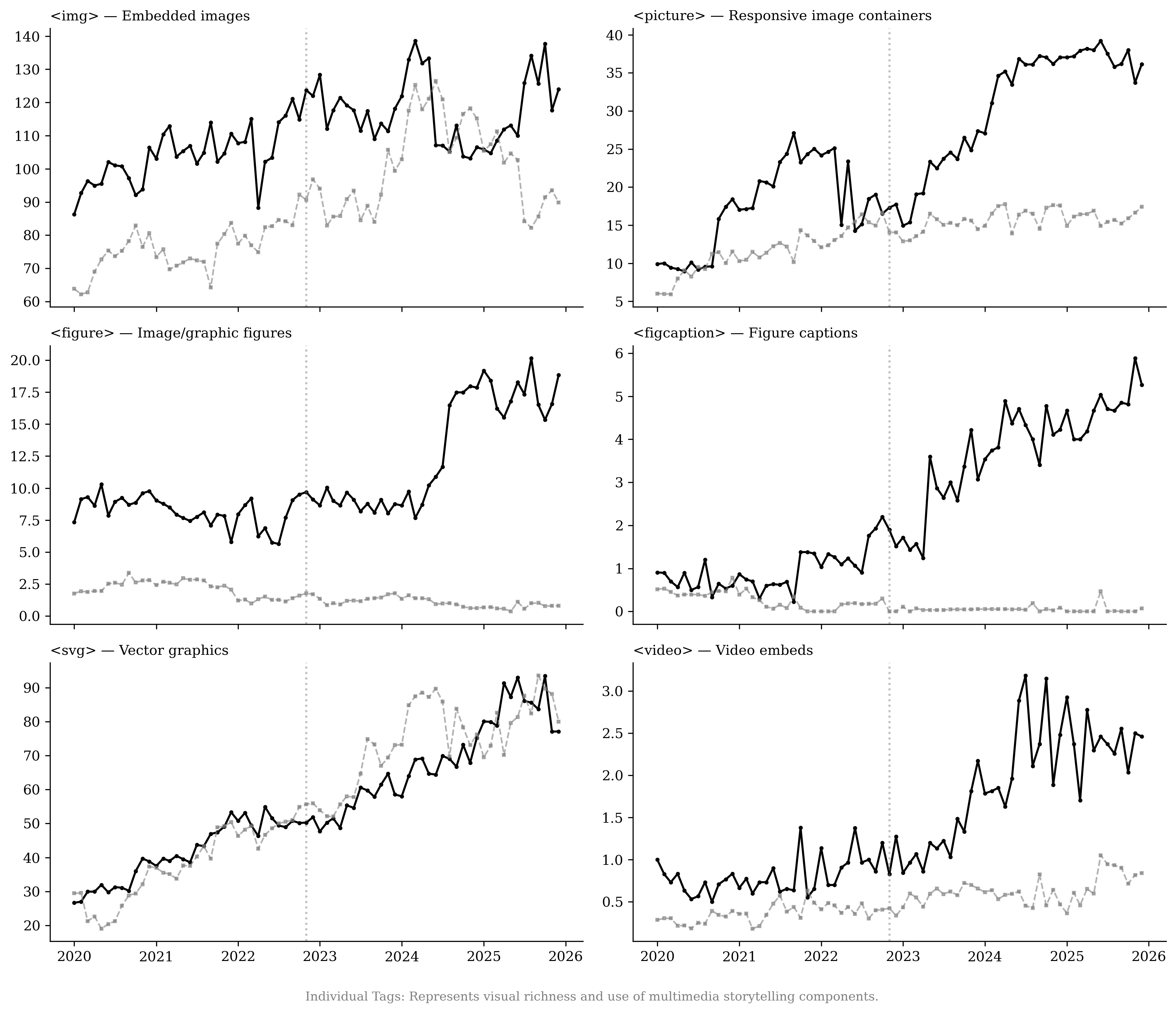}
     \label{fig:Visual_Media_Richness}
 \end{figure}
 
 \begin{figure}[!ht]
     \centering
          \caption{Ads \& Targeting}
\includegraphics[width=0.95\linewidth]{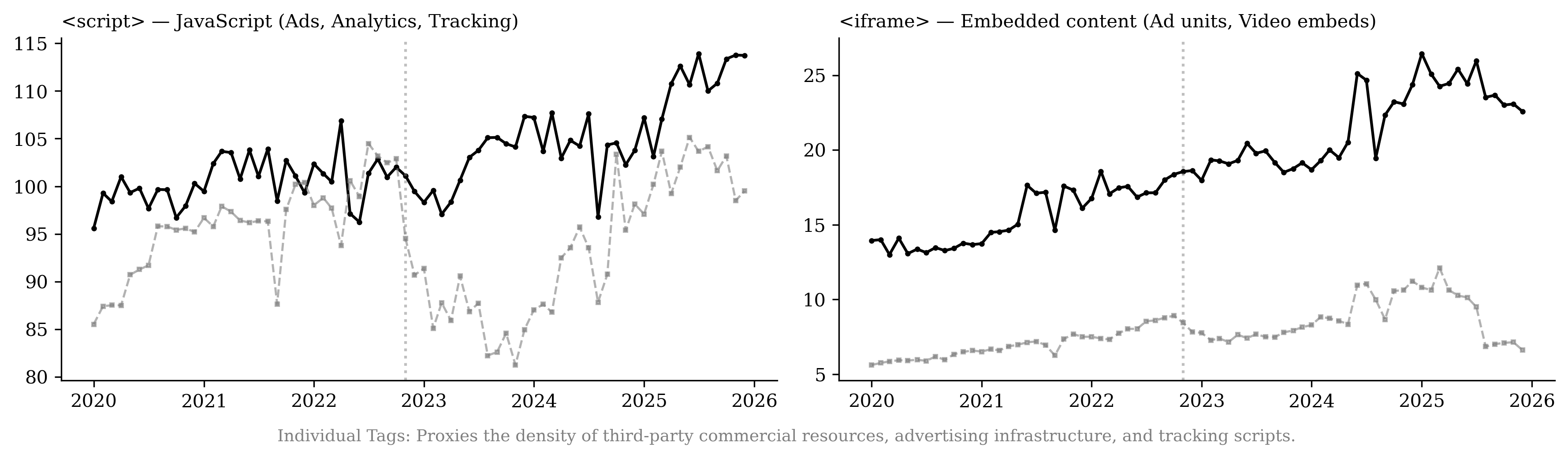}
     \label{fig:Commercial_Technical_Weight_RAW}
 \end{figure}

\begin{figure}[!ht]
\begin{center}
     \caption{Internet Archive’s Wayback Machine Content Type Unique URLs}
\label{fig:content_evolution_comparison}
    \begin{subfigure}{0.48\textwidth}
        \centering
        \caption{Text/HTML Unique URLs}
\includegraphics[width=\linewidth]{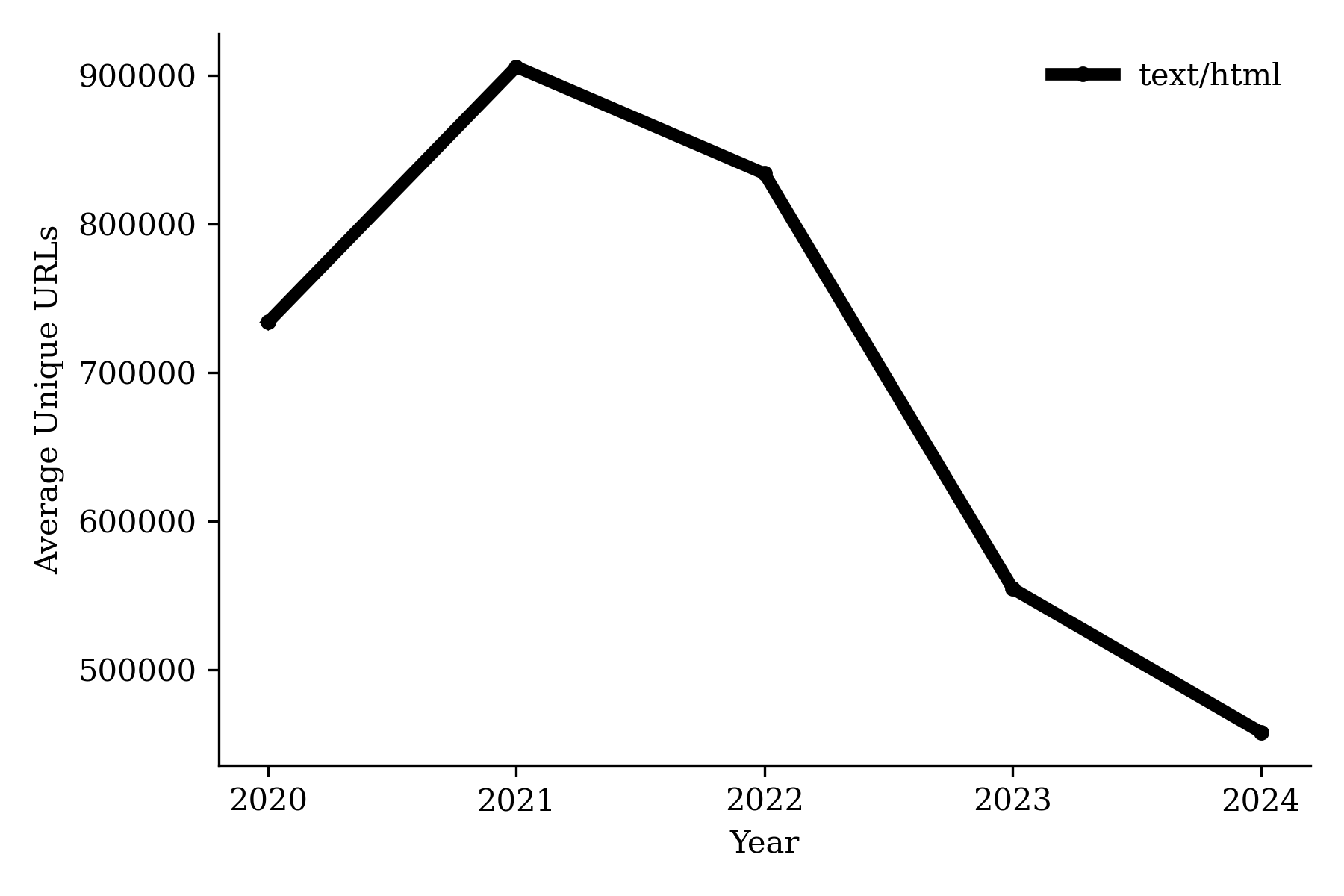}
        \label{fig:text_html_evolution}
    \end{subfigure}
    \hfill % Adds horizontal space between subfigures
    % Second Subfigure: Image/PNG Evolution
    \begin{subfigure}{0.48\textwidth}
        \centering
        \caption{Image/PNG Unique URLs}
\includegraphics[width=\linewidth]{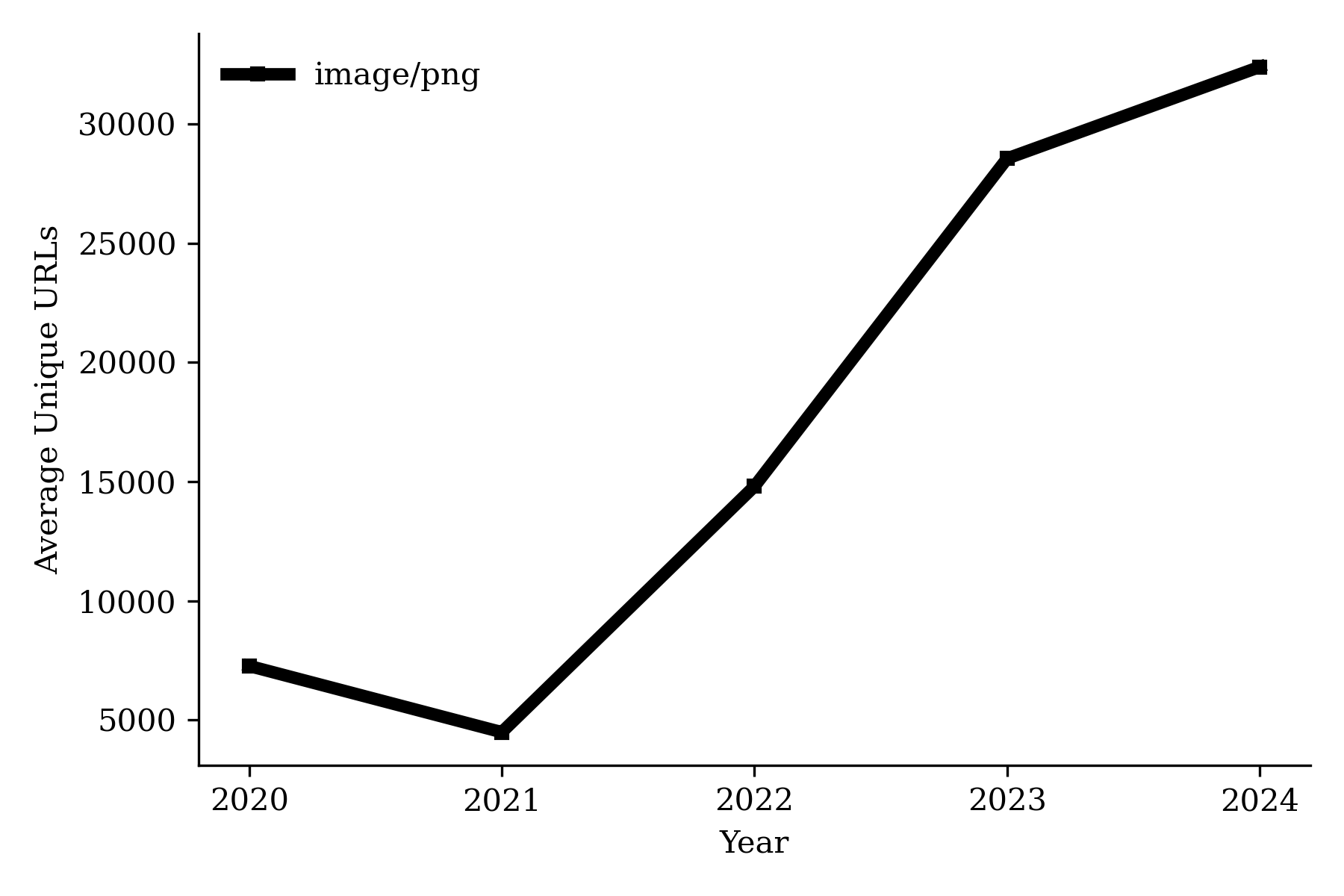}
        \label{fig:image_png_evolution}
    \end{subfigure}  
    \end{center}  
    \footnotesize
Subfigure (a) displays trends in unique URLs for text-based content, while subfigure (b) displays trends in image-based content from the Internet Archive’s Wayback Machine.
\end{figure}

\end{document}